\documentclass{aa}
\usepackage[varg]{txfonts}

\usepackage{natbib,twoopt}
\usepackage{bm}
\usepackage[breaklinks=true]{hyperref} 
\bibpunct{(}{)}{;}{a}{}{,}             
\makeatletter
  \newcommandtwoopt{\citeads}[3][][]{\href{http://adsabs.harvard.edu/abs/#3}%
    {\def\hyper@linkstart##1##2{}%
     \let\hyper@linkend\@empty\citealp[#1][#2]{#3}}}
  \newcommandtwoopt{\citepads}[3][][]{\href{http://adsabs.harvard.edu/abs/#3}%
    {\def\hyper@linkstart##1##2{}%
     \let\hyper@linkend\@empty\citep[#1][#2]{#3}}}
  \newcommandtwoopt{\citetads}[3][][]{\href{http://adsabs.harvard.edu/abs/#3}%
    {\def\hyper@linkstart##1##2{}%
     \let\hyper@linkend\@empty\citet[#1][#2]{#3}}}
  \newcommandtwoopt{\citeyearads}[3][][]%
    {\href{http://adsabs.harvard.edu/abs/#3}
    {\def\hyper@linkstart##1##2{}%
     \let\hyper@linkend\@empty\citeyear[#1][#2]{#3}}}
\makeatother

\usepackage{graphicx}
\usepackage{natbib}

\usepackage{booktabs}

\title{Abundance and temperature distributions in the hot intra-cluster gas of Abell\,4059}

\author{F. Mernier\inst{\ref{SRON},\ref{Leiden}} \and J. de Plaa\inst{\ref{SRON}} \and L. Lovisari\inst{\ref{Bonn}} \and C. Pinto\inst{\ref{Cambridge}} \and Y.-Y. Zhang\inst{\ref{Bonn}} \and J. S. Kaastra\inst{\ref{SRON},\ref{Leiden}} \and N. Werner\inst{\ref{Stanford1},\ref{Stanford2}} \and A. Simionescu \inst{\ref{Japan}}}

\institute{SRON Netherlands Institute for Space Research, Sorbonnelaan 2, 3584 CA Utrecht, The Netherlands \email{F.Mernier@sron.nl}\label{SRON} \and Leiden Observatory, Leiden University, P.O. Box 9513, 2300 RA Leiden,The Netherlands\label{Leiden} \and Argelander-Institut f\"{u}r Astronomie, Auf dem H\"{u}gel 71, D-53121 Bonn, Germany\label{Bonn} \and Institute of Astronomy, Madingley Road, CB3 0HA Cambridge, United Kingdom\label{Cambridge} \and Kavli Institute for Particle Astrophysics and Cosmology, Stanford University, 452 Lomita Mall, Stanford, CA 94305, USA\label{Stanford1} \and Department of Physics, Stanford University, 382 Via Pueblo Mall, Stanford,CA 94305-4060, USA\label{Stanford2} \and Institute of Space and Astronautical Science (ISAS), JAXA, 3-1-1 Yoshinodai, Chuo-ku, Sagamihara, Kanagawa, 252-5210 Japan \label{Japan}}

\date{Received 5 November 2014 / Accepted 13 December 2014}

\abstract{Using the EPIC and RGS data from a deep (200 ks) XMM-\textit{Newton} observation, we investigate the temperature structure ($kT$ and $\sigma_T$) and the abundances of nine elements (O, Ne, Mg, Si, S, Ar, Ca, Fe, and Ni) of the intra-cluster medium (ICM) in the nearby (z=0.046) cool-core galaxy cluster Abell 4059. Next to a deep analysis of the cluster core, a careful modelling of the EPIC background allows us to build radial profiles up to 12$'$ ($\sim$650 kpc) from the core. Probably because of projection effects, the temperature ICM is not found to be in single phase, even in the outer parts of the cluster. The abundances of Ne, Si, S, Ar, Ca, and Fe, but also O are peaked towards the core. The elements Fe and O are still significantly detected in the outermost annuli, which suggests that the enrichment by both type Ia and core-collapse SNe started in the early stages of the cluster formation. However, the particularly high Ca/Fe ratio that we find in the core is not well reproduced by the standard SNe yield models. Finally, 2-D maps of temperature and Fe abundance are presented and confirm the existence of a denser, colder, and Fe-rich ridge south-west of the core, previously observed by \textit{Chandra}. The origin of this asymmetry in the hot gas of the cluster core is still unclear, but it might be explained by a past intense ram-pressure stripping event near the central cD galaxy.}


\begin{document}

\maketitle

\titlerunning{An XMM-\textit{Newton} study of temperatures and abundances in Abell\,4059}
\authorrunning{F. Mernier et al.}

\section{Introduction}

The deep gravitational potential of clusters of galaxies retains large amounts of hot ($\sim$$10^7$--$10^8$ K) gas, mainly visible in X-rays, which accounts for no less than 80\% of the total baryonic mass. This so-called intra-cluster medium (ICM) contains not only H and He ions, but also heavier metals. Iron (Fe) was discovered in the ICM with the first generation of X-ray satellites \citep{1976MNRAS.175P..29M}; then neon (Ne), magnesium (Mg), silicon (Si), sulfur (S), argon (Ar), and calcium (Ca) were measured with \textit{ASCA} \citep[e.g.][]{1996ApJ...466..686M}. Precise abundance measurements of these elements have been made possible thanks to the good spectral resolution and the large effective area of the XMM-\textit{Newton} \citep{2001A&A...365L...1J} instruments \citep[e.g.][]{2001A&A...379..107T}. Nickel (Ni) abundance measurements and the detection of rare elements like chromium (Cr) have been reported as well \citep[e.g.][]{2006A&A...449..475W,2009ApJ...705L..62T}. Finally, thanks to its low and stable instrumental background, Suzaku is  capable of providing accurate abundance measurements in the cluster outskirts \citep[e.g.][]{2013Natur.502..656W}.

These metals clearly do not have a primordial origin; they are thought to be mostly produced by supernovae (SNe) within cluster galaxy members and have enriched the ICM mainly around $z \sim$ 2--3, i.e. during a peak of the star formation rate \citep{2006ApJ...651..142H}. However, the respective contributions of the different transport processes required to explain this enrichment are still under debate. Among them, galactic winds \citep{1978ApJ...223...47D,2009AN....330..898B} are thought to play the most important role in the ICM enrichment itself. Ram-pressure stripping \citep{1972ApJ...176....1G,2005A&A...435L..25S}, galaxy-galaxy interactions \citep{1998MNRAS.294..407G,2005A&A...438...87K}, AGN outflows \citep{2008A&A...482...97S,2009A&A...493..409S}, and perhaps gas sloshing \citep{2010MNRAS.405...91S} can also contribute to the redistribution of elements.
Studying the metal distribution in the ICM is a crucial step in order to understand and quantify the role of these mechanisms in the chemical enrichment of clusters.

Another open question is the relative contribution of SNe types producing each chemical element. While O, Ne, and Mg are thought to be produced mainly by core-collapse SNe \citep[SNcc, including types Ib, Ic, and II, e.g.][]{2006NuPhA.777..424N}, heavier elements like Ar, Ca, Fe, and Ni are probably produced mainly by type Ia SNe \citep[SNIa, e.g.][]{1999ApJS..125..439I}. The elements Si and S are produced by both types \citep[see][for a review]{2013AN....334..416D}. The abundances of high-mass elements highly depend on SNIa explosion mechanisms, while the abundances of the low-mass elements (e.g. nitrogen) are sensitive to the stellar initial mass function (IMF). Therefore, measuring accurate abundances in the ICM can help to constrain or even rule out some models and scenarios. Moreover, significant discrepancies exist between recent measurements and expectations from current favoured theoretical yields \citep[e.g.][]{2007A&A...465..345D}, and thus require  further investigation.

The temperature distribution in the ICM is often complicated and its underlying physics is not yet fully understood. For instance, many relaxed cluster cores are radiatively cooling on short cosmic timescales, which was presumed to lead to so-called cooling flows \citep[see][for a review]{1994ARA&A..32..277F}. However, the lack of cool gas (including the associated star formation) in the core revealed in particular by XMM-\textit{Newton} \citep{2001A&A...365L.104P,2001A&A...379..107T,2001A&A...365L..99K} leads to the so-called cooling-flow problem and argues for substantial heating mechanisms, yet to be found and understood. For example, heating by AGN could explain the lack of cool gas \citep[see e.g.][]{2007MNRAS.376.1547C}. Studying the spatial structure of the ICM temperature in galaxy clusters may help to solve it.

Abell 4059 is a good example of a nearby \citep[$z$=0.0460,][]{2002ApJ...567..716R} cool-core cluster. Its central cD galaxy hosts the radio source PKS 2354-35 which exhibits two radio lobes along the galaxy major axis (Taylor et al. 1994). In addition to \textit{ASCA} and \textit{ROSAT} observations \citep{1995AIPC..336..255O,1998ApJ...496..728H}, previous \textit{Chandra} studies \citep{2002ApJ...569L..79H,2004ApJ...606..185C,2008ApJ...679.1181R} show a ridge of additional X-ray emission located $\sim$20 kpc south-west of the core, as well as two X-ray ghost cavities that only partly coincide with the radio lobes. Moreover, the south-west  ridge has been found to be colder, denser, and  with a higher metallicity than the rest of the ICM, suggesting a past merging history of the core prior to the triggering of the AGN activity.

In this paper we analyse in detail two deep XMM-\textit{Newton} observations ($\sim$200 ks in total) of A\,4059, obtained through the CHEERS\footnote{CHEmical Evolution Rgs cluster Sample} project (de Plaa et al., in prep.). The XMM-\textit{Newton}  European Photon Imaging Camera (EPIC) instruments allow us to derive the abundances of O, Ne, Mg, Si, S, Ar, Ca, Fe, and Ni not only in the core, but also up to $\sim$650 kpc in the outer parts of the ICM. The XMM-\textit{Newton}  Reflection Grating Spectrometer (RGS) instruments are also used to measure N, O, Ne, Mg, Si, and Fe. This paper is structured as follows. The data reduction is described in Sect. \ref{sect:data_reduction}. We discuss our selected spectral models and our background estimation in Sect. \ref{sect:spectral_models}. We then present our temperature and abundance measurements in the cluster core, as well as their systematic uncertainties (Sect. \ref{sect:core}), measured radial profiles (Sect. \ref{sect:radial_profiles}), and temperature and Fe abundance maps (Sect. \ref{sect:maps}). We discuss and interpret our results in Sect. \ref{sect:discussion} and conclude in Sect. \ref{sect:conclusion}. Throughout this paper we assume $H_0 = 70$ km s$^{-1}$ Mpc$^{-1}$, $\Omega_m = 0.3,$ and $\Omega_\Lambda = 0.7$. At the redshift of 0.0460, 1 arcmin corresponds to $\sim$54 kpc. The whole EPIC field of view (FoV) covers $ R \simeq 0.81$ Mpc $\simeq 0.51 r_{200}$ \citep[][where $r_{200}$ is the radius within which the density of cluster reaches 200 times the critical density of the Universe]{2002ApJ...567..716R}. All the abundances are given relative to the proto-solar values from \citet{2009LanB...4B...44L}. The error bars indicate 1$\sigma$ uncertainties at a 68\% confidence level.
Unless mentioned otherwise, all our spectral analyses are done within 0.3--10 keV by using the Cash statistic \citep{1979ApJ...228..939C}.

\section{Observations and data reduction}\label{sect:data_reduction}
Two deep observations (DO) of A\,4059 were taken on  11 and 13 May 2013 with a gross exposure time of 96 ks and 95 ks respectively (here after DO\,1 and DO\,2). In addition to these deep observations, two shorter observations \citep[SO; see also][]{2011A&A...526A.105Z} are  available from the XMM-\textit{Newton} archive. The observations are summarised in Table \ref{table:observations}. Both DO and SO datasets are used for the RGS analysis while for the EPIC analysis we only use the DO datasets. In fact, the SO observations account for $\sim$20\% of the total exposure time, and consequently the signal-to-noise ratio $S/N$ would increase only by $\sqrt{1.20} \simeq 1.10,$ while the risk of including extra systematic errors and unstable fits due to the EPIC background components (Sect. \ref{sect:spectral_models} and Appendix \ref{sect:bg_modelling}) is high. The RGS extraction region is small, has a high $S/N$, and its background modelling is simpler than using EPIC; therefore, combining the DO and SO remains safe.

The datasets are reduced using the XMM-\textit{Newton} Science Analysis System (SAS) v13 and partly with the SPEX spectral fitting package \citep{1996uxsa.conf..411K} v2.04.

\subsection{EPIC}

In both DO datasets the MOS and pn instruments were operating in Full Frame mode and Extended Full Frame mode respectively. We reduce MOS\,1, MOS\,2 and pn data using the SAS tasks \texttt{emproc} and \texttt{epproc}. Next, we filter  our data to exclude soft-proton (SP) flares by building appropriate Good Time Intervals (GTI) files (Appendix \ref{subsect:gti_filtering}) and we excise visible point sources to keep the ICM emission only (Appendix \ref{subsect:rps}). We keep the single, double, triple, and quadruple events in MOS (\texttt{pattern$\le$12}). Owing to problems regarding charge transfer inefficiency for the double events in the pn detector\footnote{See the XMM-\textit{Newton} Current Calibration File Release Notes, XMM-CCF-REL-309 (Smith, Guainazzi \& Saxton 2014)}, we keep only single events in pn (\texttt{pattern$=$0}). We remove out-of-time events from both images and spectra. After the screening process, the EPIC total net exposure time is $\sim$150 ks (i.e. $\sim$$80\%$ of the initial observing time). In addition to EPIC MOS\,1 CCD3 and CCD6 which are no longer operational, CCD4 shows obvious signs of deterioration, so we discard its events from both datasets as well.

\begin{figure}
\resizebox{\hsize}{!}{
\includegraphics[trim=0.4cm 0.2cm 1.9cm 0cm, clip=true, width=\textwidth]{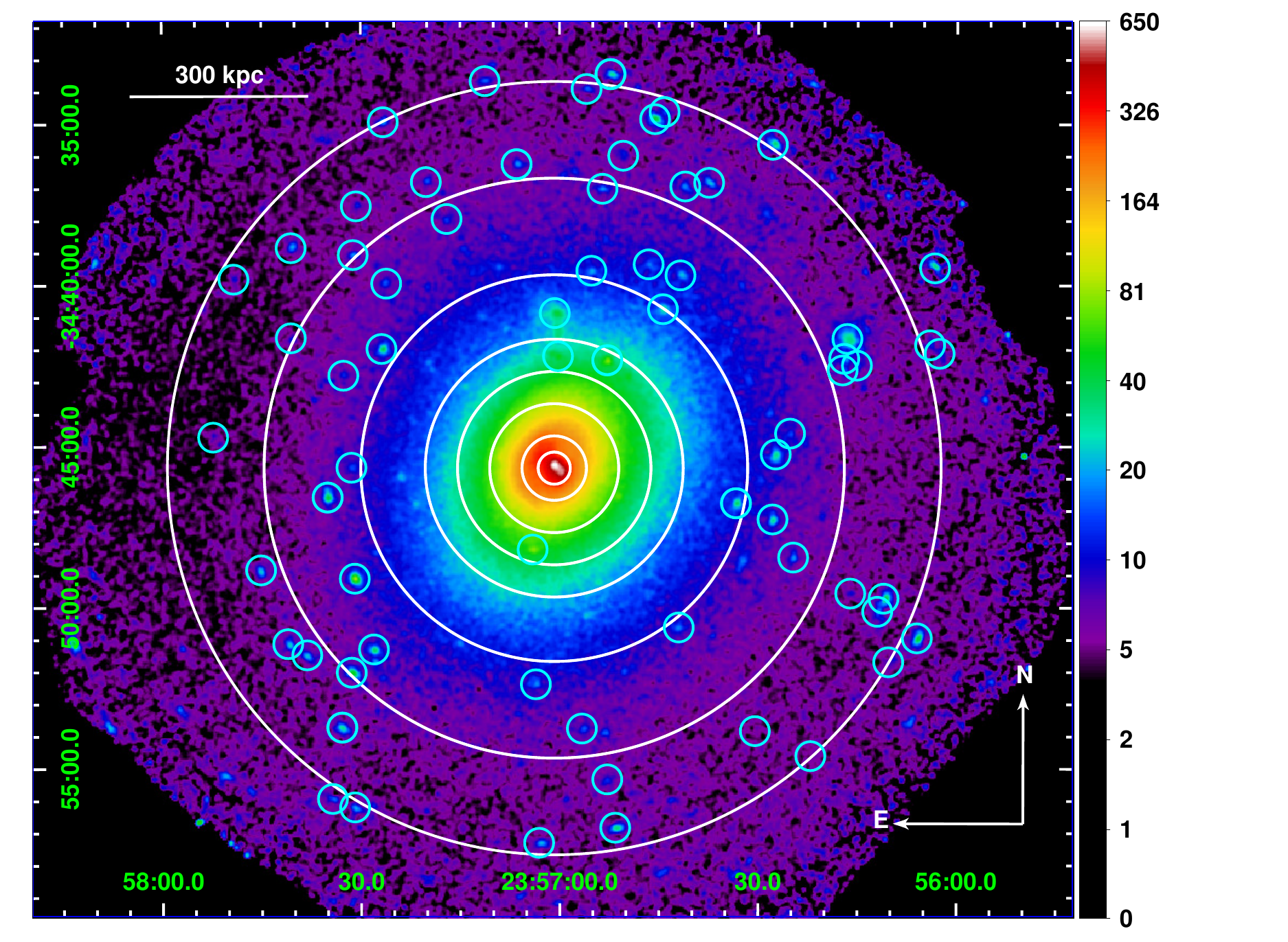}}
\caption{Exposure map corrected EPIC combined image of A\,4059, in units of number of counts. The two datasets have been merged. The cyan circles show the detected resolved point sources that we excise from our analysis. For clarity of display the radii shown here are exaggerated (excision radius $= 10''$, see Appendix \ref{subsect:rps}). The white annuli show the extraction regions that are used for our radial studies (see text and Sect. \ref{sect:radial_profiles}).}
\label{fig:A4059}
\end{figure}

\begin{figure}
\resizebox{\hsize}{!}{
\includegraphics[trim=0.4cm 0.2cm 1.9cm 0cm, clip=true, width=\textwidth]{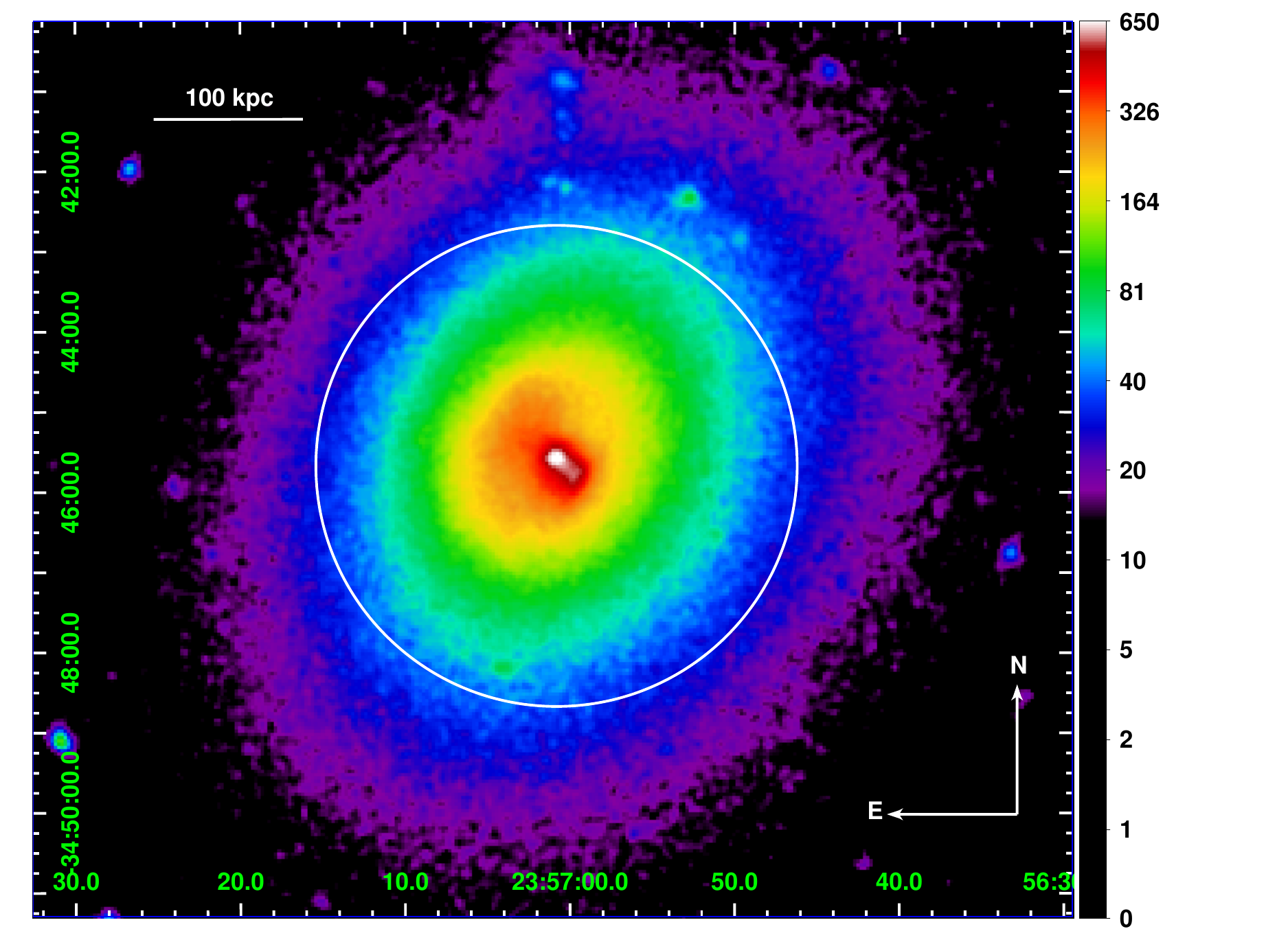}}
\caption{Close-up view from Fig. \ref{fig:A4059}, centred on the cluster core. The white circle delimitates the core region analysed in Sect. \ref{sect:core}.}
\label{fig:A4059bis}
\end{figure}

Figures \ref{fig:A4059} and \ref{fig:A4059bis} show an exposure map corrected combined EPIC image of our full filtered dataset (both detectors cover the full EPIC FoV). The peak of the X-ray emission is seen at $\sim$23h\,57$'$\,0.8$''$ RA, -34\degr\,45$'$\,34$''$ DEC.

\begin{table}
\begin{centering}
\caption{Summary of the observations of Abell\,4059. We report the total exposure time together with the net exposure time remaining after screening of the flaring background.}             
\label{table:observations}
\resizebox{\hsize}{!}{
\begin{tabular}{c l c l c c}        
\hline \hline                
ID & Obs. number & Date & Instrument & Total time & Net time \\    
 &    &      &  &  (ks)      &  (ks)  \\
\hline                        
SO\,1 &   0109950101  & 2000 11 24 & RGS & 29.3  & 20.0 \\ 
\hline
SO\,2 &   0109950201  & 2000 11 24 & RGS & 24.7  & 23.4 \\
\hline
DO\,1 &   0723800901  & 2013 05 11 & EPIC MOS\,1 & 96.4  & 71.0 \\ 
&      &  & EPIC MOS\,2 & 96.4  & 73.0 \\ 
&     &  & EPIC pn & 93.8  & 51.7 \\ 
&     &  & RGS & 97.1  & 77.1 \\ 
\hline
DO\,2 &   0723801001  & 2013 05 13 & EPIC MOS\,1 & 94.7  & 76.4 \\
&     &  & EPIC MOS\,2 & 94.7  & 77.5 \\
&     &  & EPIC pn & 92.9  & 66.4 \\
&     &  & RGS & 96.1  & 87.9 \\
\hline                                   
\end{tabular}}
\par\end{centering}
\end{table}

We extract the EPIC spectra of the cluster core from a circular region centred on the X-ray peak emission and with a radius of 3 arcmin (Fig. \ref{fig:A4059bis}). Using the same centre we  extract the spectra of eight concentric annuli, together covering the FoV within  $R \le 12$ arcmin (Fig. \ref{fig:A4059}). The core region corresponds to the four innermost annuli. The RMFs and ARFs are processed using the SAS tasks \texttt{rmfgen} and \texttt{arfgen,} respectively.
In order to look at possible substructures in temperature and metallicity, we also create EPIC maps. We divide our EPIC observations in spatial cells using the Weighted Voronoi Tesselations (WVT) adaptive binning algorithm \citep{2006MNRAS.368..497D}. We restrict the size of our full maps to $R \le 6$ arcmin. The cell sizes are defined in such a way that in every cell $S/N = 100$. The relative errors of the measured temperature and Fe abundance are then expected to be not higher than $\sim$$5\%$ and $\sim$$20\%,$ respectively (see Appendix \ref{sect:maps_simulations} for more details). Because SAS does not allow RMFs and ARFs to be processed for complex geometry regions, we extract them on 10$\times$10 square regions covering together our whole map and we attribute the raw spectra of each cell to the response files of its closest square region.
The spectra and response files are converted into SPEX format using the auxiliary program \textit{trafo}.

\subsection{RGS}

Reflection Grating Spectrometer data of all four observations are used \citep[see Table \ref{table:observations} and also][for details]{2015arXiv150101069P}. The RGS detector is centred on the cluster core and its dispersion direction extends from the north-east to the south-west.
We process RGS data with the SAS task \texttt{rgsproc}. We correct for contamination from SP flares by using the data from CCD9, where hardly any emission from the source is expected. We build the GTI files similarly to the EPIC analysis (Appendix \ref{subsect:gti_filtering}) and we process the data again with \texttt{rgsproc} by filtering the events with these GTI files. The total RGS net exposure time is 208.4 ks.
We extract response matrices and RGS spectra for the observations. The final net exposure times are given in Table \ref{table:observations}.

We subtract a model background spectrum created by the standard RGS pipeline from the total spectrum.
This is a template background file, based on the count rate in CCD9 of RGS.

We combine the RGS\,1 and RGS\,2 spectra, responses and background files of the four observations through the SAS task \texttt{rgscombine} obtaining one stacked spectrum for spectral order\,1 and one for order\,2. The two combined spectra are converted to SPEX format through \textit{trafo}. Based on the MOS\,1 image, we correct the RGS spectra for instrumental broadening as described in Appendix \ref{sect:RGS_M1image}. We include 95\% of the cross-dispersion direction in the spectrum.

\section{Spectral models}
\label{sect:spectral_models}

The spectral analysis is done using SPEX. Since there is an important offset in the pointing of the two observations, stacking the spectra and the response files of each of them may lead to bias in the fittings. Moreover, the remaining SP component is found to change from one observation to another (see Appendix \ref{sect:bg_modelling}). Therefore, the better option is to fit simultaneously the single spectra of every EPIC instrument and observation. This has been done using \textit{trafo}.

\subsection{\textit{CIE}}
\label{subsect:CIE}

We assume that the ICM is in collisional ionisation equilibrium (CIE) and we use the \textit{CIE} model in our fits (see the SPEX manual\footnote{http://www.sron.nl/spex}). Our emission models are corrected from the cosmological redshift and are absorbed by the interstellar medium of the Galaxy \citep[for this pointing $N_H \simeq 1.26 \times 10^{20}$ cm$^{-2}$ as obtained with the method of][]{2013MNRAS.431..394W}.
The free parameters in the fits are the emission measure $Y = \int n_e n_\text{H} dV$, the single-temperature $kT$, and O, Ne, Mg, Si, S, Ar, Ca, Fe, and Ni abundances. The other abundances with an atomic number $Z \ge 6$ are fixed to the Fe value.

\subsection{\textit{GDEM}}
\label{subsect:GDEM}

Although \textit{CIE} single-temperature models (i.e. isothermal) fit the X-ray spectra from the ICM reasonably well, previous papers \cite[see e.g.][]{2003ApJ...590..207P,2004A&A...413..415K,2006A&A...449..475W,2006A&A...452..397D,2009A&A...493..409S} have shown that employing a distribution of temperatures in the models provides significantly better fits, especially in the cluster cores. The strong temperature gradient in the case of cooling flows and the 2-D projection of the supposed spherical geometry of the ICM suggest that using multi-temperature models would be preferable. Apart from the \textit{CIE} model mentioned above, we also fit a Gaussian differential emission measure (\textit{GDEM}) model to our spectra. This model assumes that the emission measure $Y$ follows a Gaussian temperature distribution centred on $kT_\text{mean}$ and as defined by

\begin{equation} \label{eq:gdem}
Y(x) = \frac{Y_0}{\sigma_{T} \sqrt{2 \pi}} \exp(\frac{(x-x_\text{mean})^2}{2 \sigma^2_{T}})
,\end{equation}
where $x=\log (kT)$ and $x_\text{mean}=\log (kT_\text{mean})$ \citep[see][]{2006A&A...452..397D}.
Compared to the \textit{CIE} model, the additional free parameter from the \textit{GDEM} model is the width of the Gaussian emission measure profile $\sigma_{T}$. By definition $\sigma_T$=0 is the isothermal case.

\subsection{Cluster emission and background modelling}

We fit the spectra of the cluster emission with a \textit{CIE} and a \textit{GDEM} model successively, except for the EPIC radial profiles and maps, where only a \textit{GDEM} model is considered.

Since the EPIC cameras are highly sensitive to the particle background, a precise estimate of the local background is crucial in order to estimate ICM parameters beyond the core (i.e. where this background is comparable to the cluster emission). The emission of A\,4059  entirely fills the EPIC FoV, making a direct measure of the local background impossible. Some efforts have been made in the past to deal with this problem \citep[see e.g.][]{2009ApJ...699.1178Z,2011A&A...526A.105Z,Snowden_Kuntz}, but a customised procedure based on full modelling is more convenient in our case. In fact, an incorrect subtraction of instrumental fluorescence lines might lead to incorrect abundance estimates.

For each extraction region, several background components are modelled in the EPIC spectra in addition to the cluster emission. This modelling procedure and its application to our extracted regions are fully described in Appendix \ref{sect:bg_modelling}. We note that we do not explicitly model the cosmic X-ray background in RGS (although we did in EPIC) because any diffuse emission feature would be smeared out into a broad continuum-like component.

\section{Cluster core}
\label{sect:core}

\subsection{EPIC}\label{subsect:core_EPIC}

\begin{figure*}
        \centering
                \includegraphics[trim=1cm 0cm 0cm 0cm, clip=true, width=0.49\textwidth]{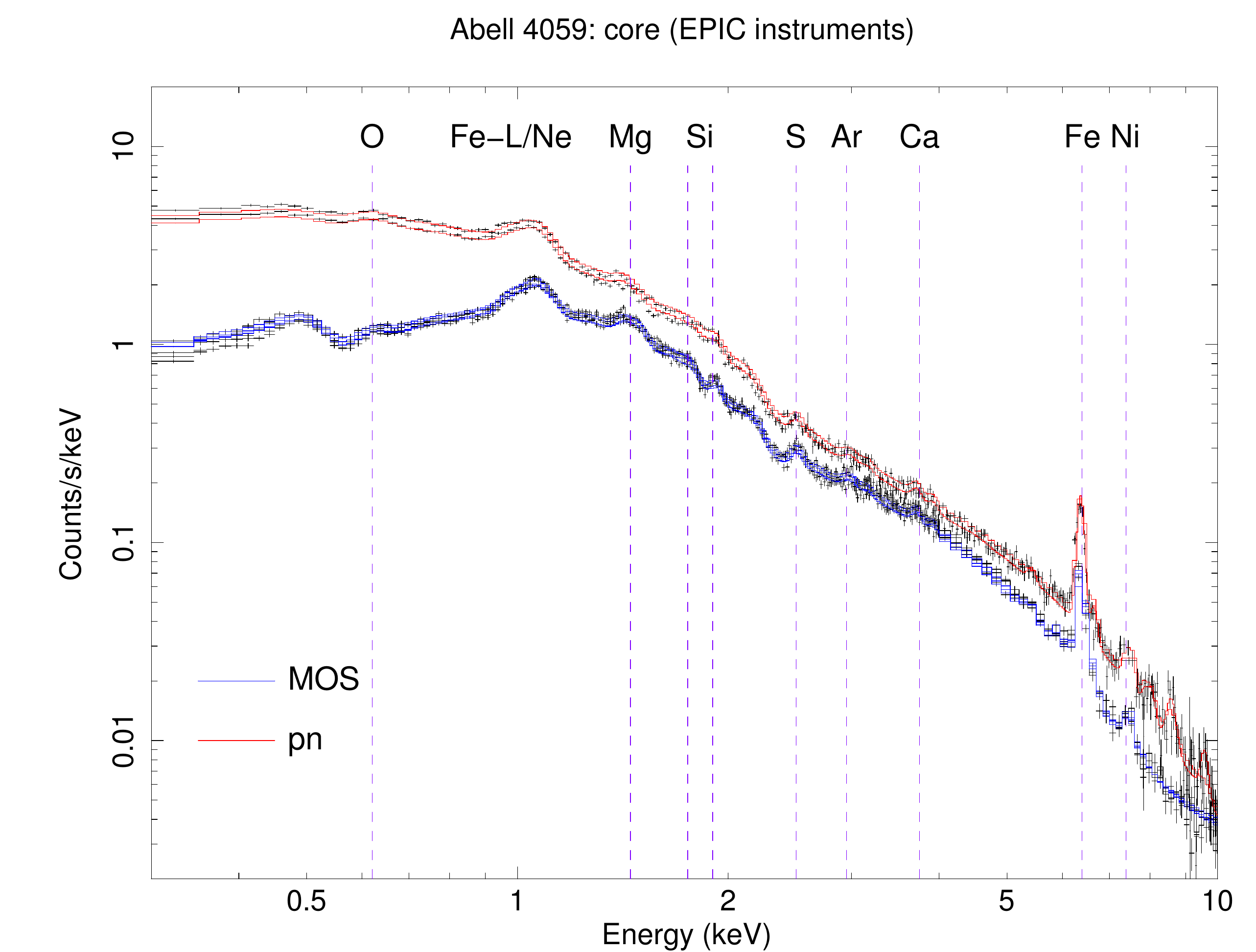}
                \includegraphics[trim=1cm 0cm 0cm 0cm, clip=true, width=0.49\textwidth]{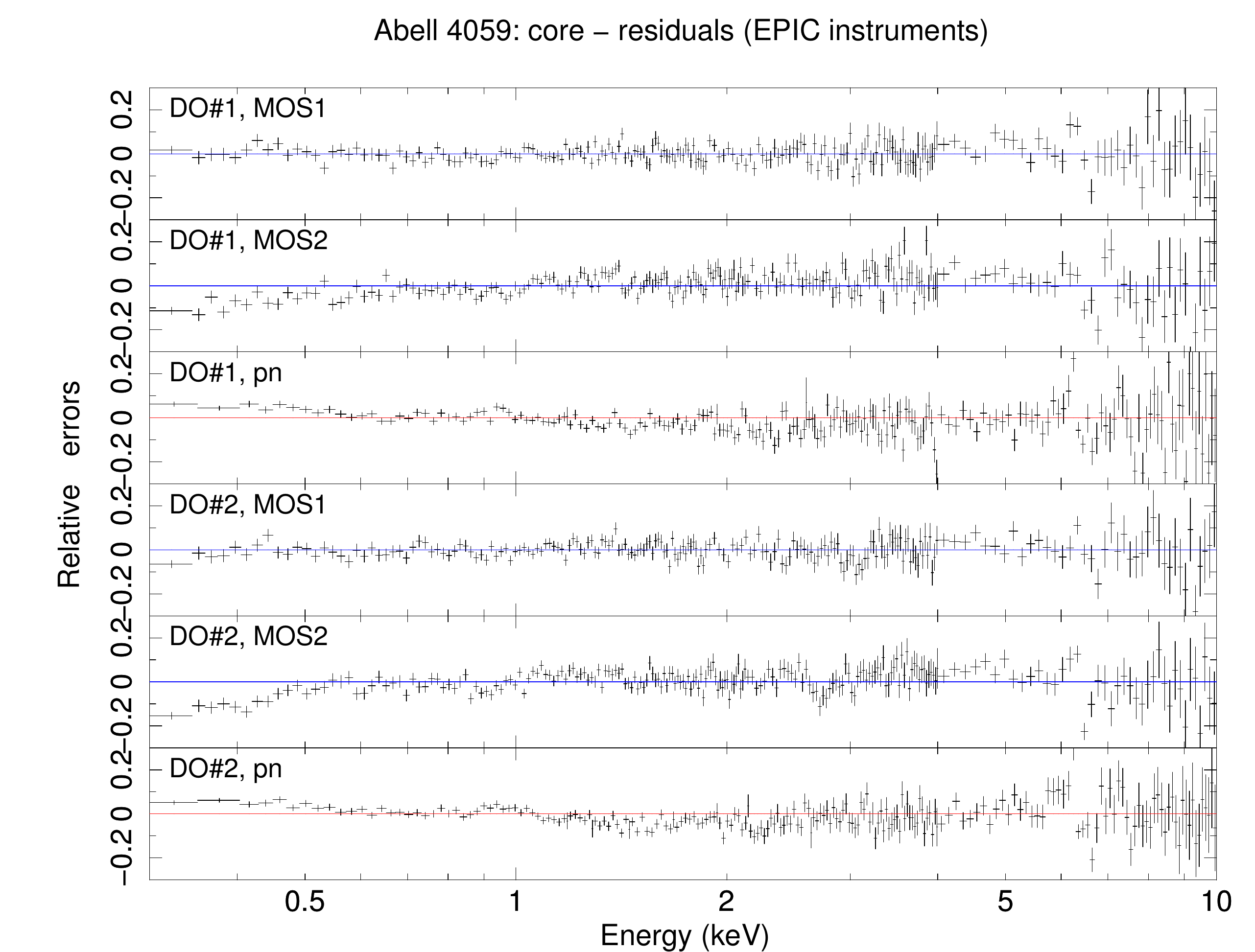}
        \caption{EPIC spectra (left) and residuals (right) of the core region (0$'$--3$'$) of Abell 4059. The two observations are displayed and fitted simultaneously with a \textit{GDEM} model. For clarity of display the data are rebinned above 4 keV by a factor of 10 and 20 in MOS and pn spectra, respectively.}
\label{fig:core_spectra}
\end{figure*}

Our deep exposure time allows us to get precise abundance measurements in the core, even using EPIC (Fig. \ref{fig:core_spectra} left). Moreover, the background is very limited since the cluster emission clearly dominates. Table \ref{table:core2} shows our results, both for the combined fits (MOS+pn) and independent fits (either MOS or pn only).

Using a multi-temperature model clearly improves the combined MOS+pn fit. Nevertheless, even by using a \textit{GDEM} model, the reduced C-stat value is still high because the excellent statistics of our data reveal anti-correlated residuals between MOS and pn, especially below $\sim$1 keV (Fig. \ref{fig:core_spectra}, right).

\begin{table*}
\caption{Best-fit parameters measured in the cluster core (circular region, $R \sim 3$ arcmin). A single-temperature (\textit{CIE}) and a multi-temperature (\textit{GDEM}) model have been successively fitted.}
\label{table:core2}
\begin{centering}
\setlength{\tabcolsep}{15pt}
\begin{tabular}{l l c@{$\pm$}l c@{$\pm$}l c@{$\pm$}l}
\hline
\hline
Parameter & Model & \multicolumn{2}{c}{MOS+pn} & \multicolumn{2}{c}{MOS only} & \multicolumn{2}{c}{pn only}\tabularnewline
\hline
C-stat / d.o.f. & \textit{CIE} & \multicolumn{2}{c}{$3719/1781$} & \multicolumn{2}{c}{$1904/1221$} & \multicolumn{2}{c}{$1109/546$} \tabularnewline
                     & \textit{GDEM} & \multicolumn{2}{c}{$3331/1780$} & \multicolumn{2}{c}{$1703/1220$} &  \multicolumn{2}{c}{$969/545$}  \tabularnewline
\hline

$Y$ ($10^{70}$ m$^{-3}$) & \textit{CIE} & $806$ & $3$ & $779.7$ & $1.8$ & $827$ & $3$ \tabularnewline
                                                & \textit{GDEM} & $821$ & $3$ & $792$ & $3$ & $845$ & $4$ \tabularnewline
\hline

$kT$ (keV) & \textit{CIE} & $3.696$ & $0.012$ & $3.837$ & $0.015$ & $3.431$ & $0.18$ \tabularnewline
\hline
$kT_\text{mean}$ (keV) & \textit{GDEM} & $3.838$ & $0.016$ & $4.03$ & $0.02$ & $3.58$ & $0.03$ \tabularnewline
$\sigma_T$ &  & $0.261$ & $0.004$ & $0.266$ & $0.007$ & $0.251$ & $0.008$ \tabularnewline
\hline

O & \textit{CIE} & $0.49$ & $0.03$ & $0.57$ & $0.04$ & $0.34$ & $0.03$ \tabularnewline
    & \textit{GDEM} & $0.46$ & $0.04$ & $0.57$ & $0.04$ & $0.33$ & $0.04$ \tabularnewline
\hline

Ne & \textit{CIE} & $1.08$ & $0.04$ & $1.09$ & $0.04$ & $1.05$ & $0.05$ \tabularnewline
      & \textit{GDEM} & $0.33$ & $0.05$ & $0.34$ & $0.06$ & $0.36$ & $0.08$ \tabularnewline
\hline

Mg & \textit{CIE} & $0.45$ & $0.04$ & $0.82$ & $0.05$ & \multicolumn{2}{c}{$< 0.04$} \tabularnewline
      & \textit{GDEM} & $0.45$ & $0.03$ & $0.78$ & $0.05$ & \multicolumn{2}{c}{$< 0.08$} \tabularnewline
\hline

Si & \textit{CIE} & $0.49$ & $0.02$ & $0.64$ & $0.03$ & $0.32$ & $0.03$ \tabularnewline
     & \textit{GDEM} & $0.51$ & $0.02$ & $0.66$ & $0.03$ & $0.35$ & $0.03$ \tabularnewline
\hline

S & \textit{CIE} & $0.46$ & $0.03$ & $0.61$ & $0.04$ & $0.25$ & $0.05$ \tabularnewline
    & \textit{GDEM} & $0.52$ & $0.03$ & $0.66$ & $0.04$ & $0.31$ & $0.05$ \tabularnewline
\hline

Ar & \textit{CIE} & $0.27$ & $0.07$ & $0.17$ & $0.15$ & $0.35$ & $0.14$ \tabularnewline
     & \textit{GDEM} & $0.41$ & $0.08$ & $0.30$ & $0.11$ & $0.54$ & $0.15$ \tabularnewline
\hline

Ca & \textit{CIE} & $0.89$ & $0.09$ & $0.91$ & $0.11$ & $0.78$ & $0.15$ \tabularnewline
      & \textit{GDEM} & $1.01$ & $0.10$ & $0.98$ & $0.13$ & $0.90$ & $0.15$ \tabularnewline
\hline

Fe & \textit{CIE} & $0.740$ & $0.008$ & $0.851$ & $0.009$ & $0.624$ & $0.009$ \tabularnewline
      & \textit{GDEM} & $0.697$ & $0.006$ & $0.803$ & $0.010$ & $0.600$ & $0.010$ \tabularnewline
\hline

Ni & \textit{CIE} & $1.04$ & $0.08$ & $1.86$ & $0.11$ & $0.34$ & $0.11$ \tabularnewline
     & \textit{GDEM} & $1.04$ & $0.07$ & $1.83$ & $0.11$ & $0.37$ & $0.10$ \tabularnewline
\hline
\end{tabular}
\par\end{centering}
\end{table*}

\begin{figure*}
        \centering
                \includegraphics[trim=1cm 0cm 0cm 0cm, clip=true, width=0.48\textwidth]{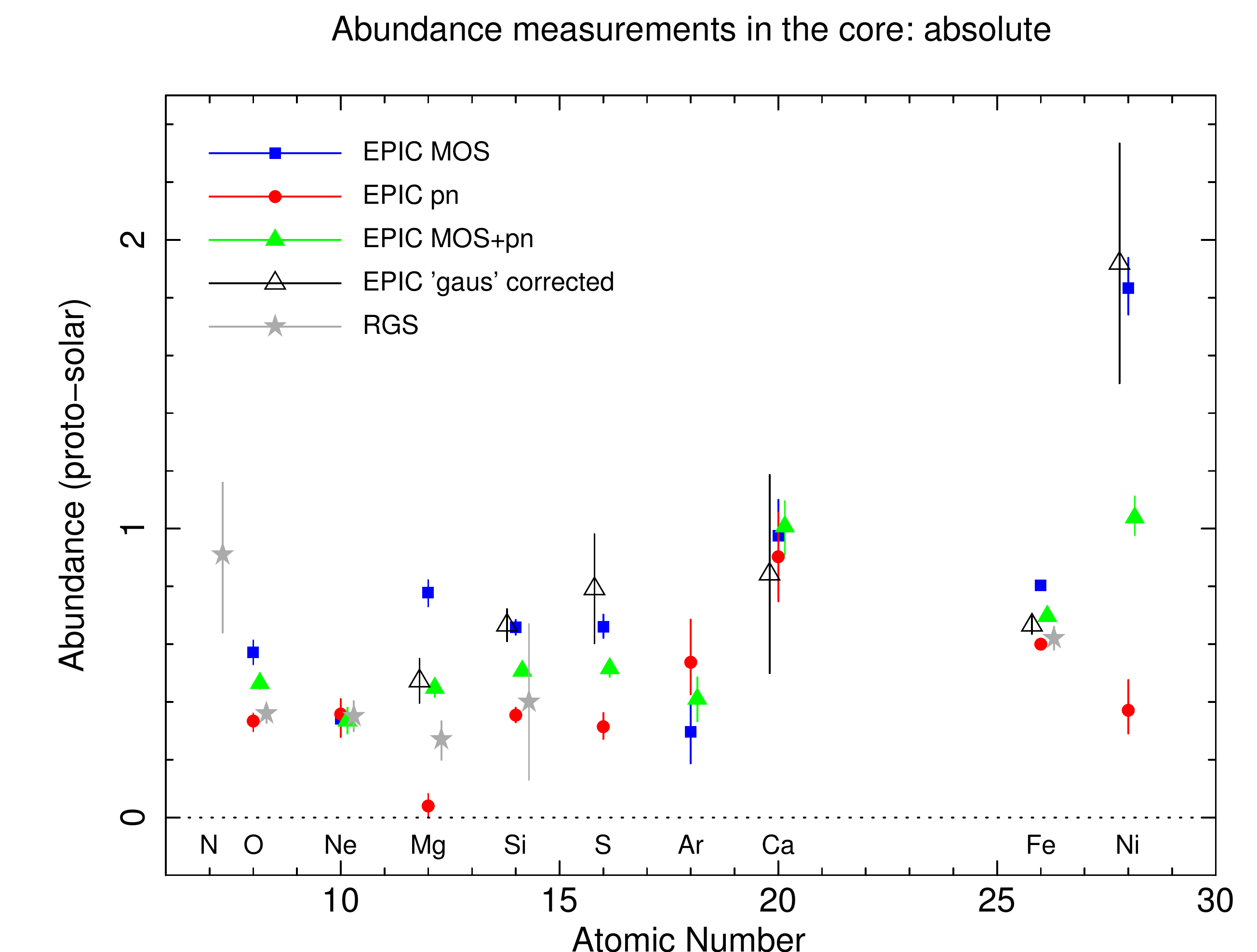}
                \includegraphics[trim=1cm 0cm 0cm 0cm, clip=true, width=0.48\textwidth]{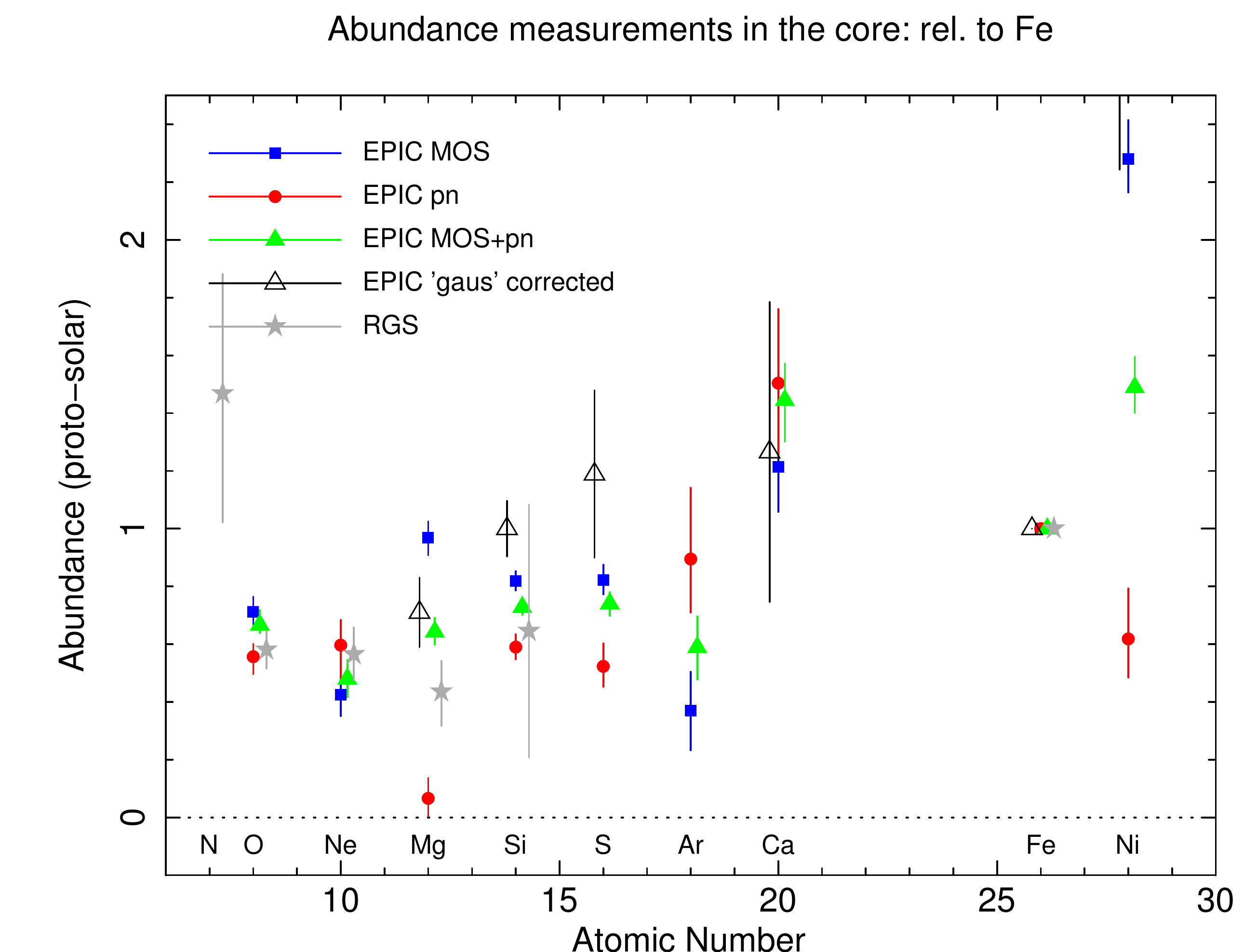}
        \caption{EPIC and RGS abundance measurements in the core of A\,4059. Left: absolute abundances. Right: Abundances relative to Fe. The black empty triangles show the mean MOS+pn abundances obtained by fitting Gaussian lines instead of the CIE models (the Gauss method; see text and Table \ref{table:EWs}). The numerical values are summarised in Table \ref{table:core_abundances}.}\label{fig:core_abundances}
\end{figure*}

When we fit the EPIC instruments independently, the reduced C-stat number decreases from 1.87 to 1.40 and 1.78 in the MOS and pn fits, respectively. Visually, the models reproduce  the spectra better as well. We also note that the temperature and abundances measurements in the core are different between the instruments (Table \ref{table:core2}). While temperature discrepancies between MOS and pn have been already reported and investigated \citep{2014arXiv1404.7130S}, here we focus on the MOS-pn abundance discrepancies. Figure \ref{fig:core_abundances} (left) illustrates these values and shows the absolute abundance measurements obtained from our \textit{GDEM} models. Except for Ne, Ar, and Ca (all consistent within 2$\sigma$), we observe systematically higher values in MOS than in pn.
Assuming (for convenience) that the systematic errors are roughly in a Gaussian distribution, we can estimate them for different abundance measurements $Z_\text{MOS}$ and $Z_\text{pn}$, having their respective statistical errors $\sigma_\text{MOS}$ and $\sigma_\text{pn}$,
\begin{equation} \label{eq:systematics}
\sigma_\text{sys} = \sqrt{ \sigma_\text{tot}^2 - \frac{\sigma_\text{MOS}^2 + \sigma_\text{pn}^2}{2} }
,\end{equation}
where $\sigma_\text{tot} = \sqrt{((Z_\text{MOS}-\mu)^2 + (Z_\text{pn}-\mu)^2)/2}$ and $\mu = (Z_\text{MOS} + Z_\text{pn})/2$.
 We obtain absolute O, Si, S, and Fe systematic errors of $\pm 25\%$, $\pm 30 \%$, $\pm 34 \%,$ and $\pm 14\%$ respectively. The MOS-pn discrepancies in Mg and Ni are too big to be estimated as reasonable systematic errors (Fig. \ref{fig:core_abundances}). No systematic errors are necessary for the absolute abundances of Ne, Ar, and Ca.

If we normalise the abundances relative to Fe in each instrument (Fig. \ref{fig:core_abundances}, right panel), O/Fe is consistent within 2$\sigma$ and Si/Fe and S/Fe within 3$\sigma$. Inversely, the discrepancies on Ar/Fe measurements slightly increase, but their statistical uncertainties are quite large because the main line ($\sim$3.1 keV) is weak. We note that the discrepancies in Mg and Ni measurements remain huge and almost unchanged. Based on the same method as above, we find that systematic errors of O/Fe, Si/Fe, and S/Fe are reduced to $\pm 8\%$, $\pm 15\%,$ and $\pm 20\%$ while the systematic errors of Ar/Fe increase to $\pm 27\%$.

\subsubsection{Equivalent widths}\label{sect:EWs}

One way of determining the origin of the discrepancies in the fitted abundance from different instruments is to derive the abundances using a more robust approach. Instead of fitting the abundances using the \textit{GDEM} model directly, we model each main emission line/complex by a Gaussian and a local continuum (hereafter the Gauss method). The \textit{GDEM} model is still used to fit the local continuum; however, only the Fe abundance is kept to its best-fit value and the other abundances are set to zero\footnote{When fitting the Fe-K line,  the Fe abundance is also set to zero.}. We then check the consistency of this method by comparing it with the abundances reported above (hereafter the GDEM method) in terms of equivalent width (EW), which we define for each line as 
\begin{equation} \label{eq:EW}
\text{EW}= \frac{F_\text{line}}{F_\text{c}(E)}
,\end{equation}
where $F_\text{line}$ and $F_\text{c}(E)$ are the fluxes of the line and the continuum at the line energy $E,$ respectively. Since the EW of a line is proportional to the abundance of its ion, in principle both methods should yield the same abundance. We compare them on the strongest lines of Mg, Si, S, Ca, Fe, and Ni in MOS and pn spectra (Table \ref{table:EWs}) and we convert the average MOS+pn EWs into abundance measurements (Fig. \ref{fig:core_abundances}). While we find consistency between the Gauss and GDEM methods for Ca and Fe-K lines both in MOS and pn, the other elements need to be further discussed.

\begin{table}
\caption{Measured equivalent widths of K-shell lines in the core (0$'$--3$'$) using the Gauss and GDEM methods independently for MOS and pn.}  
\label{table:EWs}
\resizebox{\hsize}{!}{
\begin{tabular}{l c c@{$\pm$}l c@{$\pm$}l c@{$\pm$}l c@{$\pm$}l}       
\hline\hline  
                               &  & \multicolumn{4}{c}{MOS}   & \multicolumn{4}{c}{pn} \\        
\hline
Elem.                                & Line E & \multicolumn{2}{c}{EW$_\text{GDEM}$}   & \multicolumn{2}{c}{EW$_\text{Gauss}$} & \multicolumn{2}{c}{EW$_\text{GDEM}$}   & \multicolumn{2}{c}{EW$_\text{Gauss}$} \\   
                                               & (keV) & \multicolumn{2}{c}{(eV)}  & \multicolumn{2}{c}{(eV)} & \multicolumn{2}{c}{(eV)}   & \multicolumn{2}{c}{(eV)} \\   

\hline                        

Mg & $1.44$ & $13.8$ & $0.9$ & $10.1$ & $1.2$ & $0.8$ & $0.8$ & $7.5$ & $1.7$ \\ 
Si & $2.00$ & $36.8$ & $1.7$ & $41$ & $3$ & $24$ & $2$ & $41$ & $4$ \\ 
S & $2.62$ & $39$ & $2$ & $61$ & $12$ & $23$ & $4$ & $41$ & $13$ \\ 
Ca & $3.89$ & $30$ & $4$ & $25$ & $11$ & $33$ & $5$ & $32$ & $12$ \\ 
Fe & $6.65$ & $820$ & $10$ & $776$ & $34$ & $684$ & $11$ & $652$ & $32$ \\ 
Ni & $7.78$ & $127$ & $8$ & $182$ & $33$ & $28$ & $8$ & $92$ & $26$  \\

\hline                          
\end{tabular}   }  
\end{table}

The EW of Mg obtained in pn using the Gauss method is, significantly, $\sim$9 times higher than when using the GDEM method. In the latter case, the pn continuum of the model is largely overestimated around $\sim$1.5 keV, making the Mg abundance underestimated. The elements Si and S also show significantly larger EWs in pn using the Gauss method. In terms of abundance measurements, they both agree with the MOS measurements (Fig. \ref{fig:core_abundances}). We also note that beyond $\sim$1.5 keV the MOS residuals ratio are known to be significantly higher than the pn ones \citep{2014A&A...564A..75R}, and peak near the Si line. This might also partly explain the discrepancies found for S, Si, and maybe Mg.

When using the GDEM method for pn, the Ni-K line is poorly fitted. The large difference in EW obtained when fitting it using the Gauss method  emphasises this effect. In fact, when fitting the pn spectra using a \textit{CIE} or \textit{GDEM} model, a low Ni abundance is computed by the model to compensate the issues in the calibration of the effective area around 1.0--1.5 keV (i.e. where most Ni-L lines are present). For this reason and because of  large error bars for the Ni-K line, the fit in pn ignores it.

If we fit the spectra only between 2--10 keV,  after freezing $kT$, $\sigma_T$, O, Mg, and Si obtained in our previous fits, we obtain Ni abundances of $1.61 \pm 0.35$ and $1.37 \pm 0.26$ for MOS and pn, respectively, making them consistent between each other. This clearly favours the Ni abundance measured with MOS in our previous fits. Interestingly, we also measure Fe abundances of $0.752 \pm 0.019$ and $0.676 \pm 0.017$ for MOS and pn, respectively; their discrepancies are then reduced, but still remain. Finally, we note that the pn data are shifted by $\sim$-20 eV compared to the model around the Fe-K line; this shift does not affect the abundance measurements though.

Our results on the abundance analysis in the core are summarised in Table \ref{table:core_abundances} and Fig. \ref{fig:core_abundances} and are briefly discussed in Sect. \ref{subsect:abun_core}. Because  their  uncertainties are too large, we choose not to consider Mg and Ni abundances in the rest of the paper. Moreover, although the MOS-pn discrepancies are sometimes large and make some absolute abundance measurements quite uncertain, in the following sections we are more interested in their spatial variations. By comparing combined \textit{MOS+pn} measurements only, the systematic errors we have shown here should not play an important role in this purpose.

\begin{table}
\caption{Summary of the absolute abundances measured in the core (EPIC and RGS) using a \textit{GDEM} model. The mean MOS+pn abundances obtained by fitting Gaussian lines instead of the CIE models (the Gauss method; see text and Table \ref{table:EWs}) is also included. See also Fig. \ref{fig:core_abundances}.}  
\label{table:core_abundances}
\resizebox{\hsize}{!}{
\begin{tabular}{l c@{$\pm$}l c@{$\pm$}l c@{$\pm$}l c@{$\pm$}l c@{$\pm$}l}       
\hline\hline  
Elem.                               &  \multicolumn{8}{c}{EPIC}   & \multicolumn{2}{c}{RGS} \\        
\hline
                               & \multicolumn{2}{c}{MOS}   & \multicolumn{2}{c}{pn} & \multicolumn{2}{c}{MOS+pn}   & \multicolumn{2}{c}{Gauss corr.} & \multicolumn{2}{c}{} \\   
\hline                        

N &  \multicolumn{2}{c}{$-$}   & \multicolumn{2}{c}{$-$} & \multicolumn{2}{c}{$-$}   & \multicolumn{2}{c}{$-$} & $0.9$ & $0.3$ \\ 
O &  $0.57$ & $0.04$ & $0.33$ & $0.04$ & $0.46$ & $0.04$ & \multicolumn{2}{c}{$-$} & $0.36$ & $0.03$ \\ 
Ne &  $0.34$ & $0.06$ & $0.36$ & $0.08$ & $0.33$ & $0.05$ & \multicolumn{2}{c}{$-$} & $0.35$ & $0.05$ \\ 
Mg &  $0.78$ & $0.05$ & \multicolumn{2}{c}{$< 0.08$} & $0.45$ & $0.03$ & $0.47$ & $0.08$ & $0.27$ & $0.07$ \\ 
Si &  $0.66$ & $0.03$ & $0.35$ & $0.03$ & $0.51$ & $0.02$ & $0.67$ & $0.06$ & $0.4$ & $0.3$ \\ 
S &  $0.66$ & $0.04$ & $0.31$ & $0.05$ & $0.52$ & $0.03$ & $0.79$ & $0.19$ & \multicolumn{2}{c}{$-$} \\ 
Ar &  $0.30$ & $0.11$ & $0.54$ & $0.15$ & $0.41$ & $0.08$ & \multicolumn{2}{c}{$-$} & \multicolumn{2}{c}{$-$} \\ 
Ca &  $0.98$ & $0.13$ & $0.90$ & $0.15$ & $1.01$ & $0.10$ & $0.8$ & $0.3$ & \multicolumn{2}{c}{$-$} \\ 
Fe &  $0.803$ & $0.010$ & $0.600$ & $0.010$ & $0.697$ & $0.006$ & $0.67$ & $0.03$ & $0.62$ & $0.04$ \\ 
Ni &  $1.83$ & $0.11$ & $0.37$ & $0.10$ & $1.04$ & $0.07$ & $1.9$ & $0.4$ & \multicolumn{2}{c}{$-$} \\

\hline                          
\end{tabular}   }
\end{table}

\subsection{RGS}

Our RGS analysis of the core region focuses on the $7$--$28$ {\AA} ($0.44$--$1.77$ keV) first and second order spectra of the RGS detector; RGS stacked spectra are binned by a factor of 5. We test single-, two-temperature \textit{CIE} models, and a \textit{GDEM} model for comparison. 

The models are redshifted and, to model the absorption, multiplied by a \textit{\emph{hot}} model (i.e. an absorption model where the gas is assumed to be in CIE) with a total $N_{\rm H}=1.26 \times 10^{20}\,{\rm cm}^{-2}$ \citep{2013MNRAS.431..394W}, $kT=0.5$ eV, and proto-solar abundances.

In order to take into account the emission-line broadening due to the spatial extent of the source, we have convolved the emission components by the \textit{lpro} multiplicative model in SPEX \citep{2004A&A...420..135T,2015arXiv150101069P}.

\begin{table}
\caption{RGS spectral fits of Abell 4059.}  
\label{table:rgs}
\begin{tabular}{l c@{$\pm$}l c@{$\pm$}l c@{$\pm$}l}       
\hline\hline                 
Parameter                                 & \multicolumn{2}{c}{1-T \textit{CIE}} & \multicolumn{2}{c}{2-T \textit{CIE}}   & \multicolumn{2}{c}{\textit{GDEM}} \\   
\hline                        
 C-stat/d.o.f.                      & \multicolumn{2}{c}{1274/887}       & \multicolumn{2}{c}{1244/886}         & \multicolumn{2}{c}{1268/885}         \\ 
\hline
 $Y_1$ ($10^{70}$ m$^{-3}$) & 683 & 4 & 662 & 6  & 480  & 8 \\ 
 $T_1$ (keV)                        & 2.74 & 0.08 & 2.8  & 0.1   & \multicolumn{2}{c}{}  \\ 
 $Y_2$ ($10^{70}$ m$^{-3}$) &    \multicolumn{2}{c}{}           & 4    & 1     &     \multicolumn{2}{c}{}             \\ 
 $T_2$ (keV)                                    &    \multicolumn{2}{c}{}           & 0.80 & 0.07  &        \multicolumn{2}{c}{}          \\ 
 $T_\text{mean}$ (keV)                        & \multicolumn{2}{c}{} & \multicolumn{2}{c}{}   & 3.4   & 0.2  \\ 
 $\sigma_T$                        &         \multicolumn{2}{c}{}       &       \multicolumn{2}{c}{}           & 0.26  & 0.03 \\ 
 N                                  & 0.7 & 0.2 & 0.9 & 0.3  & 0.9  & 0.3 \\ 
 O                                  & 0.32 & 0.03 & 0.35 & 0.03  & 0.36  & 0.03 \\ 
 Ne                                 & 0.40 & 0.05 & 0.43 & 0.06  & 0.35  & 0.05 \\ 
 Mg                                 & 0.26 & 0.06 & 0.32 & 0.07  & 0.27  & 0.07 \\ 
 Si                                 & 0.6 & 0.3 & 0.8 & 0.3  & 0.4  & 0.3 \\ 
 Fe                                 & 0.57 & 0.03 & 0.63 & 0.04  & 0.62  & 0.04 \\ 
\hline                          
\end{tabular}     
\end{table}

\begin{figure}
\resizebox{\hsize}{!}{
      \includegraphics[trim=1cm 0cm 0cm 0cm, clip=true]{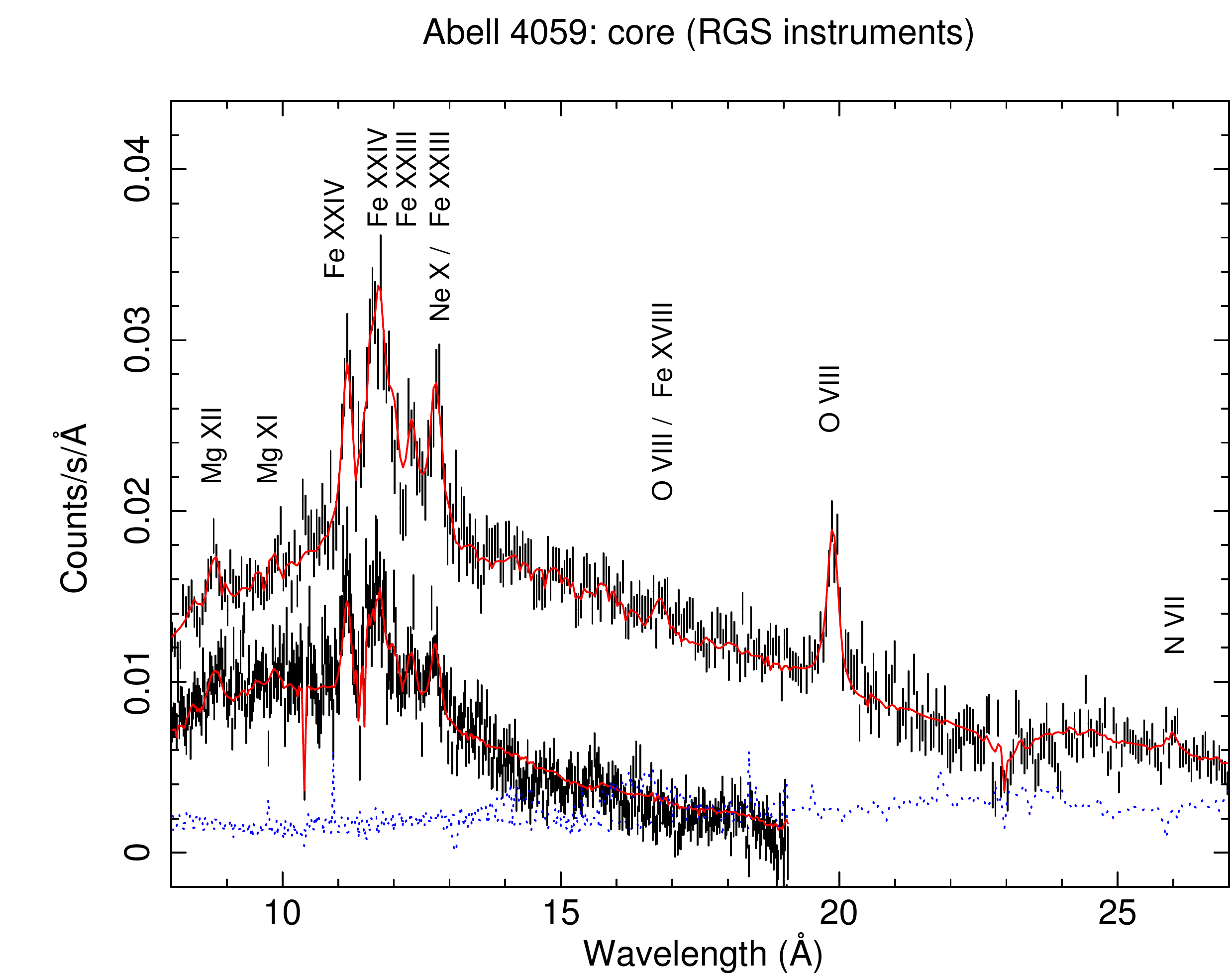}}
      \caption{RGS first and second order spectra of A\,4059 (see also Table\,\ref{table:rgs}). The spectra are fitted with a 2-T \textit{CIE} model. The subtracted backgrounds are shown in blue dotted lines. The main resolved emission lines are also indicated.}
          \label{fig:rgs}
\end{figure}

The RGS order 1 and 2 stacked spectra have been fitted simultaneously (Fig. \ref{fig:rgs}) and the results of the spectral fits are shown in Table \ref{table:rgs} and Fig. \ref{fig:core_abundances}.
The 2-T \textit{CIE} and \textit{GDEM} fits are comparable in terms of Cash statistics and the models are visually similar. Although there might be some residual emission at temperature below 1\,keV that can be reproduced by the 2-T \textit{CIE} model \citep{2013ApJ...764...46F}, using a \textit{GDEM} model is more realistic regarding temperature distribution found in the core of most clusters. The abundances are in agreement between the different models because they depend on the relative strength of the lines.

\section{EPIC radial profiles}\label{sect:radial_profiles}

We fit the EPIC spectra from each of the eight annular regions mentioned in Sect. \ref{sect:data_reduction} using a \textit{GDEM} model. We derive projected radial profiles of the temperature, temperature broadening, and abundances (Table \ref{table:radial_profiles}). In our measurements, all the cluster parameters ($Y$, $kT$, $\sigma_T$, and abundances) are coupled between the three instruments and the two datasets. Since we ignore the channels below 0.4 keV (MOS) and 0.6 keV (pn) in the outermost annulus to avoid background contamination (Appendix \ref{sect:bg_modelling}), we restrict our O radial profile within 9$'$. For the same reason, the O abundance measurement between 6$'$--9$'$ might be biased up to $\sim 25\%$ (i.e. our presumed MOS-pn systematic uncertainty for the O measurement).

\begin{table*}[!]
\begin{centering}
\caption{Best-fit parameters measured in eight concentric annuli (covering a total of $\sim$12 arcmin of FoV). The spectra of all the annuli have been fitted using a \textit{GDEM} model and adapted from our background procedure.}
\label{table:radial_profiles}
\scalebox{0.95}{
\begin{tabular}{l c@{$\pm$}l c@{$\pm$}l c@{$\pm$}l c@{$\pm$}l c@{$\pm$}l c@{$\pm$}l c@{$\pm$}l c@{$\pm$}l }
\hline 
\hline
Parameter & \multicolumn{2}{c}{$0'$ -- $0.5'$} & \multicolumn{2}{c}{$0.5'$ -- $1'$} & \multicolumn{2}{c}{$1'$ -- $2'$} & \multicolumn{2}{c}{$2'$ -- $3'$} & \multicolumn{2}{c}{$3'$ -- $4'$} & \multicolumn{2}{c}{$4'$ -- $6'$} & \multicolumn{2}{c}{$6'$ -- $9'$} & \multicolumn{2}{c}{$9'$ -- $12'$}\tabularnewline
\hline
C-stat/d.o.f. & \multicolumn{2}{c}{$2440/1482$} & \multicolumn{2}{c}{$2302/1575$} &  \multicolumn{2}{c}{$2641/1670$}  & \multicolumn{2}{c}{$2182/1658$} & \multicolumn{2}{c}{$1967/1627$} & \multicolumn{2}{c}{$2061/1703$} & \multicolumn{2}{c}{$2129/1686$} & \multicolumn{2}{c}{$2223/1671$}\tabularnewline
\hline

$Y$ ($10^{70}$ m$^{-3}$) & $82.5$ & $0.9$ & $155.9$ & $1.2$ & $314.0$ & $1.6$  & $ 240.5$ & $1.5$ & $176.1$ & $1.1$ & $256.7$ & $1.9$ & $240$ & $3$ & $150$ & $3$\tabularnewline

$kT_\text{mean}$ (keV)    & $2.84$ & $0.03$ & $3.39$ & $0.03$ & $3.69$ & $0.02$  & $4.06$ & $0.03$ & $4.16$ & $0.05$ & $4.17$ & $0.06$ & $4.21$ & $0.10$ & $3.98$ & $0.20$\tabularnewline
$\sigma_T$                        & $0.222$ & $0.008$ & $0.231$ & $0.010$ & $0.224$ & $0.012$  & $0.23$ & $0.02$ & $0.27$ & $0.02$ & $0.280$ & $0.014$ & $0.33$ & $0.02$ & $0.33$ & $0.04$\tabularnewline

O & $0.53$ & $0.08$ & $0.54$ & $0.06$ & $0.43$ & $0.04$ & $0.38$ & $0.06$ & $0.32$ & $0.07$ & $0.29$ & $0.06$ & $0.39$ & $0.08$ & \multicolumn{2}{c}{$-$}\tabularnewline

Ne & $0.63$ & $0.13$ & $0.36$ & $0.11$ & $0.41$ & $0.08$  & $0.14$ & $0.09$ & $0.11$ & $0.09$ & \multicolumn{2}{c}{$< 0.04$} & \multicolumn{2}{c}{$< 0.04$} & \multicolumn{2}{c}{$< 0.29$}\tabularnewline

Mg & $0.51$ & $0.09$ & $0.51$ & $0.07$ & $0.44$ & $0.05$  & $0.42$ & $0.07$ & $0.45$ & $0.09$ & $0.23$ & $0.08$ & $0.18$ & $0.10$ & \multicolumn{2}{c}{$< 0.34$}\tabularnewline

Si & $0.78$ & $0.05$ & $0.59$ & $0.04$ & $0.50$ & $0.03$  & $0.32$ & $0.04$ & $0.32$ & $0.05$ & $0.08$ & $0.05$ & $0.07$ & $0.05$ & \multicolumn{2}{c}{$< 0.03$}\tabularnewline

S & $0.69$ & $0.08$ & $0.55$ & $0.06$ & $0.57$ & $0.05$  & $0.36$ & $0.06$ & $0.29$ & $0.07$ & $0.09$ & $0.07$ & \multicolumn{2}{c}{$< 0.13$} & $0.41$ & $0.17$\tabularnewline

Ar & $0.8$ & $0.2$ & $0.65$ & $0.16$ & $0.40$ & $0.13$  & $0.40$ & $0.16$ & \multicolumn{2}{c}{$< 0.42$} & $0.2$ & $0.2$ & \multicolumn{2}{c}{$< 0.07$} & $0.8$ & $0.5$\tabularnewline

Ca & $1.8$ & $0.3$ & $1.2$ & $0.2$ & $1.12$ & $0.15$  & $0.77$ & $0.19$ & $0.5$ & $0.3$ & $0.7$ & $0.2$ & $0.41$ & $0.36$ & \multicolumn{2}{c}{$< 1.34$}\tabularnewline

Fe & $0.88$ & $0.03$ & $0.75$ & $0.02$ & $0.653$ & $0.013$  & $0.46$ & $0.02$ & $0.38$ & $0.02$ & $0.31$ & $0.02$ & $0.20$ & $0.02$ & $0.17$ & $0.04$\tabularnewline

Ni & $1.11$ & $0.17$ & $1.28$ & $0.14$ & $0.97$ & $0.12$  & $0.72$ & $0.15$ & $0.68$ & $0.18$ & $0.27$ & $0.18$ & \multicolumn{2}{c}{$< 0.25$} & \multicolumn{2}{c}{$< 0.07$}\tabularnewline
\hline
\end{tabular}}
\par\end{centering}
\end{table*}

\begin{figure*}[!]
        \centering
                \includegraphics[width=0.32\textwidth]{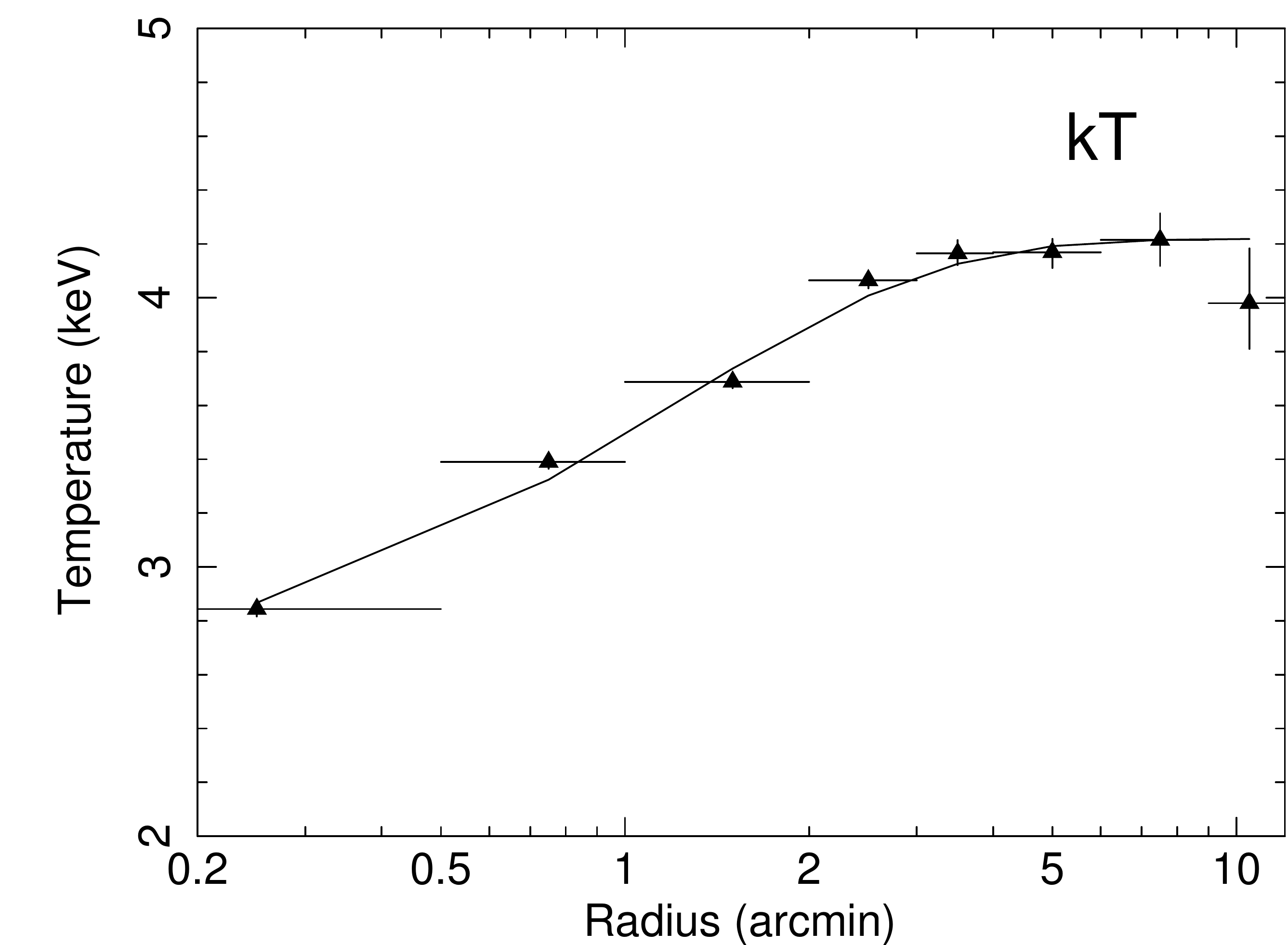}
                \includegraphics[width=0.32\textwidth]{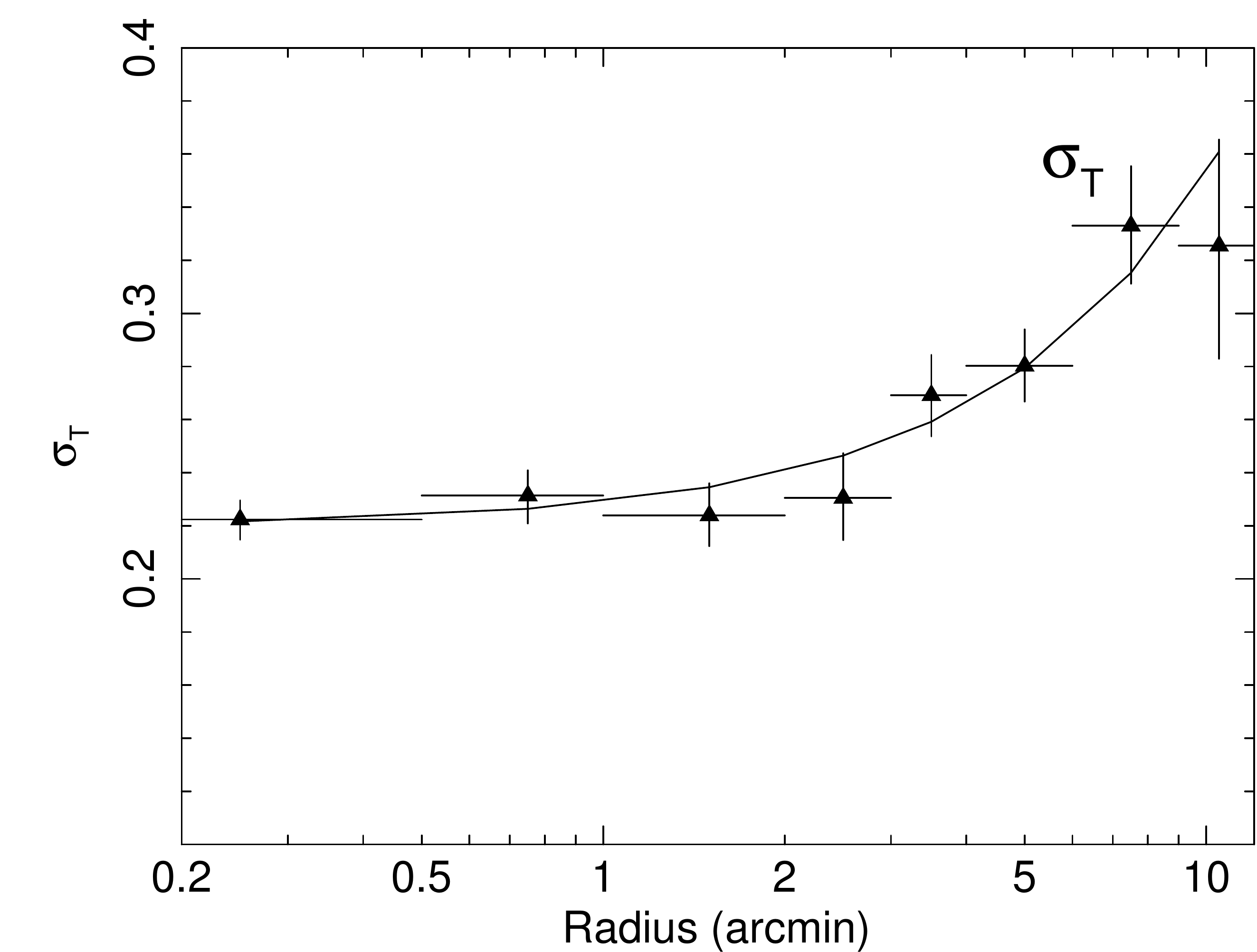}
                \includegraphics[width=0.32\textwidth]{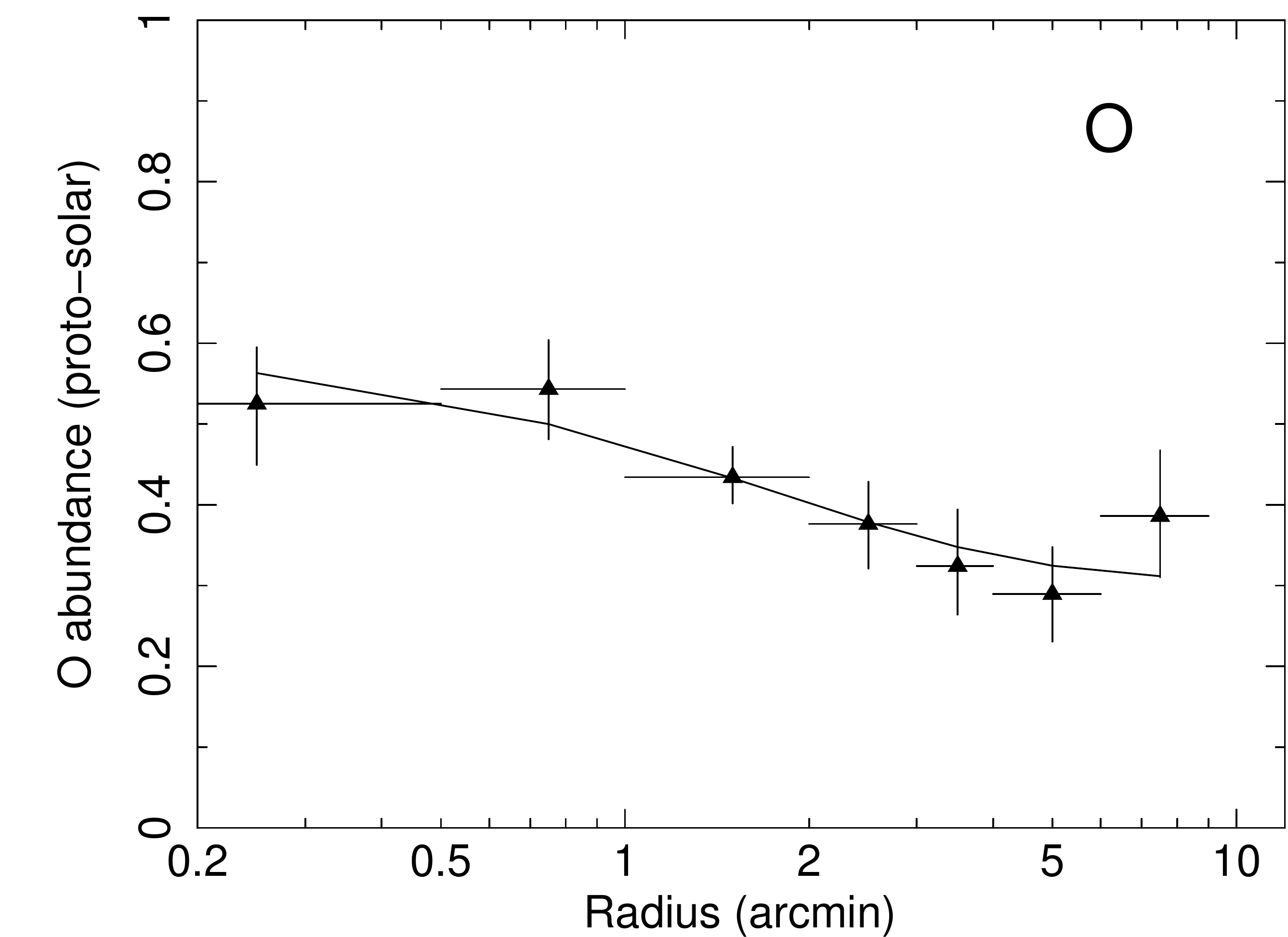}
\\
                \includegraphics[width=0.32\textwidth]{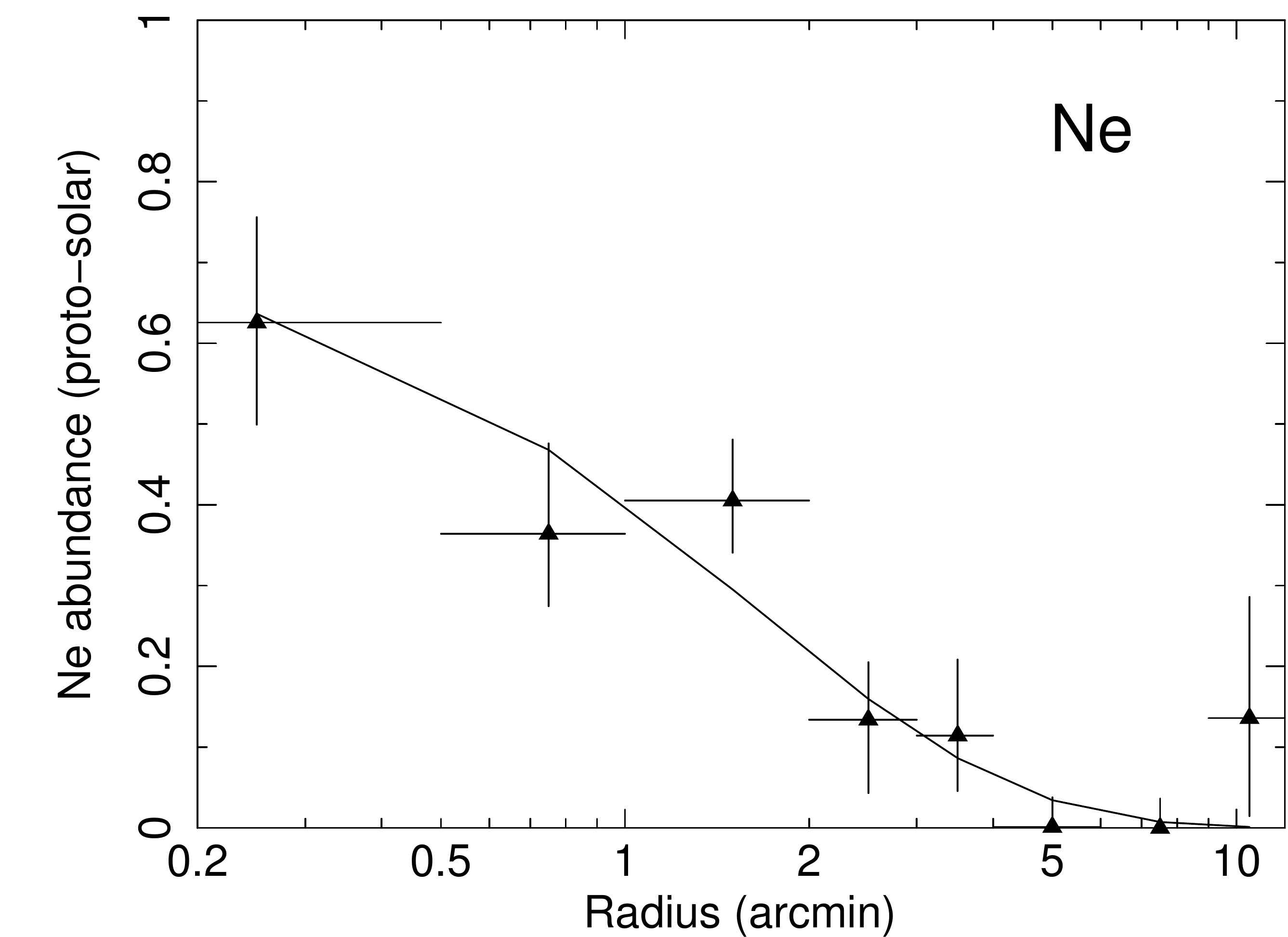}
                \includegraphics[width=0.32\textwidth]{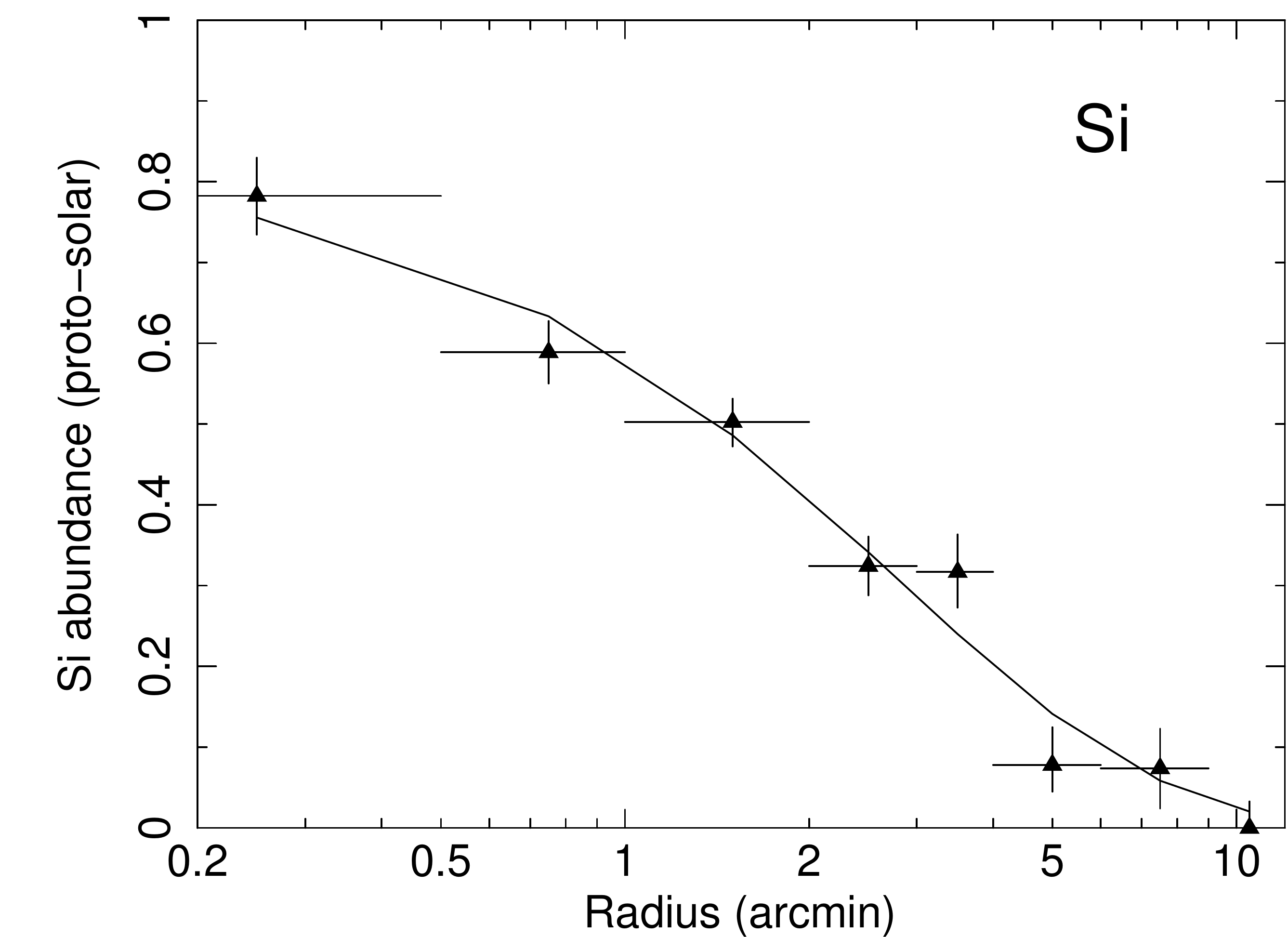}
                \includegraphics[width=0.32\textwidth]{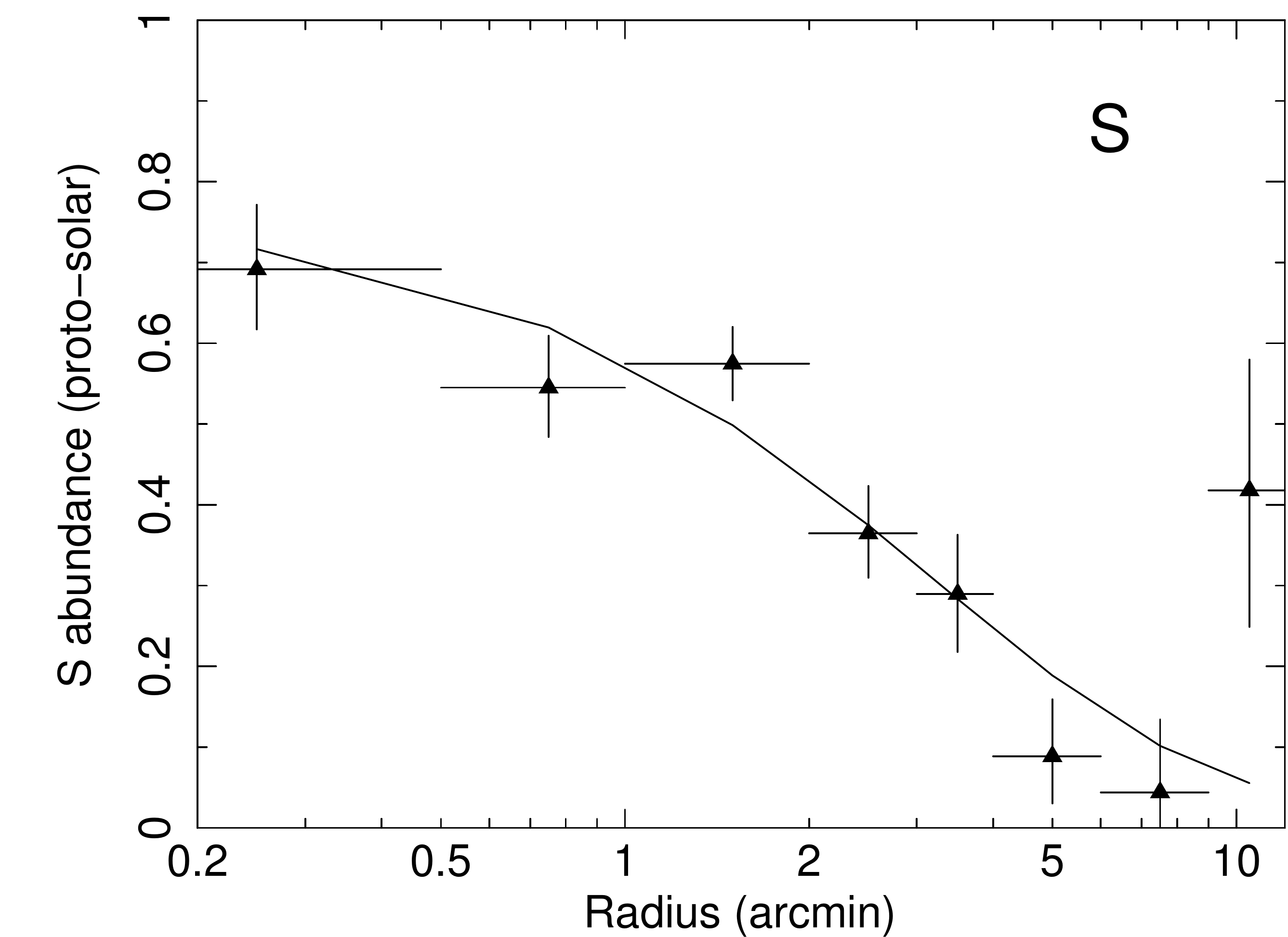}
\\
                \includegraphics[width=0.32\textwidth]{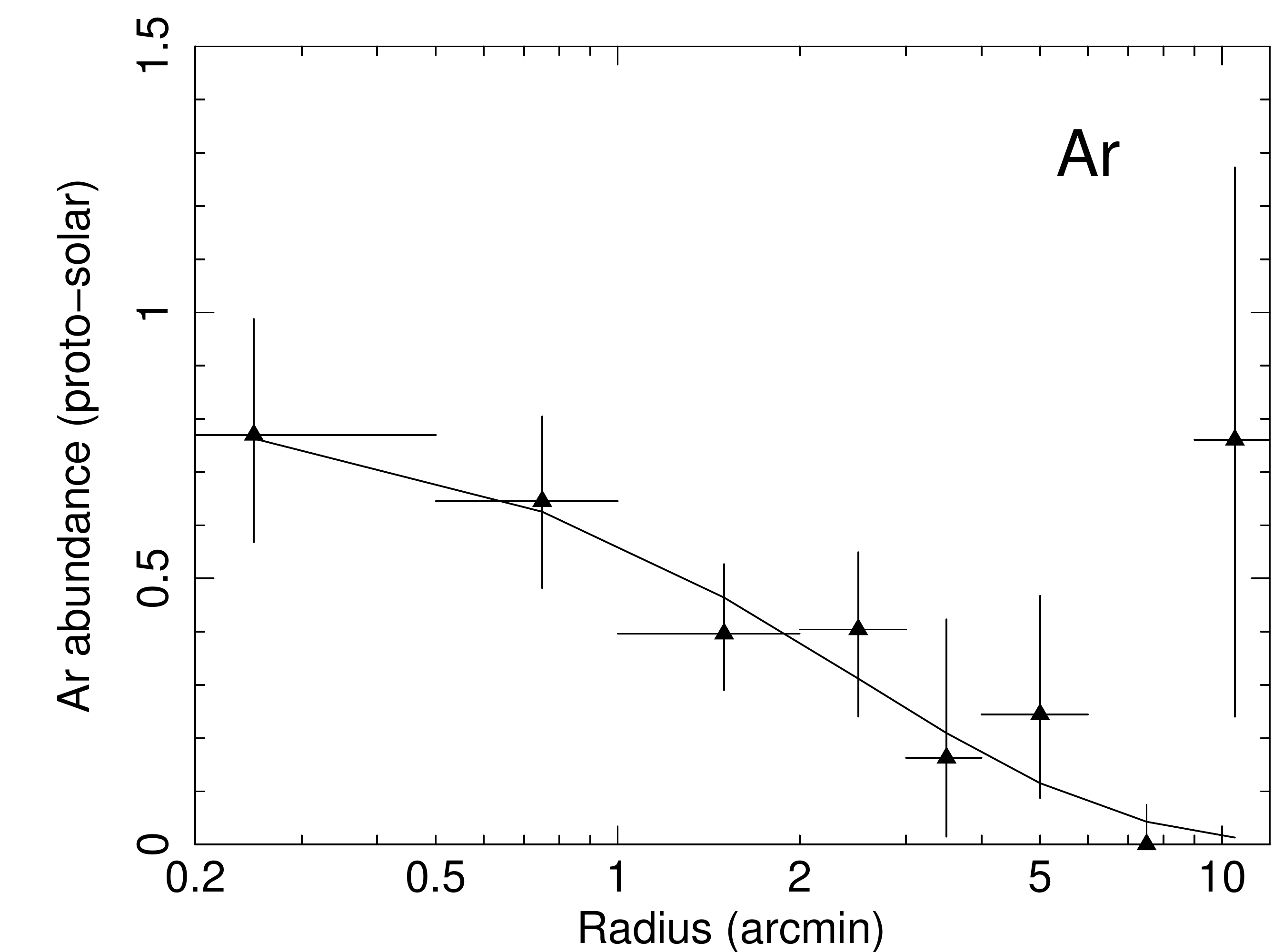}
                \includegraphics[width=0.32\textwidth]{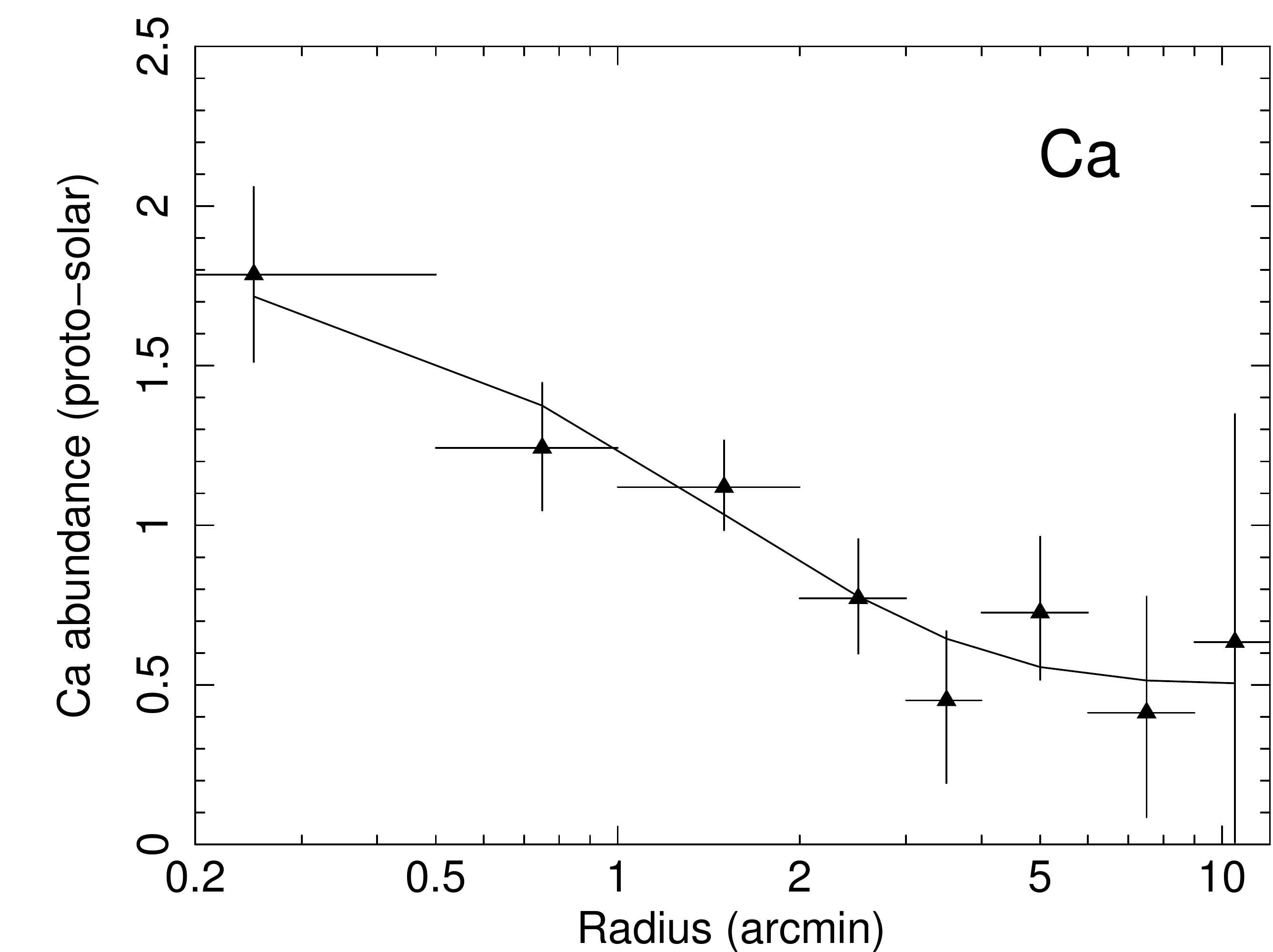}
                \includegraphics[width=0.32\textwidth]{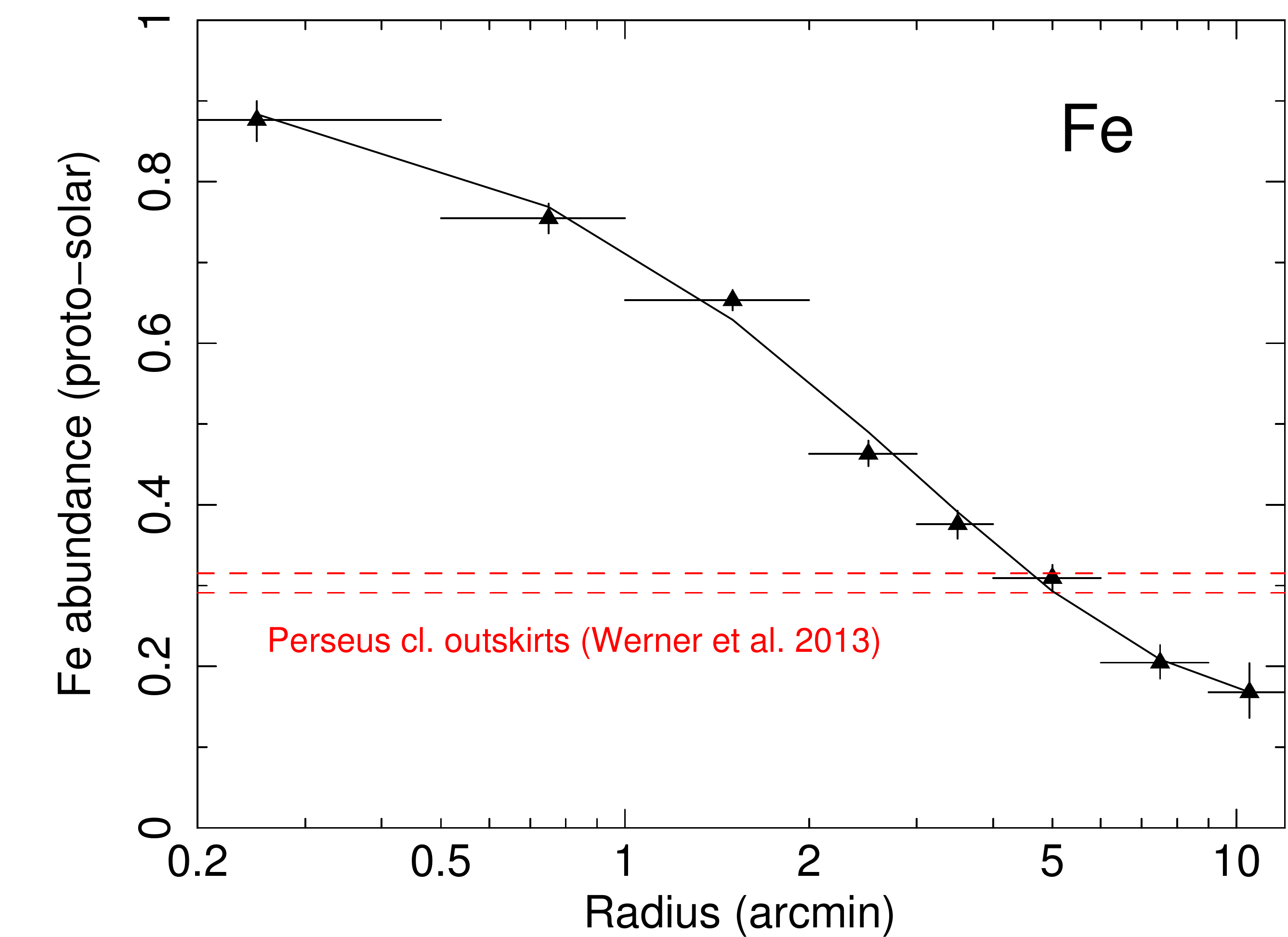}

        \caption{EPIC radial profiles of Abell 4059. The datapoints show our best-fit measurements (Table \ref{table:radial_profiles}). The solid lines show our best-fit empirical distributions (Table \ref{table:radial_models}). The spectra of all the annuli have been fitted using a \textit{GDEM} model and adapted from our background modelling. We note the change of abundance scale for Ar and Ca.}
\label{fig:radial_profiles}
\end{figure*}

In order to quantify the trends that appear in our profiles, we fit them with simple empirical distributions. For temperature and abundance profiles, 
\begin{equation} \label{eq:empir_distrib1}
kT(r) = D_\infty + \alpha \exp (-r/r_0)
\end{equation}
\begin{equation} \label{eq:empir_distrib2}
Z(r) = D_\infty + \alpha \exp (-r/r_0)
\end{equation}
and for $\sigma_T$ radial profile,
\begin{equation} \label{eq:empir_distrib3}
\sigma_T(r) = D_\infty + \alpha r^{\gamma}.
\end{equation}
Table \ref{table:radial_models} shows the results of our fitted trends. Figure \ref{fig:radial_profiles} shows the radial profiles and their respective best-fit distributions.

\begin{table}[!]
\caption{Best-fit parameters of empirical models for our radial profiles. For the meaning of $\alpha$, $r_0$, $\gamma$, and $D_\infty$, see Eqs. \ref{eq:empir_distrib1}, \ref{eq:empir_distrib2}, and \ref{eq:empir_distrib3} in the text. Unless mentioned ($^{CIE}$), the empirical models follow the \textit{GDEM} measurements of Table \ref{table:radial_profiles}.}
\label{table:radial_models}
\resizebox{\hsize}{!}{   
\begin{tabular}{l c@{$\pm$}l c@{$\pm$}l c@{$\pm$}l c@{$\pm$}l r}
\hline 
\hline
Param. & \multicolumn{2}{c}{$\alpha$} & \multicolumn{2}{c}{$r_0$} & \multicolumn{2}{c}{$\gamma$} & \multicolumn{2}{c}{$D_\infty$} & $\chi^2$/d.o.f.\tabularnewline
\hline

$kT_\text{mean}$      & $-1.66$ & $0.04$ & $1.21$ & $0.08$ & \multicolumn{2}{c}{$-$} & $4.22$ & $0.04$ & $17.28/4$ \tabularnewline
$\sigma_T$                           & $0.009$ & $0.010$ & \multicolumn{2}{c}{$-$} & $1.2$ & $0.3$ & $0.220$ & $0.016$ & $3.79/4$ \tabularnewline
$kT^{CIE}$  & $-1.61$ & $0.04$ & $1.04$ & $0.07$ & \multicolumn{2}{c}{$-$} & $4.05$ & $0.03$ & $22.11/4$ \tabularnewline

O & $0.29$ & $0.07$ & \multicolumn{2}{c}{$1.76_{-0.4}^{+1.1}$} & \multicolumn{2}{c}{$-$} & $0.31$ & $0.03$ & $7.75/3$ \tabularnewline
O & \multicolumn{2}{c}{$-$} & \multicolumn{2}{c}{$-$} & \multicolumn{2}{c}{$-$} & $0.41$ & $0.02$ & $14.22/5$ \tabularnewline

Ne & $0.74$ & $0.12$ & $1.63$ & $0.3$ & \multicolumn{2}{c}{$-$} & \multicolumn{2}{c}{$ < 0.019$} & $4.88/4$ \tabularnewline

Si & $0.83$ & $0.03$ & $2.83$ & $0.2$ & \multicolumn{2}{c}{$-$} & \multicolumn{2}{c}{$< 0.02$} & $7.28/4$ \tabularnewline

S & $0.75$ & $0.06$ & $3.3$ & $0.6$ & \multicolumn{2}{c}{$-$} & \multicolumn{2}{c}{$< 0.02$} & $11.72/4$ \tabularnewline

Ar & $0.84$ & $0.18$ & \multicolumn{2}{c}{$2.5_{-0.6}^{+1.0}$} & \multicolumn{2}{c}{$-$} & \multicolumn{2}{c}{$< 0.07$} & $3.52/4$ \tabularnewline
Ar & \multicolumn{2}{c}{$-$}  & \multicolumn{2}{c}{$-$}  & \multicolumn{2}{c}{$-$} & $0.25$ & $0.04$ & $26.52/6$ \tabularnewline

Ca & $1.43$ & $0.3$ &  \multicolumn{2}{c}{$1.5_{-0.4}^{+1.6}$} & \multicolumn{2}{c}{$-$} & \multicolumn{2}{c}{$< 0.64$} & $2.24/4$ \tabularnewline
Ca & \multicolumn{2}{c}{$-$} &  \multicolumn{2}{c}{$-$} & \multicolumn{2}{c}{$-$} & $0.96$ & $0.13$ & $22.12/6$ \tabularnewline

Fe & $0.80$ & $0.02$ & $2.96$ & $0.3$ & \multicolumn{2}{c}{$-$} & $0.14$ & $0.03$ & $9.01/4$ \tabularnewline

Fe$^{CIE}$ & $0.82$ & $0.03$ & $3.06$ & $0.3$ & \multicolumn{2}{c}{$-$} & $0.18$ & $0.03$ & $11.39/4$ \tabularnewline

\hline
\end{tabular}}

\end{table}

\begin{figure}
\resizebox{\hsize}{!}{
\includegraphics[trim=1.2cm 0.3cm 0cm 0cm, clip=true, width=0.49\textwidth]{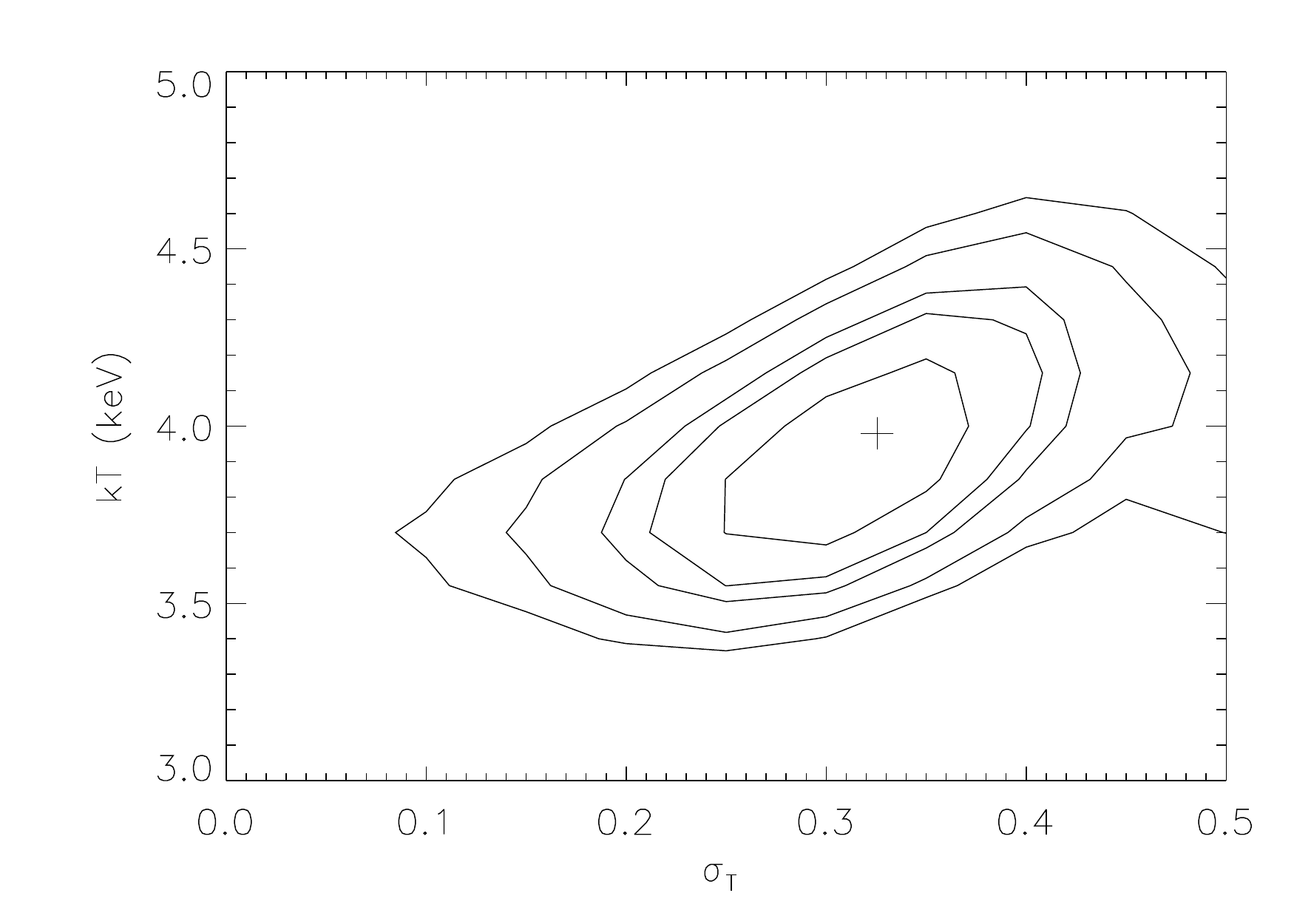}}
\caption{Error ellipses comparing the temperature $kT$ with the broadening of the temperature distribution $\sigma_T$ in the 9$'$--12$'$ annulus spectra. Contours are drawn for 1, 2, 3, 4, and 5$\sigma$. The "$+$" sign shows the best-fit value.}
\label{fig:steps_ann7}
\end{figure}

The temperature profile reveals a significant drop from $\sim$2.5$'$ to the innermost annuli, confirming the presence of a cool-core. Beyond $\sim$2.5$'$, the temperature stabilises around $kT \sim 4.2$ keV. 
More surprisingly, after a plateau around $0.22$ from the core to $\sim$2.5$'$, $\sigma_T$ increases up to $0.33 \pm 0.04$ in the outermost annulus. This increase is  significant in our best-fit distribution. In this outer region, we show that $kT$ and $\sigma_T$ are slightly correlated (Fig. \ref{fig:steps_ann7}); however, the radial profiles of $kT$ and $\sigma_T$ show different trends. Moreover, constraining $\sigma_T$=$0$ in the outermost annulus clearly deteriorates the goodness of the fit (Fig. \ref{fig:steps_ann7}), meaning that the $\sigma_T$ increase is probably genuine.

Our analysis reveals a slightly decreasing O radial profile. Even if fully excluding a flat trend is hard based on our data, the exponential model (Eq. \ref{eq:empir_distrib2}) gives a better fit than a constant model $Z(r) = D_\infty$ (Table \ref{table:radial_models}). A decrease from $0.54 \pm 0.06$ to $0.29 \pm 0.06$ is observed between 0.5$'$--6$'$ as well. Finally, O is still strongly detected in the outermost annuli. We note, however, that additional uncertainties should be taken into account (see  above). In fact, the O measurement near the edge of the FoV may also be slightly affected by the modelling of the Local Hot Bubble (Appendix \ref{sect:bg_modelling}) through its flux and its assumed O abundance.

As mentioned earlier, Ne is hard to constrain, but is detected. Its abundance drops to zero outside the core while it is found to be more than half its proto-solar value within 0.5 arcmin. Profiles of Si and S abundances also decrease, typically from $\sim$0.8 to very low values in the outermost annuli. In every annulus the Si and S measurements are quite similar; this is also confirmed by the best-fit trends which exhibit consistent parameters between the two profiles. The Ar radial profile is harder to interpret because of its large uncertainties, but the trend suggests the same decreasing profile as observed for Si and S.

The Ca radial profile shows particularly high abundances in general, significantly peaked toward the core where it reaches $1.8 \pm 0.3$ times the proto-solar value and $2.0 \pm 0.3$ times the local Fe abundance.
Finally, we show that Fe abundance is also significantly peaked within the core and decreases toward the outskirts, where our fitted model suggests a flattening to $0.14 \pm 0.03$.

We note that our radial analysis focuses on the projected profiles only. Although deprojection can give a rough idea about the 3-D behaviour of the radial profiles, they are based on the assumption of a spherical symmetry, which is far from being the case in the innermost parts of A\,4059 (Sect. \ref{sect:maps}). Moreover, the deprojected abundance radial profiles are not thought  to deviate significantly from the projected ones \citep[see e.g.][]{2006A&A...449..475W}. Based on the analysis of \citet{2004A&A...413..415K}, we estimate that the contamination of photons into incorrect annuli as a result of the EPIC point-spread function (PSF) changes our Fe abundance measurements by      $\sim$2$\%$ and $\sim$4$\%$ in the first and second innermost annuli, respectively, which is not significant regarding our 1$\sigma$ error bars. The choice of a \textit{GDEM} model should take into account both the multi-temperature features due to projection effects and the possible PSF contamination in the $kT$ radial profile.

\section{Temperature, $\sigma_T$, and Fe abundance maps}\label{sect:maps}

Using a \textit{GDEM} model, we derive temperature and abundance maps from the EPIC data of our two deep observations.
The long net exposure time ($\sim$140 ks) for A\,4059 allows  the distribution of $kT$, $\sigma_T$, and Fe abundance to be mapped within 6$'$. As in  the radial analysis, all the EPIC instruments and the two datasets are fitted simultaneously.

\begin{figure*}
        \centering
                \includegraphics[width=0.33\textwidth]{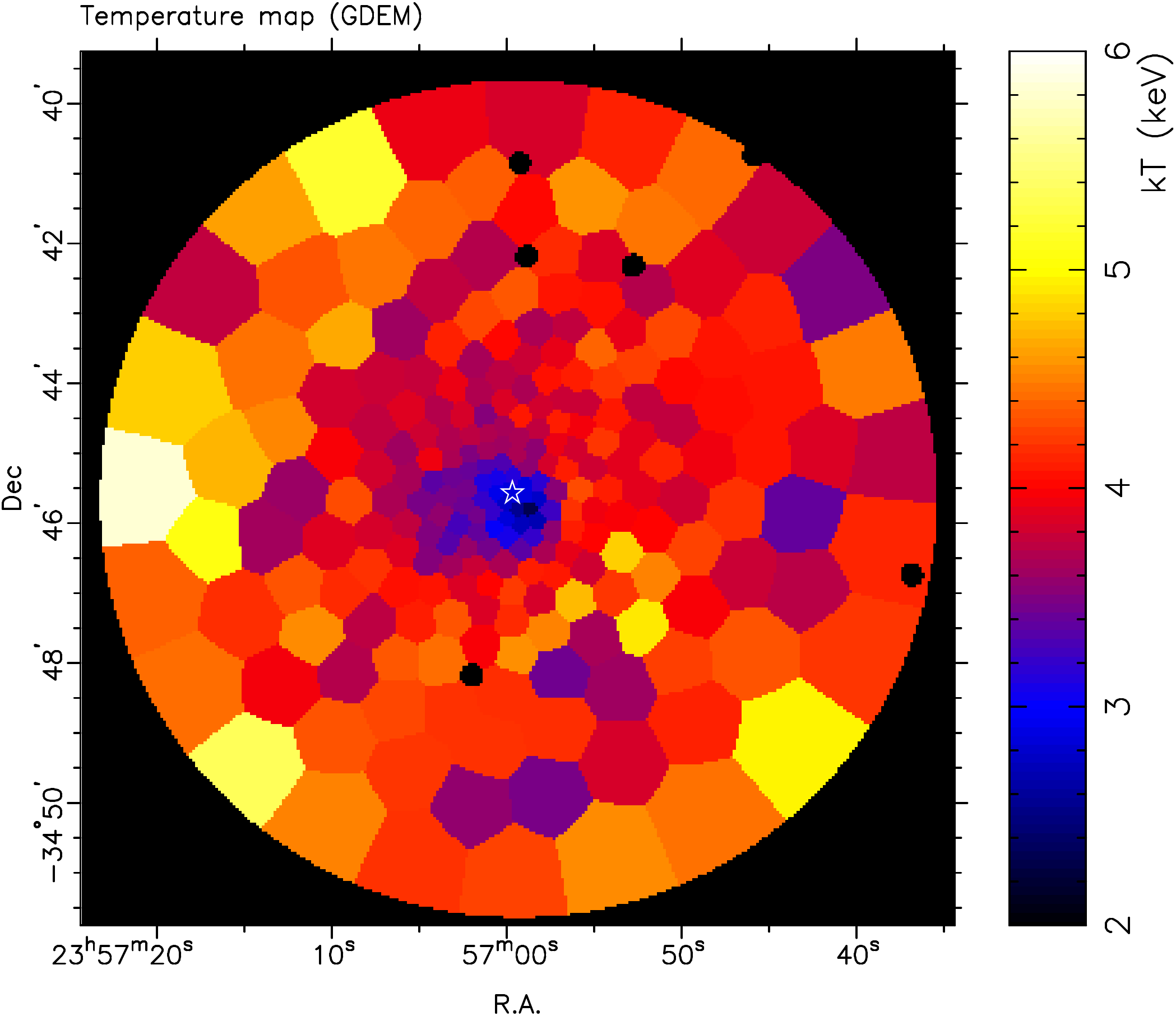}
                \includegraphics[width=0.33\textwidth]{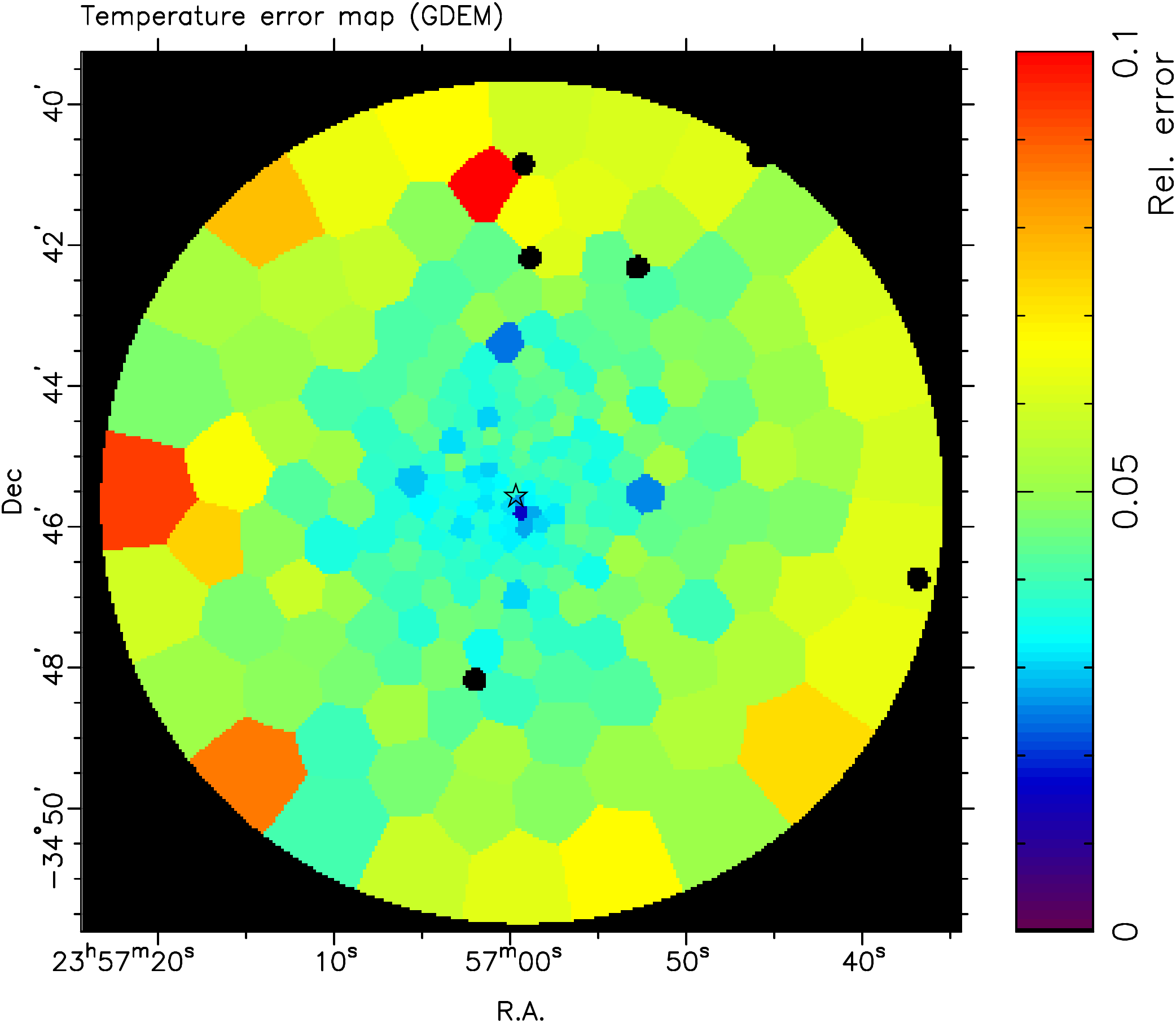}
                \includegraphics[width=0.33\textwidth]{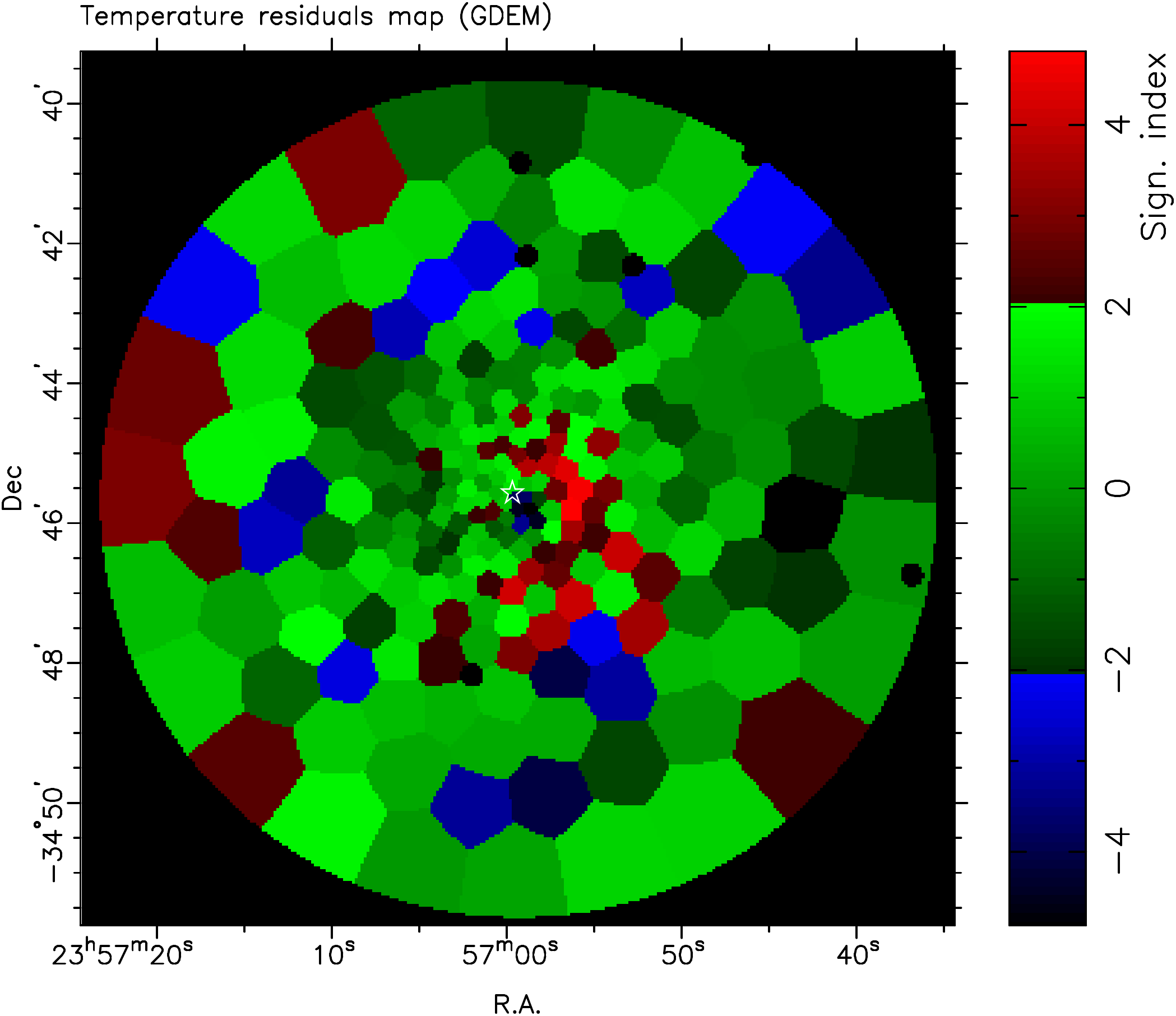}
\\
                \includegraphics[width=0.33\textwidth]{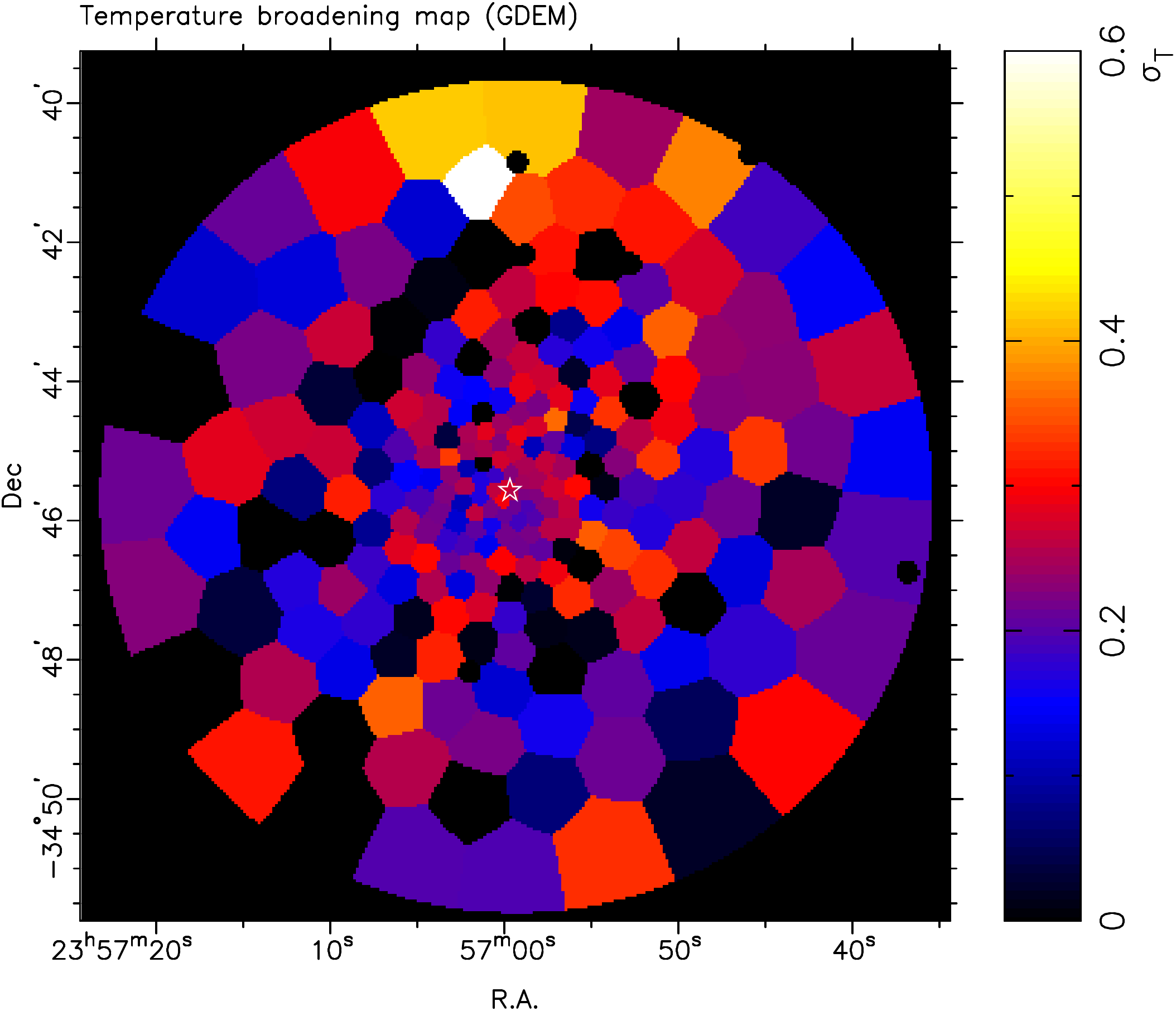}
                \includegraphics[width=0.33\textwidth]{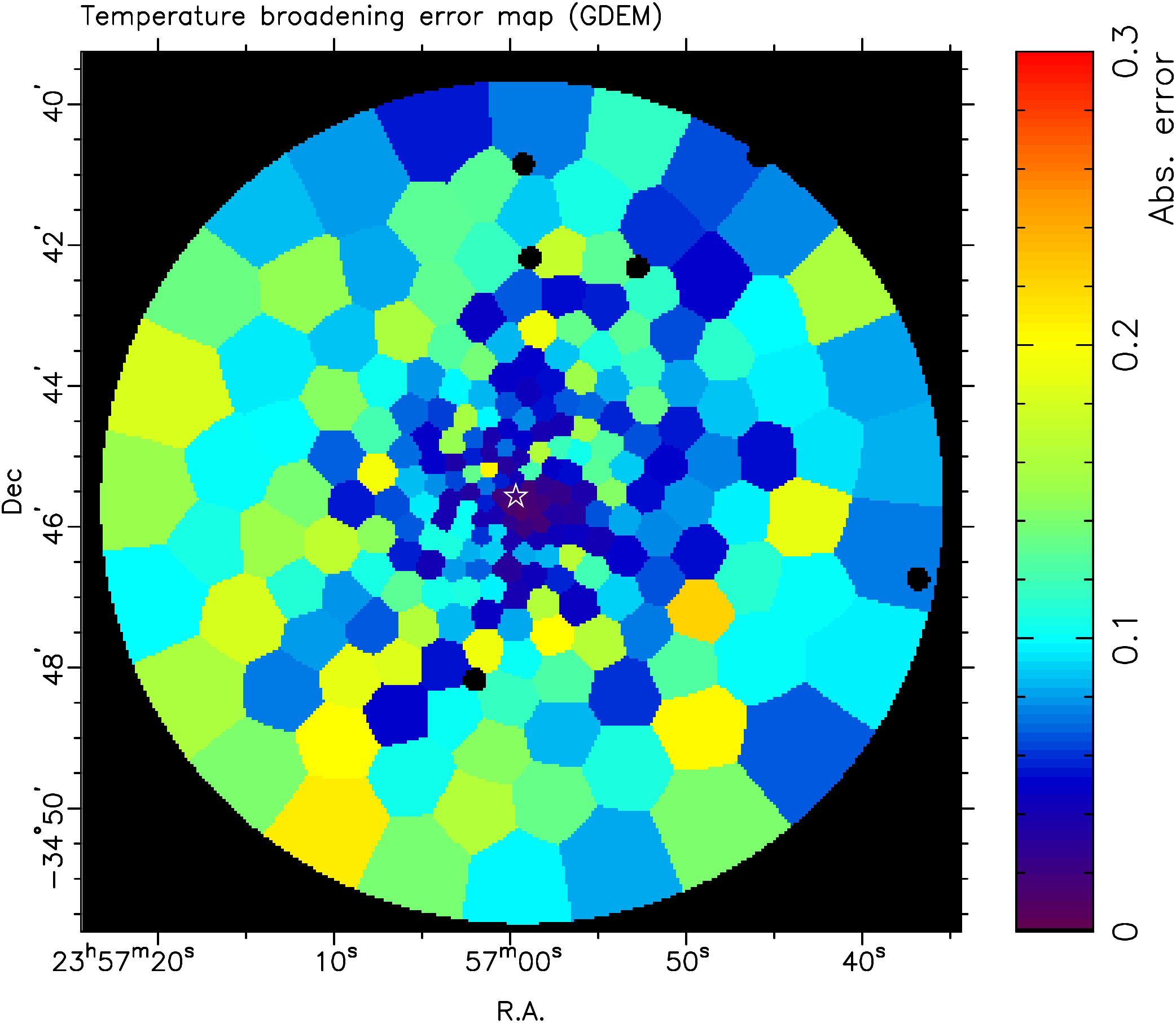}
                \includegraphics[width=0.33\textwidth]{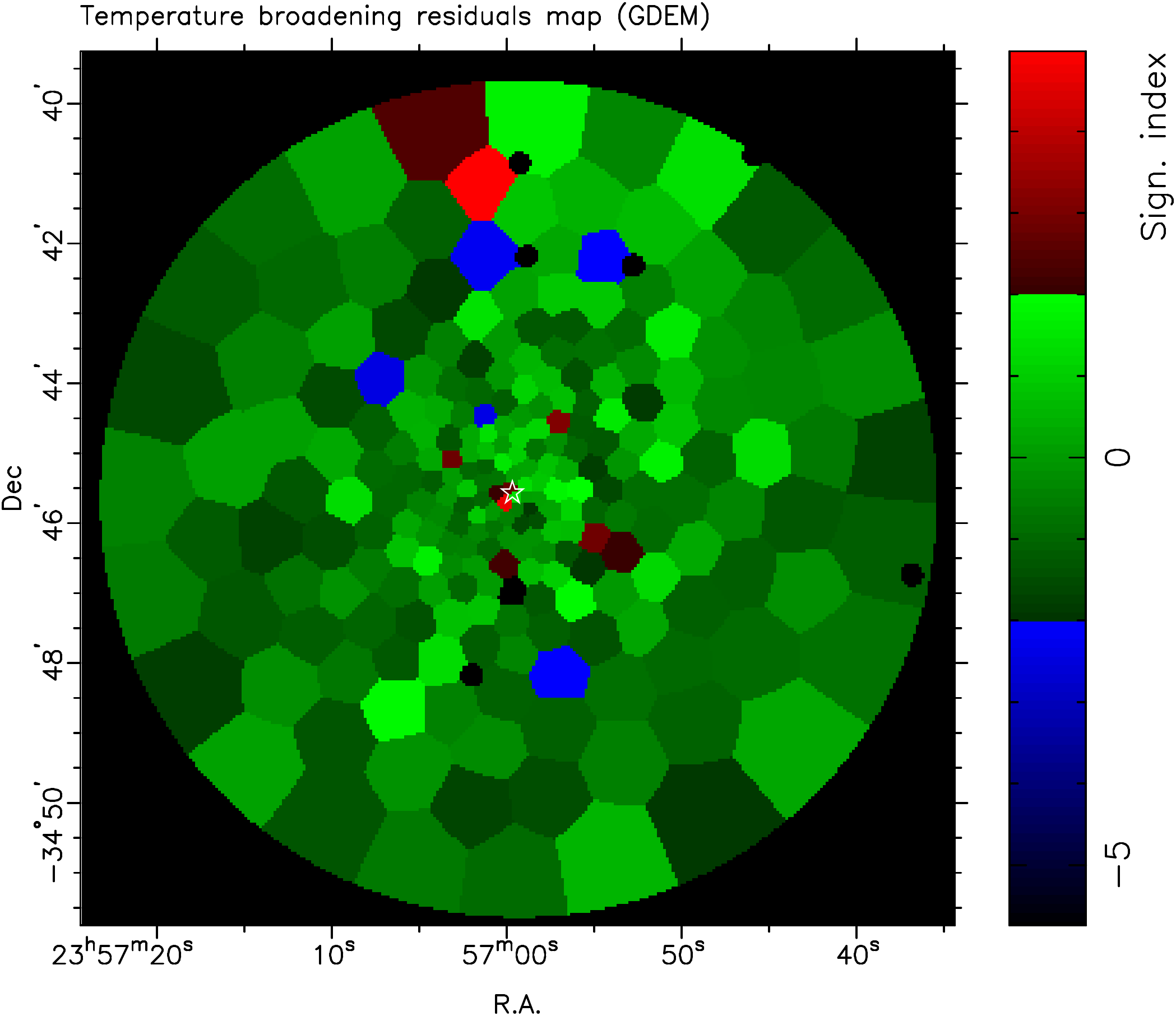}
\\
                \includegraphics[width=0.33\textwidth]{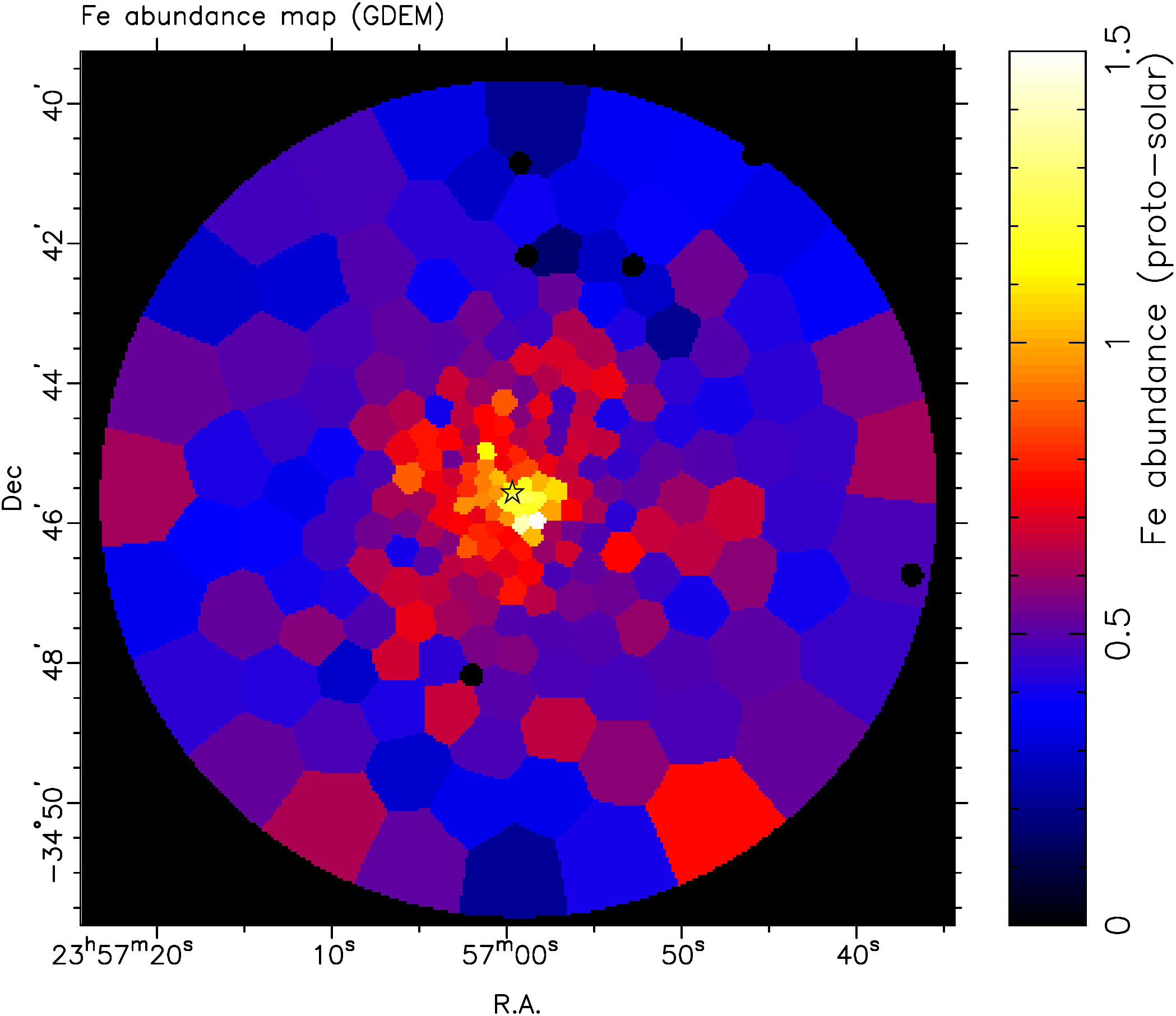}
                \includegraphics[width=0.33\textwidth]{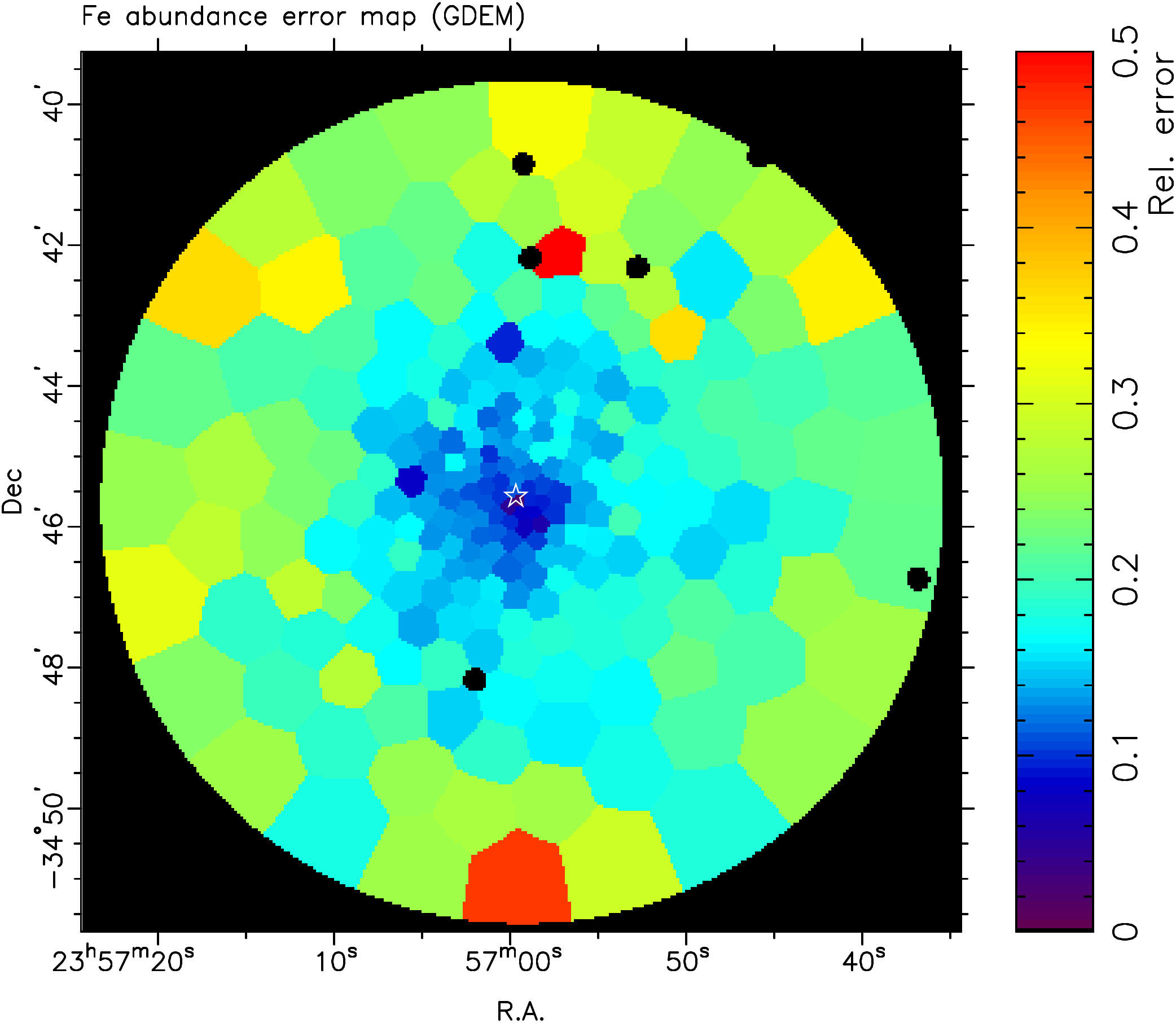}
                \includegraphics[width=0.33\textwidth]{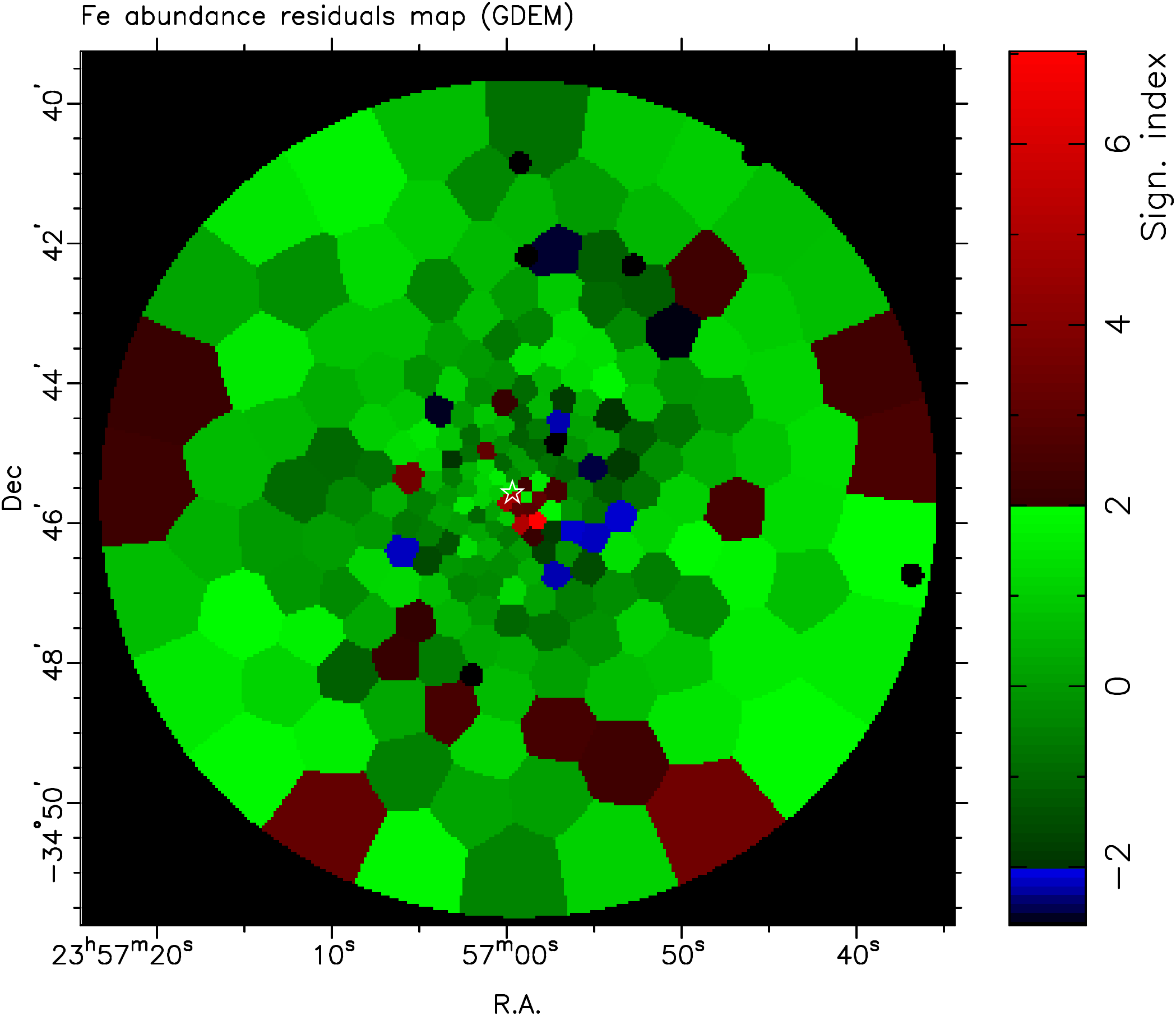}
        \caption{From upper to lower panels: $kT$, $\sigma_T$ and Fe abundance maps of A\,4059. The left panels show the basic maps (using a \textit{GDEM} model). The middle panels show their corresponding absolute ($\Delta \sigma_T$) or relative ($\Delta T/T$; $\Delta \text{Fe}/\text{Fe}$) errors. The right panels show their corresponding residuals (see text). In the centre of each map, the (black or white) star shows the peak of X-ray emission. All the maps cover $R \le 6$ arcmin of FoV.}\label{fig:maps}
\end{figure*}

In order to emphasise the impact of the statistical errors on the maps and to possibly reveal substructures, we create so-called residuals maps following the method of \citet{2011A&A...528A..60L}. In each cell, we subtract from each measured parameter the respective value estimated from our modelled radial profile (Fig. \ref{fig:radial_profiles}) at the distance $r$ of the geometric centre of the cell. The significance index is defined as being this difference divided by the error on the measured parameter. The $kT$, $\sigma_T$, and Fe abundance maps and their respective error and residuals maps are shown in Fig. \ref{fig:maps}.

The $kT$ map reveals the cool core of the cluster in detail. It appears to be asymmetric and to have a roughly conic shape extending from the north to the east and pointing toward the south-west. Along this axis, the temperature gradient is steeper to the south-west than to the north-east  of the core. Most of the relative errors obtained with the \textit{CIE} model (not shown here) are within 2--5\%, which is in agreement with our expectations (Appendix \ref{sect:maps_simulations}); however, they slightly increase with radius. This trend is stronger when using the \textit{GDEM} model, and the errors are somewhat larger.
A very local part ($\sim$5 cells) of the core is up to 8$\sigma$ cooler than our modelled temperature profile. This coldest part is offset $\sim$25$''$ SW from the X-ray peak emission. This contrasts with the western part of the core, which shows a significantly hotter bow than the average $\sim$55$''$ away from the X-ray peak emission. We also note that some outer cells are found significantly (>2$\sigma$) colder or hotter than the radial trend.

The $\sigma_T$ map confirms the positive $\sigma_T$ measurements in most of the cells outside the core, typically within 0.1--0.4. Globally, $\sigma_T$ is consistent with that measured from the $\sigma_T$ radial profile. We note that outside the core the errors are  inhomogeneous and are sometimes hard to estimate precisely.

The Fe map also shows that the core is asymmetric. As it is  in the $kT$ map, the abundance gradient from the core toward the south-west is steeper than toward the north-east. The highest Fe emitting region is found to be $\sim$25$''$ SW offset from the X-ray peak emission and coincides with the coldest region. In this offset SW region, Fe is measured to be more than 7$\sigma$ over-abundant.

We note that the smallest cells ($\sim$12$"$) have a size comparable to the EPIC PSF ($\sim$6$"$ FWHM); a contamination from leaking photons between adjacent cells might thus slightly affect our mapping analysis. However, the PSF has a smoothing effect on the spatial distributions, and gradients may be only stronger than they actually show in the map. This does not affect our conclusion of important asymmetries of temperature and Fe abundance in the core of A\,4059.

\section{Discussion}\label{sect:discussion}

We determined the temperature distribution and the elemental abundances of O, Ne, Si, S, Ar, Ca, and Fe in the core region ($\le 3'$) of A\,4059 and in eight concentric annuli centred on the core. In addition, we built 2-D maps of the mean temperature ($kT$), the temperature broadening ($\sigma_T$), and the Fe abundance. Because of the large cross-calibration uncertainties, Mg and Ni abundances are not reliable in these datasets using EPIC, and we  prefer to measure the Mg abundance using RGS instead.

\subsection{Abundance uncertainties and SNe yields}\label{subsect:abun_core}

As shown in Table \ref{table:core2}, the Ne abundance measured using EPIC depends strongly on the choice of the modelled temperature distribution. The main Ne lines are hidden in the Fe-L complex, around $\sim$1 keV. This complex contains many strong Fe lines and is extremely sensitive to  temperature. A slight change in the temperature distribution will thus significantly affect the Ne abundance measurement, making it not very reliable using EPIC \citep[see also][]{2006A&A...449..475W}. For the same reason, Fe abundances of single- and multi-temperature models might change slightly but already cause a significant difference between both models. 

Most of the discrepancies in the abundance determination between the EPIC instruments  come  from an incorrect estimation of the lines and/or the continuum in pn (Sect. \ref{subsect:core_EPIC}). Cross-calibration issues between MOS and pn have been already reported \citep[see e.g.][]{2007A&A...465..345D,2014arXiv1404.7130S}, but their deterioration has probably increased over time despite current calibration efforts \citep{2014A&A...564A..75R}. Our analysis using the Gauss method (Table \ref{table:EWs} and Fig. \ref{fig:core_abundances}) suggests that in general MOS is more reliable than pn in our case, even though MOS might slightly overestimate some elements as well (e.g. Mg, S, or even Fe). In all cases, this latest method is the most robust one with which to estimate the abundances in the core using EPIC.

\begin{figure*}
        \centering
                \includegraphics[width=0.48\textwidth]{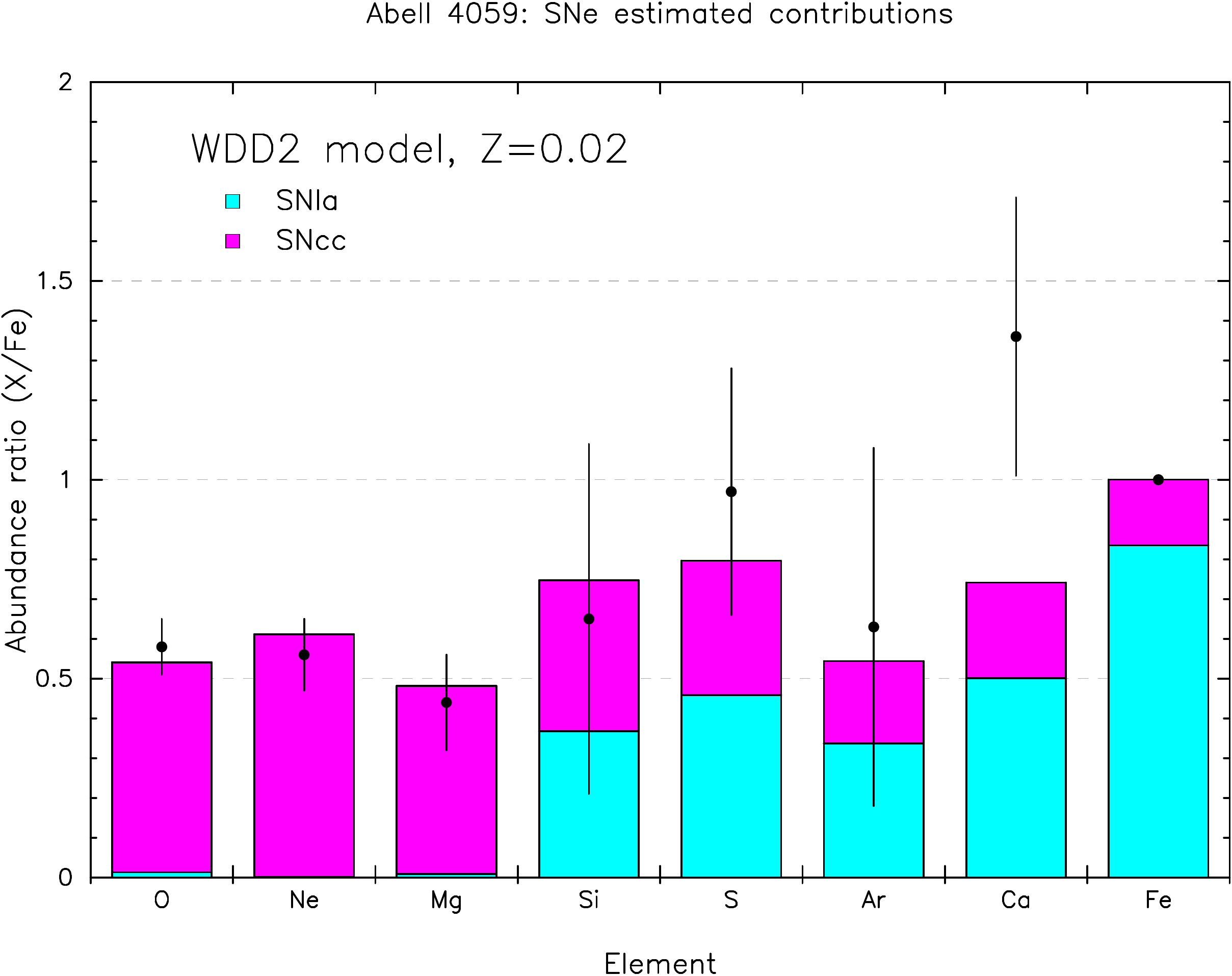}
                \includegraphics[width=0.48\textwidth]{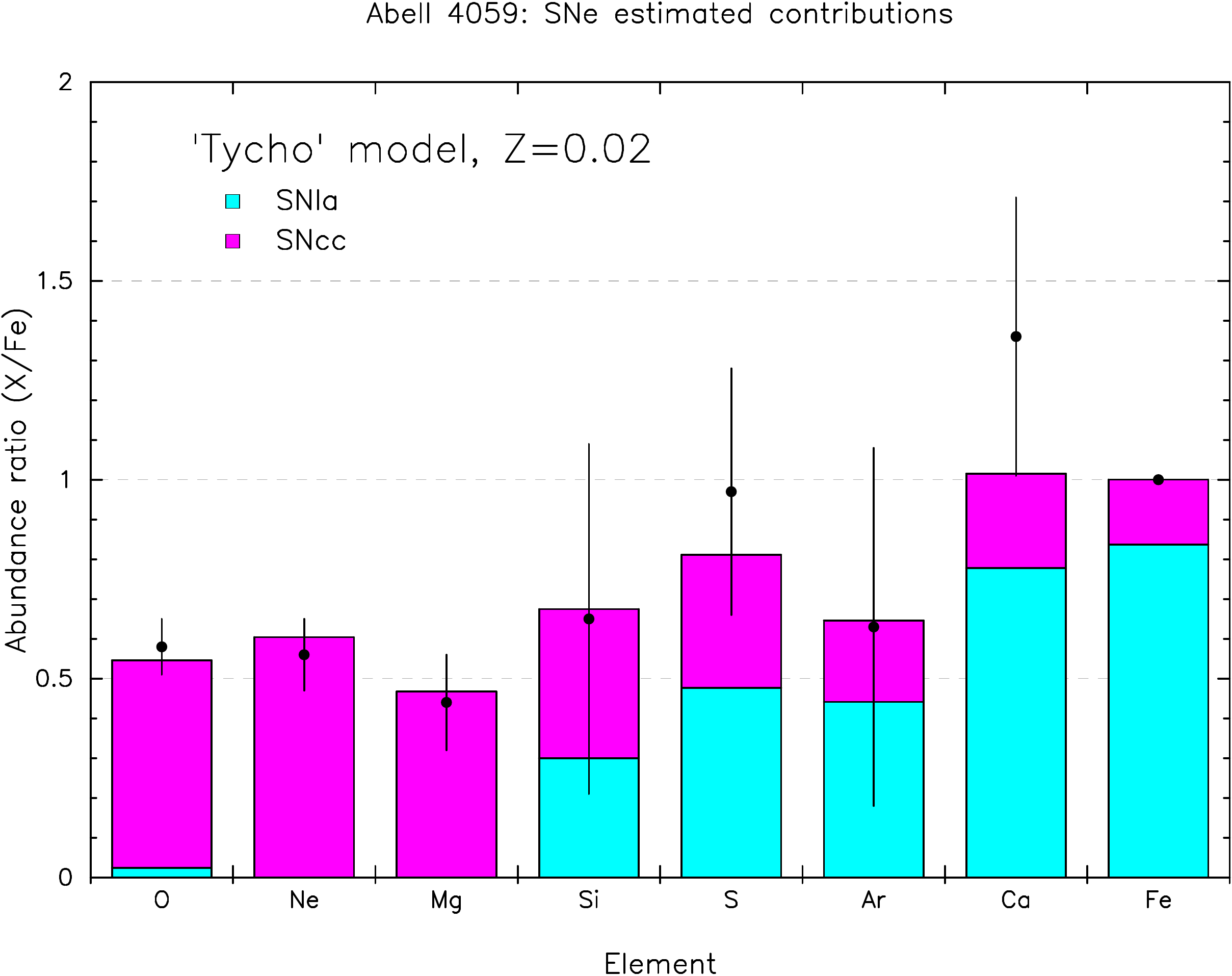}
        \caption{Comparison of our EPIC abundance measurements with standard SNe yield models. Left panel: WDD2 delayed-detonation SNIa model \citep{1999ApJS..125..439I}. Right panel: empirically modified delayed detonation SNIa model from the yields of the Tycho supernova \citep{2006ApJ...645.1373B}. The two models are computed with a Salpeter IMF and an initial metallicity of $Z=0.02$ \citep{2006NuPhA.777..424N}.}\label{fig:SNe_models}
\end{figure*}

Another interesting result is our detection of very high Ca/Fe abundances in the core. This trend has been already reported by \citet{2006A&A...452..397D} in S\'ersic 159-03 \citep[see also][]{2007A&A...465..345D}. Within 0.5$'$ the combined EPIC measurements give a Ca/Fe ratio of $2.0 \pm 0.3$. This is even higher than measured within 3$'$ (Ca/Fe $= 1.45 \pm 0.14$). 
Following the approach of \citet{2007A&A...465..345D} and assuming a Salpeter IMF, we select different SNIa models \citep[soft deflagration versus delayed-detonation,][]{1999ApJS..125..439I} as well as different initial metallicities affecting the yields from SNcc population \citep{2006NuPhA.777..424N}. We fit the constructed SNe models to our measured abundances in the core (O, Ne, Mg, and Si from RGS; Ar and Ca from EPIC; Fe from the Gauss method). We find that a WDD2 model, taken with $Z$$=$$0.02$ and a Salpeter IMF, reproduce our measurements best, with $(\chi^2/\text{d.o.f.})_\text{WDD2} = 4.28/6$ (Fig. \ref{fig:SNe_models}).
Although the fit is reasonable in terms of reduced $\chi^2$, it is unable to explain the high Ca/Fe value that we found. Based again on \citet{2007A&A...465..345D}, we also considered a delayed-detonation model that fitted the Tycho SNIa remnant best \citep{2006ApJ...645.1373B}. The fit is improved ($(\chi^2/\text{d.o.f.})_\text{Tycho} = 1.77/5$), but the model barely reaches the lower error bar of our measured Ca/Fe. Assuming that the problem is not fully solved even by using the latest model, we can raise two further hypotheses that might explain it:
\begin{enumerate}

\item Calcium abundance measurements might suffer from additional systematic uncertainties. Our analysis (Sects. \ref{subsect:core_EPIC} and \ref{sect:radial_profiles}) shows, however, that MOS and pn Ca/Fe measurements are consistent within the entire core (3$'$). Moreover, the continuum and EW of Ca lines ($\sim$3.9 keV) are correctly estimated by our \textit{CIE} models. Because of current efforts to limit them, uncertainties in the atomic database can contribute only partly. Finally the effective area at the position of this line is smooth and no instrumental-line feature is known around $\sim$3.9 keV.

\item Some SNe subclasses, so far ignored, might contribute to the metal enrichment in the ICM. For example, the so-called calcium-rich gap transients as a possible subclass of SNIa, are expected to produce a large amount of Ca even outside galaxies, making the transportation of Ca in the ICM much easier \citep{2014ApJ...780L..34M}.
\end{enumerate}

\subsection{Abundance radial profiles}\label{subsect:origins_metals}

All the abundance radial profiles decrease with radius. Interestingly, O shows a slight decrease (confirmed by our empirical fitted distribution), even though a flat profile cannot be fully excluded. This decreasing trend has been observed in other clusters, such as Hydra A \citep{2009A&A...495..721S}, A2029, and Centaurus \citep{2011A&A...528A..60L}. However, the observations of A\,496 \citep{2011A&A...528A..60L} and A\,1060 \citep{2007PASJ...59..299S} suggest a flatter profile. The O distribution is less clear in S\'ersic 159-03 \citep{2006A&A...452..397D,2011A&A...528A..60L}.

Moreover, only O and Fe profiles show abundances significantly higher than zero in the outermost annuli. The Fe profile is clearly peaked to the core, and agrees with typical slopes found in many other clusters \citep[e.g.][]{2009A&A...495..721S,2011A&A...528A..60L}. Moreover, its apparent plateau in the outer regions may suggest a constant Fe abundance in the ICM even outside $r_{500}$, as recently observed by \textit{Suzaku} in Perseus \citep{2013Natur.502..656W} and other clusters \citep[e.g.][]{2008A&A...487..461L,2011A&A...527A.134M}. As seen in Fig. \ref{fig:radial_profiles}, the Fe abundance found in the outskirts of Perseus ($0.303 \pm 0.012$, in proto-solar abundance units) is higher than what we find for A\,4059, even when accounting for the systematic uncertainties estimated from the core in Sect. \ref{subsect:core_EPIC}. This constant Fe abundance found in other cluster outskirts and this work suggest that the bulk of the enrichment at least by SNIa started in the early stages of the cluster formation.

In the previous cluster analyses where O appeared to be flat, the increase of O/Fe with radius is usually justified by arguing a very early population of SNIa and SNcc, starting after an intense star formation around $z \sim 2$--$3$ \citep{2006ApJ...651..142H} and undergoing a very efficient mixing all over the potential well, followed by a delayed population of SNIa responsible for the Fe peaked profile, and produced preferably in the central galaxy members in which a strong ram-pressure stripping is assumed \citep[see also discussion for S\'ersic 159-03 from][]{2006A&A...452..397D}. It has also been suggested that ram-pressure stripping could shape the Fe peak profile between $z=1$ and $z=0$ \citep{2005A&A...435L..25S}. However, \citet{2014A&A...567A.102D} suggest that the bulk of the Fe peak was already in place before $z=1$ in most clusters, meaning that at least SNIa type products started to get a centrally peaked distribution early on in the cluster formation. In fact, Fe seems to follow the near-infrared light profile of the central cD galaxies much better at $z=1$ than at $z=0$, suggesting that most of the current mixing mechanisms tend to spread out the metals in the ICM. 

The decreasing O radial profile measured in this work suggests that the same kind of scenario is likely for SNcc type products. Although its best-fit slope of the profile appears to be flatter than the slope of the Fe radial profile (Table \ref{table:radial_models}), the O/Fe radial values are still compatible with a constant distribution (except possibly for the 6'-9' annulus, where systematics might affect the O measurements). Consequently, it is not necessary to invoke a delayed population of SNIa and/or SNcc occurring after $z=1$, although it might contribute to a minor part of the metals found in the core. At $z \sim 2$--$3$ the central cD galaxy and its surrounding galaxy members were already actively star-forming  and could have produced the bulk of all metals observed in the core, probably injected into the ICM through galactic winds. More recently, ram-pressure stripping could have also played a minor role  in the enrichment of the core, for example to explain the asymmetry found on the maps (see below).

Assuming a flat and positive distribution of Fe and O beyond the FoV, the mixing of the metals is likely very efficient in the outskirts, where the entropy is high. In the core however, the entropy was already very stratified early on without any major mergers to disturb it, and the mixing mechanisms could be less efficient there.

While O and Fe are detected far from the core and this favours  an early initial enrichment from SNIa and SNcc types, puzzlingly we do not detect significant abundances of Ne and Si in the outermost annuli. This result is less striking in the S and Ar radial measurements, even though our fitted trends give small upper limits for $D_\infty$. Nevertheless, abundance measurements in the outer parts of the FoV can also suffer from additional systematic uncertainties related to the background contribution. These uncertainties may explain our lack of clear detection of Ne, Si, S, and Ar in the outermost annuli. Finally, we note the similarity between the Si and S profiles, already reported in the cD galaxy M\,87 by \citet{2011MNRAS.418.2744M}.

In addition to these radial trends, our maps show local regions of anomalously rich Fe abundance in the core. This is particularly striking in the south-west ridge, where the Fe abundance is $>7$$\sigma$ higher than the average trend from its corresponding radial profile. Since no galaxy can be associated with this particular region, it is hard to explain its enrichment with galactic winds. As previously reported and discussed by \citet{2008ApJ...679.1181R}, it is possible that an important part of the metals in the core comes from one early starburst galaxy that passed very close to the cD central galaxy before the onset of the central AGN. In this case ram-pressure stripping could probably have played a dominant role in the enrichment within $\sim$0.5 arcmin after the initial enrichment seen through the radial profiles. This possible scenario is also discussed in the next section.

\subsection{Temperature structures and asymmetries}\label{subsect:cl_hist}

Although the ICM appears  homogeneous and symmetric at large scale, the inner part appears to be more asymmetric (Fig. \ref{fig:A4059bis}). As already observed in the past by \textit{Chandra} \citep{2002ApJ...569L..79H,2008ApJ...679.1181R}, the south-west ridge is clearly visible as an additional peaked X-ray emission near the core, and a diffuse tail from the core toward the north-east can also be  detected.

Evidence of asymmetries are also found in our spectral analyses. Although our radial $kT$ profile looks similar to other cool-core clusters, our $kT$ and Fe abundance maps show clear inhomogeneities in the ICM structure of A\,4059.  Compared to the 2-D maps previously measured using \textit{Chandra} \citep{2008ApJ...679.1181R}, the $S/N$ of the cells in our EPIC maps are $\sim$3.3 and $\sim$2.5 times greater for $kT$ and the Fe abundance, respectively, allowing us to confirm these substructures with a higher precision and over a larger FoV.

First, like the Fe abundance, the temperature gradient is steeper within the south-west ridge than north-east of the core. The central core (including the south-west ridge) is also significantly colder ($\sim$2.3 keV) and the south-west ridge has a higher Fe abundance ($\sim$1.5) than the rest of the core within $0.5'$. These results confirm the previous study by \citet{2008ApJ...679.1181R} who also found strong asymmetry in the core of A\,4059 using \textit{Chandra}. Their pressure map shows neither asymmetry nor discontinuity in the core, even around the south-west ridge. From both \textit{Chandra} and XMM-\textit{Newton} studies, it is clear that this ridge plays a role in the metal enrichment of the core (see also Sect. \ref{subsect:origins_metals}) and must be closely linked to the history of the cluster \citep{2008ApJ...679.1181R}. The hotter bow region found W of the core is likely related to it. Based on the \textit{Chandra} images \citep{2008ApJ...679.1181R}, sloshing seems an unlikely  explaination for  the origin of the ridge. Indeed, it appears to be a second brightness peak separated from the core, and its particular morphology is very different from the typical spiral regular pattern of sloshing fronts \citep[see e.g.][]{2013ApJ...773..114P,2014arXiv1410.1955I}. Another scenario is that this local cool, dense, and Fe-rich asymmetry was already present before the triggering of central AGN radio-activity; it was formed by a gas-rich late-type galaxy that plunged very close to the central cD galaxy. An intense starburst caused by  its interactions with the dense local ICM occurred and it lost an important part of its metals as a result of  the strong gravitational interaction coupled with intense ram-pressure stripping.

\begin{figure}
\resizebox{\hsize}{!}{
\includegraphics{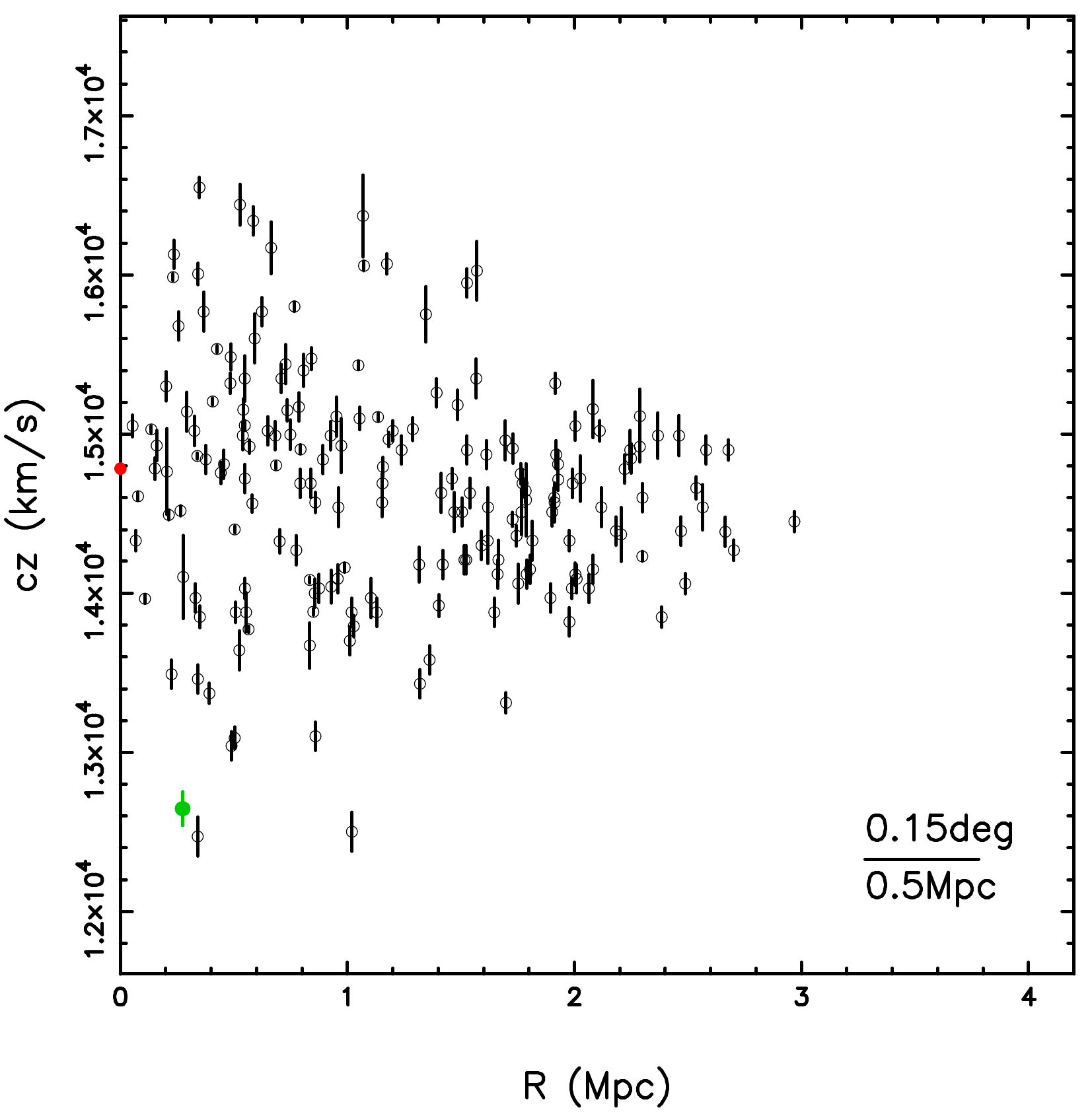}}
\caption{Line-of-sight velocity versus projected distance from the central cD galaxy for the member galaxies with optical spectroscopic redshifts in A\,4059 taken from \citet{2011A&A...526A.105Z}. The central cD galaxy is shown in red. The location of spiral galaxy ESO 349-G009 (green) in the caustic indicates that it belongs to the cluster.
\label{fig:cluster_members}}
\end{figure}

\begin{figure}
\resizebox{\hsize}{!}{
\includegraphics[trim=0.5cm 0.05cm 1.5cm 0cm, clip=true, width=\textwidth]{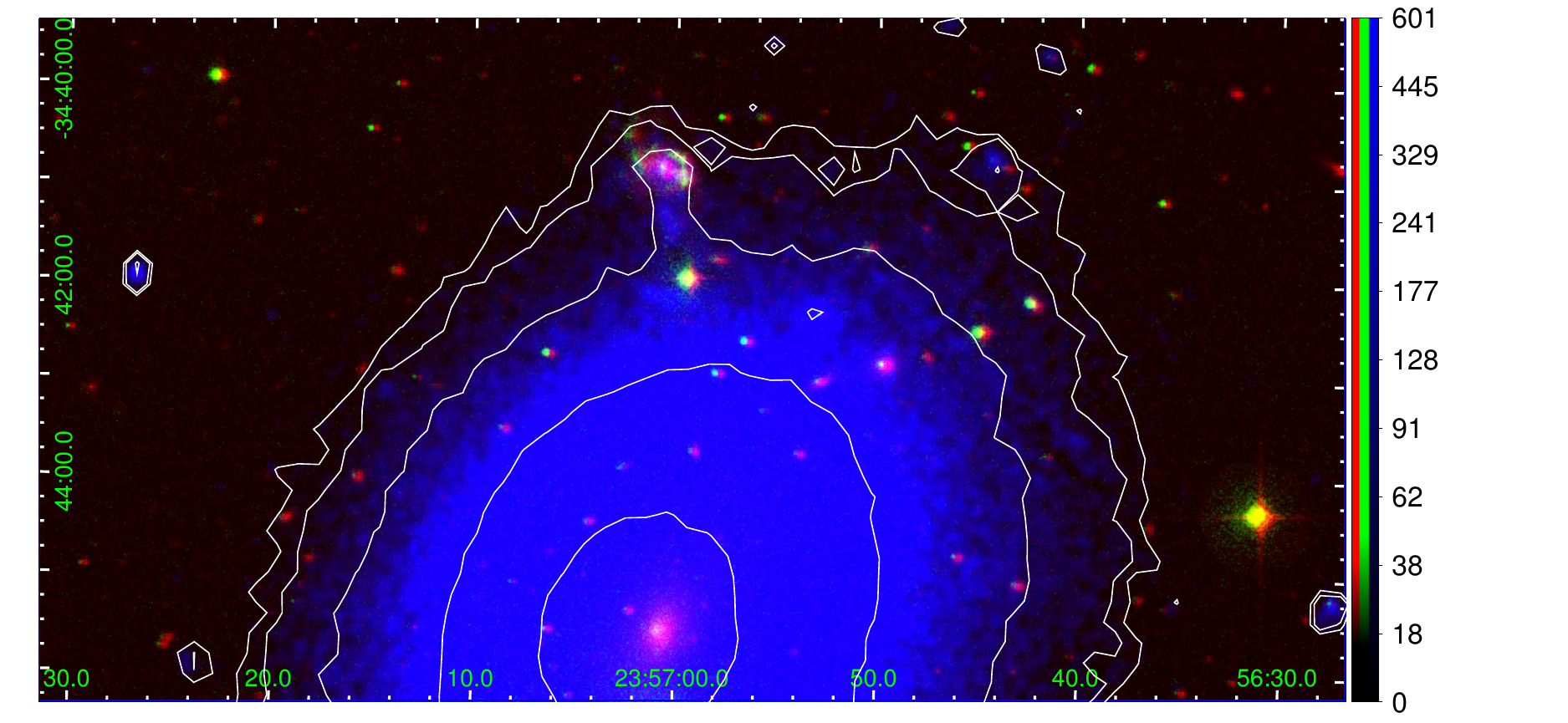}}
\caption{RGB mosaic of the central-north part of A\,4059. Red: optical filter (UK Schmidt telescope, public data). Green: UVM2 filter (OM instrument). Blue and contours: X-rays (EPIC MOS2+pn). The spiral galaxy ESO 349-G009 and the central cD galaxy are in the top and the bottom of the image respectively.
\label{fig:RGB_mosaic}}
\end{figure}

\citet{2008ApJ...679.1181R} estimated that such a galaxy should be within $300 v_3$ kpc of the cluster core. They suggested the bright spiral galaxy ESO 349-G009 as being a good candidate, although they were not sure whether this object belongs to A\,4059. Looking at the caustic taken from \citet[][see individual galaxy redshifts in the references therein, e.g. \citealt{2005ASPC..329..283A}]{2011A&A...526A.105Z}, we can confirm that this is indeed the case (Fig. \ref{fig:cluster_members}). The galaxy is located in the front part of the cluster and moves with a high radial velocity compared to the cD galaxy ($\Delta v \simeq 1800$ km/s). Assuming that this scenario is correct and that the movement of this galaxy near the central cD galaxy was essentially along the line of sight, the absence of an obvious metal tail from ram-pressure stripping on the plane of the FoV is naturally explained. Moreover, the X-ray isophotes joining ESO 349-G009 and the cluster ICM (Fig. \ref{fig:RGB_mosaic}) show an interaction between them and might suggest that the galaxy is escaping from the core. The UV light detected in its arms using the XMM-\textit{Newton} OM instrument (e.g. UVM2 filter) reveals that the galaxy still has a high star formation rate. The gas mass of the ridge ($5 \times 10^9$ $M_\sun$) is a small percentage of the total stellar mass of ESO 349-G009 \citep{2008ApJ...679.1181R}.

Finally, both the radial profile and map reveal a constant or increasing trend of $\sigma_T$ with radius. This is likely explained by projection effects such as the increased effective length along our line of sight. For cooling core clusters, this effective length increases as a function of radius, and a longer effective length will mix more temperatures along the line of sight. A still broad range of temperatures in the local ICM beyond the core cannot be fully excluded, but seems more unlikely. Indeed, although the few outer local colder or hotter cells found in the $kT$ residuals map (Fig. \ref{fig:maps}, top-right) might argue in favour of this second explanation, the temperature (and thus $\sigma_T$) measurements in the outer map cells are very sensitive to the background modelling, and are thus affected by these additional systematic uncertainties.

\section{Conclusions}\label{sect:conclusion}

In this paper we have studied a very deep XMM-\textit{Newton} observation ($\sim$140 ks of net exposure time) of the nearby cool-core cluster A\,4059. Several temperature and abundance parameters have been derived from the spectra both in the core and in eight concentric annuli; moreover, we were able to derive $kT$, $\sigma_T$, and Fe abundance maps. We conclude the following:

\begin{itemize}
\item   The temperature structure shows the cool-core and in addition increasing deviations from apparent isothermality in and out of the core.

\item   The abundances of O, Ne, Si, S, Ar, Ca, and Fe all are peaked toward the core, and we report the presence of Fe and O beyond $\sim$0.3 Mpc from the core. This suggests that the enrichment from SNIa and SNcc started early on in the cluster formation, probably through galactic winds in the young galaxy members.

\item   The EPIC image as well as the temperature and Fe abundance maps reveal strong asymmetries in the cluster core. We confirm a colder and Fe-richer ridge south-west of the core, previously found by \textit{Chandra}, perhaps due to an intense ram-pressure stripping event. Therefore, in addition to an early enrichment through galactic winds, ram-pressure stripping might have greatly contributed to a more recent enrichment of the inner core.

\item   The Ca/Fe abundance ratio in the core is particularly high ($1.45 \pm 0.14$ using a combined EPIC fit), even accounting for systematic uncertainties. If we assume the Ca/Fe abundance of the entire core to be genuine, it is unlikely explained by current standard SNe yields models. Recently proposed calcium-rich gap transient SNIa might be an interesting alternative with which to explain the high Ca abundance generally found in the ICM.

\item   Because of cross-calibration issues, the EPIC MOS and pn detectors measure significantly different values of  temperature and most abundances. Although this leads to systematic uncertainties on their absolute values, the discrepancies are generally smaller when considering abundances relative to Fe. Moreover, it should not affect relative differences between spectra from different regions if  the same instrument(s) are used. Fitting a Gaussian line and a local continuum instead of CIE models is a robust method to measure more reliable abundances.
\end{itemize}

\begin{acknowledgements}
This work is a part of the CHEERS (CHEmical Evolution Rgs cluster Sample) collaboration. We would like to thank its members and the anonymous referee for their feedback and discussions. FM thanks Huub R\"ottgering and Darko Donevski for useful discussions. LL acknowledges support by the DFG through grant LO 2009/1-1. YYZ acknowledges support by the German BMWi through the Verbundforschung under grant 50\,OR\,1304. This work is based on observations obtained with XMM-\textit{Newton}, an ESA science mission with instruments and contributions directly funded by ESA member states and the USA (NASA). The SRON Netherlands Institute for Space Research is supported financially by NWO, the Netherlands Organisation for Space Research.
\end{acknowledgements}

\bibliography{A4059}{}
\bibliographystyle{aa}




\newpage

\appendix

\section{Detailled data reduction}\label{sect:details_reduction}
\subsection{GTI filtering}\label{subsect:gti_filtering} 

In order to reduce the soft-proton (SP) background, we build good time intervals (GTI) using the light curves in the 10--12 keV band for MOS and 12--14 keV band for pn. We fit the count-rate histograms from the light curves of each instrument, binned in 100 s intervals, with a Poissonian function and we reject all time bins for which the number of counts lies outside the interval $\mu \pm 2\sigma$ (i.e. $\mu \pm 2\sqrt{\mu}$), where $\mu$ is the fitted average of the distribution. We repeat the same screening procedure and threshold (so-called 2$\sigma$-clipping) for 10 s binned histograms in the 0.3--10 keV band because \citet{2004A&A...419..837D} reported episodes of particularly soft background flares. In order to get a qualitative estimation of the residual SP flare contamination, we use the \texttt{Fin\_over\_Fout} algorithm which compares the count rates in and out of the FoV of each detector \citep{2004A&A...419..837D}. We found that in both observations MOS\,1 displays a $F_\text{in}/F_\text{out}$ ratio higher than 1.3, meaning that the observations have been significantly contaminated by SP events. This value is still reasonable though, and a look at the filtered light curve lead us to  keep the MOS\,1 datasets. Furthermore, a careful modelling of convenient SP spectral components are used in our spectral fittings as well (see Appendix \ref{sect:bg_modelling}).

\subsection{Resolved point sources excision}\label{subsect:rps} 

The point sources in our FoV contribute to the total flux and may bias the astrophysical results that we aim to derive from the cluster emission. Therefore, they should be discarded. We detect all the resolved point sources (RPS) with the SAS task \texttt{edetect\_chain} and we proceed with a second check by eye in order to discard erroneous detections and possibly include a few missing candidates. It is common practice in extended source analysis to excise bright point sources from the EPIC data. We note, however, that many sources have fluxes below the detection limit $S_\text{cut}$ and an unresolved component might remain (Appendix \ref{subsect:UPS}). 

A remaining problem is how to choose the excision radius in the best way. A very small excision radius may leave residual flux from the excised point sources while a very large radius may cut out a significant fraction of the cluster emission leading to decreased $S/N$. We define $A_\text{eff}$ as the extraction region area for the cluster emission when the point sources are excised with a radius $r_\text{s}$,

\begin{equation} \label{eq:RPS1}
A_\text{eff} = A\left(1-\pi r_\text{s}^2 \int_{S_\text{cut}}^{\infty} \left(\frac{dN}{dS}\right)dS\right)
,\end{equation}where $N$ is the number of sources, $S$ is the flux, and $A$ is the full detection area.

Since we are dealing with a Poissonian process, $S/N$ can be estimated as $S/N = \frac{C}{\sqrt{C+B}}$, where $C$ and $B$ are the number of counts of the cluster emission and the total background, respectively. The value of $C$ depends on the extraction area and can thus be written $C=C^*A_\text{eff}$, where $C^*$ is the local surface brightness of the cluster (counts/$''$), and $B$ can be divided into the instrumental or hard particle (HP) background $I$, an unresolved point sources (UPS) component, and the remaining excised point source flux outside the excision region. The total background can be thus written as

\begin{equation} \label{eq:RPS2}
B = I + \int_{0}^{S_\text{cut}} S \left(\frac{dN}{dS}\right) dS + (1-EEF(r_\text{s})) \int_{S_\text{cut}}^{\infty} S \left(\frac{dN}{dS}\right)dS
,\end{equation}
where $EEF(r_\text{s})$ is the encircled energy fraction of the PSF as a function of radius.
We can finally write the total $S/N$ as
\begin{equation} \label{eq:RPS3}
S/N = \frac{C^*\sqrt{A\left(1-\pi r_\text{s}^2 \int_{S_\text{cut}}^{\infty} \left(\frac{dN}{dS}\right)dS\right)}}{\sqrt{C^*+I + \int_{0}^{S_\text{cut}} S \left(\frac{dN}{dS}\right) dS + (1-EEF(r_\text{s})) \int_{S_\text{cut}}^{\infty} S \left(\frac{dN}{dS}\right)dS}}
.\end{equation}

The optimum $S/N$ can be then computed as a function of $r_\text{s}$ and $S_\text{cut}$ (Eq. \ref{eq:RPS3}). In Appendix \ref{subsect:UPS} we discuss the origin of $dN/dS$. We find and adopt an optimised radius for RPS excision in our dataset of $\sim$10$''$.

\subsection{RGS spectral broadening correction from MOS\,1 image}\label{sect:RGS_M1image}

Because the RGS spectrometers are slitless and the source is spatially extended in the dispersion direction, the RGS spectra are broadened. The effect of the broadening of a spectrum by the spatial extent of the source is given by
\begin{equation} \label{eq:RGS_broadening}
\Delta\lambda = \frac{0.138}{m} \Delta\theta{\AA}
,\end{equation}
where $m$ is the spectral order and $\theta$ is the offset angle in arcmin (see the XMM-\textit{Newton} Users Handbook).

The MOS\,1 DET\,Y direction is parallel to the RGS dispersion direction. Therefore, we extract the brightness profile of the source in the dispersion direction from the MOS\,1 image and use this to account for the broadening following the method described by \citet{2004A&A...420..135T}. This method is implemented through the \texttt{Rgsvprof} task in SPEX. As an input of this task, we choose a width of 10$'$ around the core and along the dispersion axis, in which the cumulative brightness profile is estimated.
In order to correct for continuum and background, we use a MOS\,1 image extracted within 0.5–1.8 keV (i.e. the RGS energy band).
This procedure is applied to both observations and we average the two spatial profiles obtaining a single profile that will be used for the stacked RGS spectrum.

\section{EPIC background modelling}\label{sect:bg_modelling}

We split the total EPIC background into two categories, divided further into several components:
\begin{enumerate}
        \item Astrophysical X-ray background (AXB), from the emission of astrophysical sources and thus folded by the response files. The AXB includes the Local Hot Bubble (LHB), the galactic thermal emission (GTE), and the UPS.
        \item Non-X-ray background (NXB), consisting of soft or hard particles hitting the CCD chips and considered  as photon events. For this reason, they are \emph{\emph{not}} folded by the response files. The NXB contains the SP and the HP backgrounds.
\end{enumerate}

In total, five components are thus carefully modelled.

\subsection{Hard particle background}\label{subsect:HP}

High energy particles are able to reach the EPIC detectors from every direction, even when the filter wheel is closed. Besides continuum emission, they also produce instrumental fluorescence lines which should be carefully modelled. Moreover, for low $S/N$ areas, we observe a soft tail in the spectra due to the intrinsic noise of the detector chips. A good estimate of the HP background can be obtained by using Filter Wheel Closed (FWC) data which are publicly available on the XMM-\textit{Newton} SOC webpage\footnote{http://xmm.esac.esa.int}. We select FWC data that were taken on 1 October 2011 and 28 April 2011 with an exposure time of 53.7 ks and 35.5 ks for MOS and pn, respectively. We removed the MOS\,1 events from CCD3, CCD4, and CCD6 to be consistent with our current dataset.

Instead of subtracting directly the FWC events from our observed spectra, modelling  the HP background directly  allows a much more precise estimate of the instrumental lines fluxes, which are known to vary across the detector \citep{Snowden_Kuntz}.

We fit the individual FWC MOS and pn continuum spectra with a broken power law $F(E) = Y E^{-\Gamma} e^{\eta(E)}$ where $\eta(E)$ is given by 
\begin{equation} \label{eq:broken_pl}
\eta(E) = \frac{r\xi + \sqrt{r^2 \xi^2 + b^2 (1-r^2)}}{1-r^2}
\end{equation}
with $\xi = \ln (E/E_0)$ and $r = \frac{\sqrt{1 + (\Delta\Gamma)^2} - 1}{|\Delta\Gamma|}$ (see SPEX manual). In this model, the independent parameters are $A$, $\Gamma$ (spectra index), $\Delta\Gamma$ (spectral index break), $E_0$ (break energy), and $b$ (break strength). Unlike the instrumental lines, this continuum does not vary strongly across the detector. Tables \ref{table:complete-bg} and \ref{table:lines} show the best-fit parameters that we found for the entire FoV extraction area and the modelled instrumental lines, respectively. In addition to the broken power-law, each instrumental line is modelled with a narrow (FWHM $\leq 0.3$) Gaussian function. Although a delta function is more realistic, a in this case allowing a slight broadening optimises the correction for the energy redistribution on the instrumental lines.

\begin{table}
\caption{Best-fit parameters of the HP component, estimated from the full FoV of FWC observations. An equal sign (=) means that the MOS\,2 value is coupled with the MOS\,1 value.}
\label{table:complete-bg}
\resizebox{\hsize}{!}{
\begin{tabular}{l c@{$\pm$}l c@{$\pm$}l c@{$\pm$}l}
\hline 
\hline
Parameters & \multicolumn{2}{c}{MOS\,1} & \multicolumn{2}{c}{MOS\,2} & \multicolumn{2}{c}{pn}\tabularnewline
\hline
 $Y$ ($10^{46}$ ph/s/keV)  & $87.3$ & $1.2$ & $133.6$ & $1.6$ & $478$ & $117$ \tabularnewline

 $\Gamma$ & $0.33$ & $0.01$ & \multicolumn{2}{c}{=} & $1.37$ & $0.70$ \tabularnewline

 $\Delta\Gamma$ & $-0.18$ & $0.02$ & \multicolumn{2}{c}{=} & $-1.08$ & $0.25$ \tabularnewline

 $E_\text{break}$ (keV) & $3.49$ & $0.25$ & \multicolumn{2}{c}{=} & $1.05$ & $0.53$ \tabularnewline

 $b$ & \multicolumn{2}{c}{$\leq 0.01$} & \multicolumn{2}{c}{=} & $0.39$ & $0.17$ \tabularnewline
\hline

\end{tabular}}
\par

\end{table}

\begin{table}
\caption{Fluorescent instrumental lines produced by the hard particles. The centroid energies are adapted from \citet{Snowden_Kuntz} and \citet{Iakubovskyi_PhD} (MOS except Si K) and from our best-fit model (pn + MOS Si K).}
\label{table:lines}
\begin{centering}
\setlength{\tabcolsep}{12pt}
\begin{tabular}{cc|cc}

\hline
\hline
\multicolumn{2}{c|}{MOS} & \multicolumn{2}{c}{pn} \tabularnewline
\hline
Energy (keV) & Line & Energy (keV) & Line\tabularnewline
\hline
1.49 & Al K$\alpha$ & 1.48 & Al K$\alpha$ \tabularnewline
1.75 &  Si K$\alpha$ & 4.51 & Ti K$\alpha$ \tabularnewline
5.41 &  Cr K$\alpha$ & 5.42 & Cr K$\alpha$ \tabularnewline
5.90 &  Mn K$\alpha$ & 6.35 & Fe K$\alpha$ \tabularnewline
6.40 &  Fe K$\alpha$ & 7.47 & Ni K$\alpha$ \tabularnewline
7.48 &  Ni K$\alpha$ & 8.04 & Cu K$\alpha$ \tabularnewline
8.64 &  Zn K$\alpha$ & 8.60 & Zn K$\alpha$ \tabularnewline
9.71 &  Au L$\alpha$ & 8.90 & Cu K$\beta$ \tabularnewline
 &  & 9.57 & Zn K$\beta$ \tabularnewline

\hline

\end{tabular}
\par\end{centering}
\end{table}

\subsection{Unresolved point sources}\label{subsect:UPS}

An important component of the EPIC background is the contribution of UPS to the total X-ray background. Its flux can be estimated using the so called $\log N$--$\log S$ curve derived from blank field data. This curve describes how many sources are expected at a certain flux level. The source function has the form of a derivative ($dN/dS$) and can be integrated to estimate the number of sources in a certain flux range,
\begin{equation} \label{eq:UPS1}
N(<S) = \int_S^\infty \left(\frac{dN'}{dS'}dS'\right)
,\end{equation}
where $N$ is the number of sources and $S$ is the low-flux limit.

The most common bright UPS are AGNs, but  galaxies and hot stars contribute as well. Based on the \textit{Chandra} deep field, \citet{2012ApJ...752...46L} find that AGNs are the most dominant in terms of number counts, but in the 0.5--2 keV band the galaxy counts become higher than the AGN counts below a few times $10^{-28}$ W m$^{-2}$ deg$^{-2}$. The assumed spectral model of this component is a power-law with a photon index of $\Gamma$$=$$1.41$ \citep[see e.g.][]{2003A&A...408..431M,2004A&A...419..837D}. In reality, the power-law index may vary slightly between 1.4--1.5, given the uncertainties in the different surveys and estimations \citep{2009A&A...493..501M}. 
Based on the \textit{Chandra} Deep Field South (CDF-S) data, \citet{2012ApJ...752...46L} define the $(dN/dS)$ relations for each source category as follows:

\begin{equation} \label{eq:UPS2}
\begin{centering}
\frac{dN}{dS}^\text{AGN} = \begin{cases} K^\text{AGN}(S/S_\text{ref})^{-\beta_1^\text{AGN}} & (S\le f_b^\text{AGN}) \\ K^\text{AGN}(f_b/S_\text{ref})^{\beta_2^\text{AGN}-\beta_1^\text{AGN}}(S/S_\text{ref})^{-\beta_2^\text{AGN}} & (S> f_b^\text{AGN}) \end{cases}
\end{centering}
,\end{equation}
\begin{equation} \label{eq:UPS3}
\begin{centering}
\frac{dN}{dS}^\text{gal} = K^\text{gal}(S/S_\text{ref})^{-\beta^\text{gal}}
\end{centering}
,\end{equation}
\begin{equation} \label{eq:UPS4}
\begin{centering}
\frac{dN}{dS}^\text{star} = K^\text{star}(S/S_\text{ref})^{-\beta^\text{star}}
\end{centering}
.\end{equation}

Each relation describes a power law with a normalisation constant $K$ and a slope $\beta$. Since the ($dN/dS$) relation of AGNs shows a break, there is an additional $\beta_2$ parameter and a break flux $f_b$. The reference flux is defined as $S_\text{ref} \equiv 10^{-14}$ erg cm$^{-2}$ s$^{-1}$. The best-fit parameters for the studied energy bands are listed in Table 1 of \citet{2012ApJ...752...46L}.

The relations above can be used to estimate the flux from sources that are not detected in our EPIC observations. The UPS component also holds for the deepest \textit{Chandra} observations. \citet{2006ApJ...645...95H} found a detection limit of $1.4\times 10^{-16}$ in a 1 Ms CDF-S observation and estimated the unresolved flux to be $(3.4\pm1.7)\times 10^{-12}$ erg cm$^{-2}$ s$^{-1}$ deg$^{-2}$ in the 2--8 keV band. Since \textit{Chandra} has a much lower confusion limit and a narrower PSF, we do not expect EPIC to reach this detection limit even in a deep cluster observation. It is therefore not necessary to know the $\log N$--$\log S$ curve below this flux limit to obtain a reasonable estimate for the unresolved flux.

In the flux range from $1.4\times 10^{-16}$ up to the EPIC flux limit, we can calculate the flux using the $\log N$--$\log S$ relation. The total unresolved flux $\Omega_\text{UPS}$ for the 2--8 keV band is then calculated using
\begin{equation} \label{eq:UPS5}
\Omega_\text{UPS} = 3.4 \times10^{-12} + \int_{1.4\times 10^{-16}}^{S_\text{cut}} S'\left(\frac{dN}{dS'}\right) dS'\text{ erg cm$^{-2}$ s$^{-1}$ deg$^{-2}$}.
\end{equation}

Using the Eqs. \ref{eq:UPS2}, \ref{eq:UPS3}, and \ref{eq:UPS4} for $\frac{dN}{dS}$ in the integral above, the unresolved flux calculation is straightforward. Given the detection limit of our observations $S_\text{cut} = 3.83 \times 10^{-15}$ W m$^{-2}$, we find a total UPS flux of $8.07 \times 10^{-15}$ W m$^{-2}$ deg$^{-2}$. This value can be used to constrain the normalisation of the power-law component describing the AXB background in cluster spectral fits. We note that this method does not take the cosmic variance into account \citep[see e.g.][]{2003AN....324...24M}, which means that the normalisation may still be slightly biased.

\subsection{Local Hot Bubble and Galactic thermal emission}\label{subsect:LHB_GTE}

The LHB component is thought to originate from a shock region between the solar wind and our local interstellar medium \citep{2008ApJ...674..209K}, while the GTE is the X-ray thermal emission from the Milky Way halo. At soft energies (below $\sim$1 keV), the flux of these two foreground components is significant. They are both known to vary spatially across the sky, but we assume that they do not change significantly within the EPIC FoV. Both components are modelled with a \textit{CIE} component where we assume the abundances to be proto-solar. Both temperatures are left free, but are expected to be within 0.1--0.7 keV. The GTE component is absorbed by a gas with hydrogen column density ($N_H = 1.26 \times 10^{20}$ cm$^{-2}$), while the LHB component is not.

\subsection{Residual soft-proton component}

Even after  filtering  soft flare events from our raw datasets,   a quiescent level of SP remains that might affect the spectra, especially at low $S/N$ and above $\sim$1 keV. It is extremely hard to precisely estimate the normalisation and the shape of its spectrum since SP quiescent events strongly vary with detector position and time \citep{Snowden_Kuntz}. They may also depend on the attitude of the satellite. For these reasons, blank sky XMM-\textit{Newton} observations are not good enough for our deep exposures. The safest way to deal with this issue is to model the spectrum by a single power law \citep{Snowden_Kuntz}. Using a broken power law might be slightly more realistic, but the number of free parameters is then too high to make the fits stable. Although the spectral index $\Gamma$ of the power law is unfortunately unpredictable and may be different for MOS and pn instruments and  between different observations. Since \citet{Snowden_Kuntz} reported spectral indices between $\sim$0.1--1.4, we allow the $\Gamma$ parameter in our fits to vary within this range.

\subsection{Application to our datasets}

We apply the procedure described above for each component on our two observations of A\,4059. We extract an annular region with inner and outer radii of 6$'$ and 12$',$ respectively, and centred on the cluster core (Fig. \ref{fig:A4059}, the  outer two annuli), assuming that all the background components described above contribute to the detected events covered by this area. In order to get a better estimation of the foreground thermal emission (GTE and  LHB), we fit a \textit{ROSAT} PSPC spectrum from \citet{2011A&A...526A.105Z} simultaneously with our EPIC spectra. This additional observation covers an annulus centred to the core and with inner and outer radii of 28$'$ ($\sim$$r_{200}$) and 40$'$ ($\sim$$r_{200} +12'$), respectively, avoiding instrumental features and visible sources. We note that in this fit we also take the UPS contribution into account. Depending on the extraction area, all the normalisations (except for the UPS component, Appendix \ref{subsect:UPS}) are left free, but are properly coupled between each observation and instrument.

\begin{figure}
\resizebox{\hsize}{!}{
\includegraphics{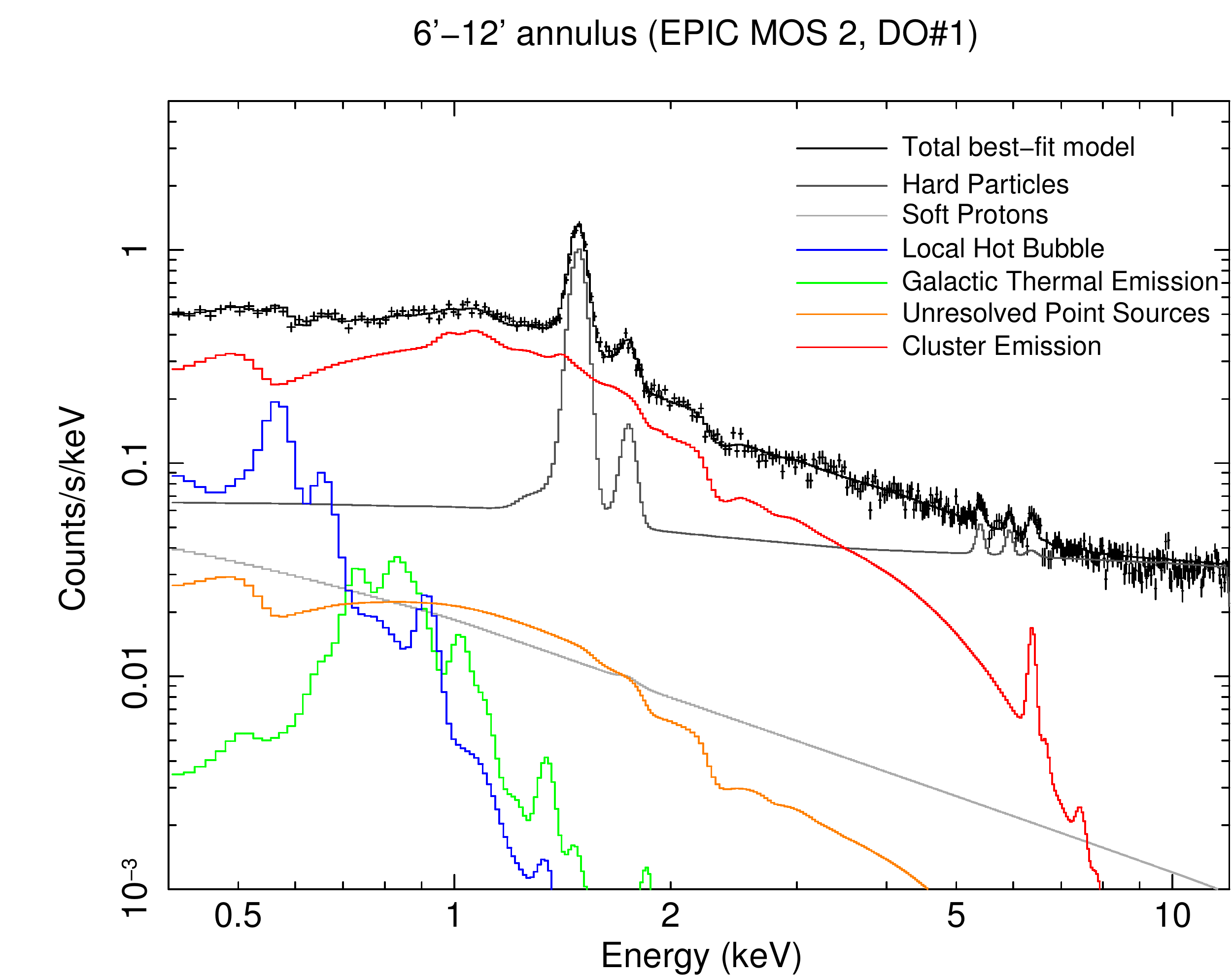}}
\caption{EPIC MOS2 spectrum of the 6$'$--12$'$ annular region around the core (see text). The solid black  line represents the total best-fit model. Its individual modelled components (background and cluster emission, solid coloured  lines) are also shown.}
\label{fig:background-modelling}
\end{figure}

Table \ref{table:background-modelling} shows the different background values that we found for the extracted annulus. Figure \ref{fig:background-modelling} shows the result for the MOS\,2 spectrum at the first observation, its best fit model, and the contribution of every modelled component. As expected, the NXB contribution is more important at high energies. Above $\sim$5 keV, the cluster emission is much smaller than the HP background. Consequently and as already reported, the temperature and abundances measured by EPIC are harder to estimate in the outer parts of the FoV.

\begin{table}
\caption{Best-fit parameter values of the total background estimated in the 6$'$--12$'$ annular region around the core (see text). A simple asterisk ($*$) means that the value reaches the upper or lower fixed range. An equal sign ($=$) means that the corresponding parameters from DO\,1 and DO\,2 are coupled together.}
\label{table:background-modelling}
\resizebox{\hsize}{!}{
\begin{tabular}{l l l c@{$\pm$}l c@{$\pm$}l}
\hline 
\hline
Bkg & Parameter &  Instrument & \multicolumn{2}{c}{DO\,1} & \multicolumn{2}{c}{DO\,2}\tabularnewline
comp. &  &  & \multicolumn{2}{c}{ } & \multicolumn{2}{c}{ }\tabularnewline
\hline
SP & Norm. ($10^{46}$ ph/s/keV) & MOS\,1 & $46$&$10$  & $30.3$&$4.5$  \tabularnewline
      &                                                & MOS\,2 & $18.1$&$9.4$  & $14.5$&$3.4$  \tabularnewline
      &                                                & pn & $22.8$&$4.2$  & $15.70$&$1.07$  \tabularnewline
      & $\Gamma$                            & MOS & $1.18$&$0.08$  & $1.63$&$0.11$  \tabularnewline
      &                                                & pn & $0.29$&$0.09$  & $0.10$&$_{0.00*}^{0.02}$  \tabularnewline

\hline
GTE & $Y$ ($10^{69}$ m$^{-3}$) & MOS+pn & $26.4$&$4.7$  & \multicolumn{2}{c}{$=$} \tabularnewline
 & $kT$ (keV)                                   &                     & $0.54$&$0.08$  & \multicolumn{2}{c}{$=$} \tabularnewline
\hline
LHB & $Y$ ($10^{69}$ m$^{-3}$) & MOS+pn & $311.8$&$5.1$  & \multicolumn{2}{c}{$=$} \tabularnewline
 & $kT$ (keV)                                   &                     & $0.168$&$0.002$  & \multicolumn{2}{c}{$=$} \tabularnewline
\hline
UPS & Norm. ($10^{49}$ ph/s/keV) & MOS+pn & \multicolumn{2}{c}{$58.29$ (fixed)}  & \multicolumn{2}{c}{$=$} \tabularnewline
      & $\Gamma$                              &                     & \multicolumn{2}{c}{$1.41$ (fixed)}  & \multicolumn{2}{c}{$=$}  \tabularnewline
\hline

\end{tabular}}
\par

\end{table}

\begin{figure*}
        \centering
                \includegraphics[width=0.49\textwidth]{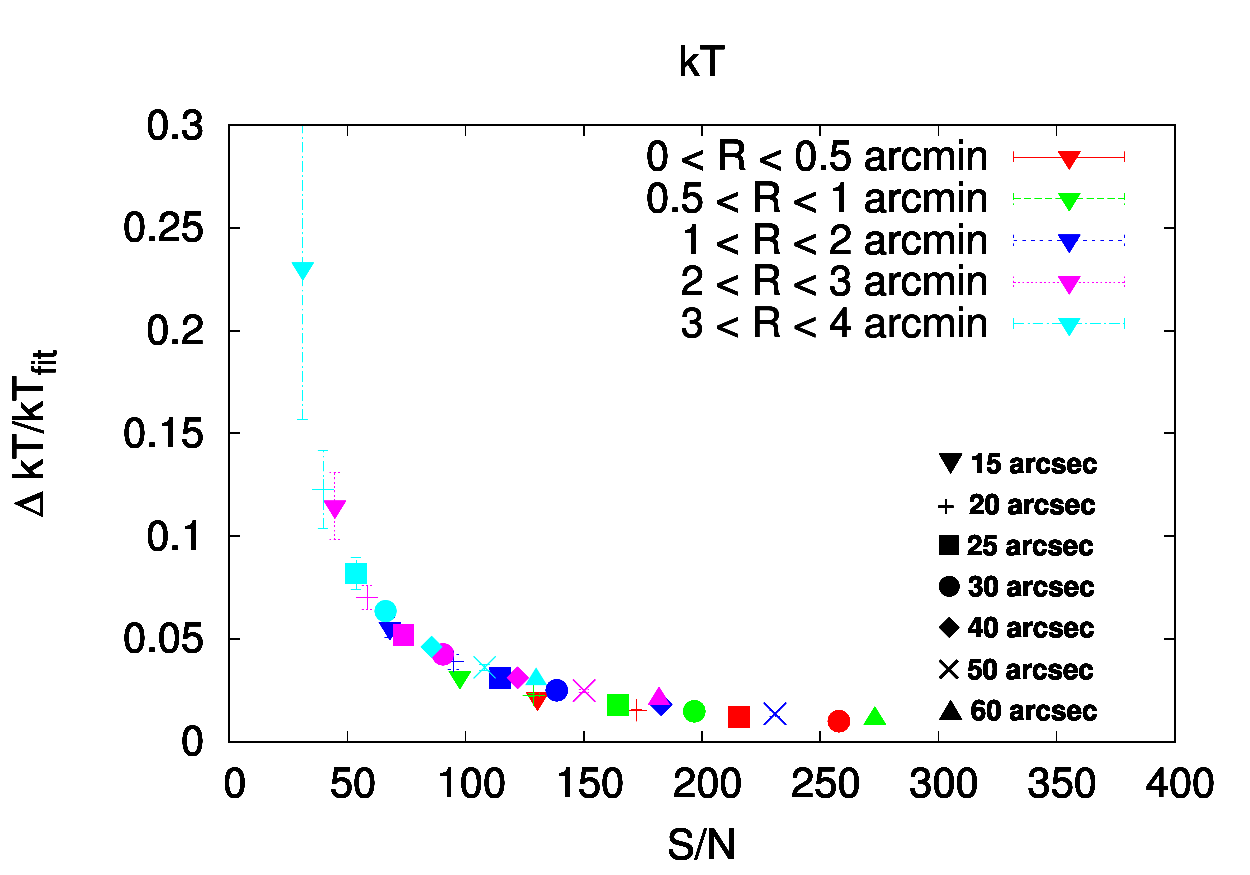}
                \includegraphics[width=0.49\textwidth]{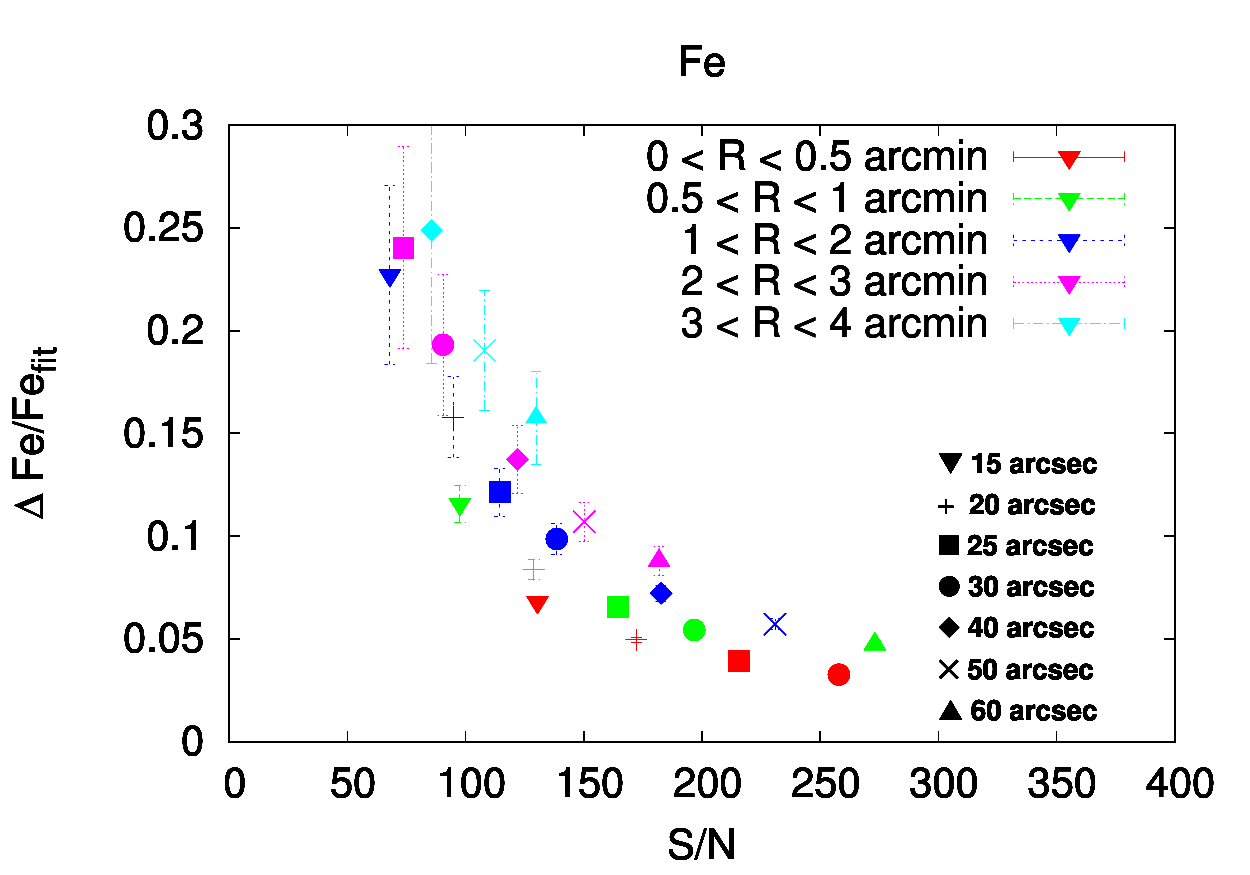}
        \caption{Expected relative errors on the temperatures and abundances. Different cell sizes (symbols) are simulated within the  inner five annuli (colours).}\label{fig:simul-Lorenzo}
\end{figure*}

Finally, we apply and adapt our best background model to the core region (Sect. \ref{sect:core}) and the eight concentric annuli (Sect. \ref{sect:radial_profiles}). The normalisation of every background component has been scaled and corrected for vignetting if necessary. From the background parameters, only the normalisations of the HP component (initially evaluated from the 10--12 keV band, where negligible cluster emission is expected), as well as those of the instrumental fluorescent lines, are kept free for all the spectra.  In the outermost annulus (9$'$--12$'$) we ignore the channels below 0.4 keV (MOS) and 0.6 keV (pn) to avoid low energy instrumental noise. For the same reason we ignore the channels below 0.4 keV (MOS) and 0.5 keV (pn) in the second outermost annulus (6$'$--9$'$). The background is also applied to and adapted for the analysis of the spectra of each map cell (Sect. \ref{sect:maps}).

\section{$S/N$ requirement for the maps}\label{sect:maps_simulations}

Despite their good statistics, we want to optimise the use of our data and find the best compromise between the required spatial resolution of our maps (Sect. \ref{sect:maps}) and $S/N$. The former is necessary when searching for inhomogeneities and $kT$/metal clumps (i.e. the smaller the better), the latter to ensure that the associate error bars are small enough to make our measurement significant. Clearly, these variables depend on the properties of the cluster and on the exposure time of our observations.

We perform a set of simulations to determine what the best combination of $S/N$ and spatial resolution is for the case of A\,4059. For every annulus (i.e. the ones determined in Sect. \ref{sect:radial_profiles}) we simulate a spectrum with input parameters (i.e. $kT$, O, Ne, Mg, Si, S, Ca, Fe, Ni, and the normalisation) corresponding to the ones determined in the radial profiles analysis. The AXB and the HP background are added to the total spectrum by using the properties derived in Appendix \ref{sect:bg_modelling}. We allow their respective normalisations to vary within $\pm 3$\% in order to take into account spatial variations on the FoV. Starting from the value we derived for the radial profile, we rescale the normalisation of the simulated spectrum to the particular spatial resolutions we are interested in (here we test 15$''$, 20$''$, 25$''$, 30$''$, 40$''$, 50$'',$ and 60$''$). We then fit the spectrum as done for the real data and for all the annuli and spatial resolutions we calculate the relative errors on the temperature and Fe abundance as a function of $S/N$. The median values of 300 realisations are shown in Fig. \ref{fig:simul-Lorenzo} with their 1$\sigma$ errors.

A $S/N$ of 100 is required to measure the abundance with a relative error lower than $\sim$$20\%$. With this  choice the temperature will be also determined with a very good accuracy, i.e. relative errors always lower than $\sim$$5\%$.

\end{document}